\newcommand{\pa}{\partial}
\newcommand{\be}{\begin{equation}}
\newcommand{\ee}{\end{equation}}
\newcommand{\bea}{\begin{eqnarray}}
\newcommand{\eaa}{\end{eqnarray}}

\newcommand{\nt}{\notag\\}

\newcommand{\cG}{{\cal G}}

\renewcommand{\a}{\alpha}

\newcommand{\da}{{\dot\alpha}}
\newcommand{\db}{{\dot\beta}}
\newcommand{\bl}{{\tilde\lambda}}

\newcommand{\tx}{{\tilde x}}

\renewcommand{\b}{\beta}

\newcommand{\la}{\lambda}

\newcommand{\q}{\theta}
\newcommand{\bq}{\bar\theta}

\newcommand{\ep}{\epsilon}

\newcommand{\cN}{{\cal N}}
\newcommand{\cD}{{\cal D}}
\newcommand{\cO}{{\cal O}}
\newcommand{\cA}{{\cal A}}

\newcommand{\cL}{{\cal L}}
\newcommand{\cT}{{\cal T}}

\newcommand{\tr}{{\rm tr}}

\newcommand{\p}[1]{(\ref{#1})}
\newcommand{\bt}[1]{{\bar t}}
\newcommand{\ts}{\textstyle}

\newcommand{\half}{{\ts \frac{1}{2}}}
\newcommand \vev [1] {\langle{#1}\rangle}

\newcommand \ket [1] {|{#1}\rangle}
\newcommand \bra [1] {\langle {#1}|}
\newcommand\lr[1]{{\left({#1}\right)}}
\newcommand{\ft}[2]{{\textstyle\frac{#1}{#2}}}

\newcommand{\R}{  \widehat{\mathcal{A}} }

\documentclass[12pt,a4paper]{article}
\input epsf
\usepackage{amsmath,amsfonts,epsfig,color,latexsym}
\usepackage{amssymb}
\usepackage[english, USenglish]{babel}
\usepackage{epsfig,psfrag}
\usepackage{a4wide}
\usepackage{rotating}
\usepackage{cite}

\csname @addtoreset\endcsname{equation}{section}
\begin{document}




\thispagestyle{empty}

\null\vskip-52pt \hfill
\begin{minipage}[t]{45mm}
CERN-PH-TH/2011-060
DCPT--11/09\\
  IPhT--T11/036 \\
\end{minipage}

\vskip1.8truecm
\begin{center}
\vskip 0.2truecm {\Large\bf
The super-correlator/super-amplitude duality: Part I}

\vskip 1.8truecm

{\bf    Burkhard Eden$^{a}$, Paul Heslop$^{a}$, Gregory P. Korchemsky$^{b}$, Emery Sokatchev$^{c,d,e}$ \\
}

\vskip 0.6truecm
$^{a}$ {\it  Mathematics department, Durham University, 
Science Laboratories,
 \\
South Rd, Durham DH1 3LE,
United Kingdom \\
 \vskip .2truecm
$^{b}$ Institut de Physique Th\'eorique\,\footnote{Unit\'e de Recherche Associ\'ee au CNRS URA 2306},
CEA Saclay, \\
91191 Gif-sur-Yvette Cedex, France\\
\vskip .2truecm $^{c}$ Physics Department, Theory Unit, CERN,\\ CH -1211, Geneva 23, Switzerland \\
\vskip .2truecm $^{d}$ Institut Universitaire de France, \\103, bd Saint-Michel
F-75005 Paris, France \\
\vskip .2truecm $^{e}$ LAPTH\,\footnote[2]{Laboratoire d'Annecy-le-Vieux de Physique Th\'{e}orique, UMR 5108},   Universit\'{e} de Savoie, CNRS, \\
B.P. 110,  F-74941 Annecy-le-Vieux, France
                       } \\
\end{center}

\vskip 1truecm 
\centerline{\bf Abstract} 
\medskip
\noindent

We extend the recently discovered duality between MHV amplitudes and the
light-cone limit of correlation functions of a particular type of local
scalar operators to generic non-MHV amplitudes in planar $\cN=4$ SYM
theory. We consider the natural generalization of the bosonic correlators to
super-correlators of stress-tensor multiplets and show, in a number of
examples, that their light-cone limit exactly reproduces the square of
the matching super-amplitudes. Our correlators are computed at Born level.
If all of their points form a light-like polygon,
the correlator is dual to the tree-level amplitude.  If a subset of points
are not on the polygon but are integrated over, they become Lagrangian
insertions generating the loop
corrections to the correlator. In this case the duality with amplitudes
holds
at the level of the integrand. We build up the superspace formalism needed
to formulate the duality and present the explicit example of the $n-$point
NMHV tree amplitude as the dual of the lowest nilpotent level in the
correlator. 
\newpage

\thispagestyle{empty}

 {\small \tableofcontents}

\newpage
\setcounter{page}{1}\setcounter{footnote}{0}



\section{Introduction}

One of the most remarkable recent developments in the AdS/CFT
correspondence \cite{Maldacena:1997re} was the
duality between scattering amplitudes in the $\cN=4$ super-Yang-Mills (SYM) theory and Wilson loops. It relates the all-order planar $n-$gluon MHV scattering amplitude, depending on the particle light-like momenta $p_i$ (with $p^2_i=0$ and $\sum_{i=1}^n p_i=0$), to a Wilson loop defined on a light-like polygonal contour. The latter is obtained by the so-called T-duality transformation from  momenta to dual coordinates:
\begin{equation}\label{moco}
    p_i = x_i - x_{i+1} \equiv x_{i,i+1}\,, \qquad x^2_{i,i+1}=0\,, \qquad x_{i+n} \equiv x_i\,.
\end{equation}
The Wilson loop contour has its cusps at the points $x_i$ and has light-like sides $[x_i, x_{i+1}]$. The duality was first { observed} at strong coupling
\cite{am07} and soon afterwards also at weak coupling 
\cite{Drummond:2007aua,Brandhuber:2007yx,Drummond:2007cf,Drummond:2007au,Drummond:2008aq,Bern:2008ap}. 

{This duality revealed an important hidden symmetry of the MHV  
amplitudes of dynamical origin, the so-called dual conformal symmetry
\cite{Drummond:2006rz,Drummond:2007aua}. The new  
symmetry of the MHV amplitudes can be understood through their duality with light-like Wilson loops, which have a conventional conformal symmetry acting in Minkowski space-time. 

Recently, the MHV amplitudes/Wilson loops duality has been promoted to a triality relation by bringing into consideration bosonic correlation functions of scalar operators, defined in the special limit in which the adjacent operators become light-like separated. As was
shown in \cite{AEKMS}, the asymptotic behavior of the correlator in
this limit is controlled by the light-like Wilson loop squared. Moreover, in \cite{Eden:2010zz, Eden:2010ce} it was argued that the integrand
defining the loop corrections to the correlator coincides with the
four-dimensional integrand of the MHV amplitude.

Detailed studies of the super-amplitudes describing the
scattering of all kinds of particles (gluons, gluinos and scalars) in
the $\cN=4$ SYM theory \cite{Drummond:2008vq}, and of the string sigma model \cite{BM} led to the discovery of a larger, dual {\it superconformal} symmetry. { This symmetry was proven at tree-level in~\cite{Brandhuber:2008pf}. It is broken beyond tree-level~\cite{Korchemsky:2009hm,Bargheer:2009qu}  as will be discussed in more detail later.}

Unlike the dual conformal symmetry of the MHV amplitudes, this new symmetry of the super-amplitudes could not be traced back to the light-like Wilson loop, which is a purely bosonic object. 

It is natural to ask if some kind of duality or even triality relations exist also for the {\it super}-amplitudes, i.e. the full collection of MHV, NMHV, etc. amplitudes, and what are the
corresponding supersymmetric objects in the dual field theory?  In two
recent papers \cite{Mason:2010yk,Simon} the supersymmetric extension of the bosonic light-like Wilson loop was considered as a possible dual model for super-amplitudes.

In this paper and in the twin paper \cite{twin} we propose the
supersymmetrization of the recently discovered duality
between MHV amplitudes and the light-cone limit of correlation
functions of a particular type of local scalar operators  \cite{Eden:2010zz,
 Eden:2010ce}. These scalar operators are the bottom component of
the so-called $\cN=4$ stress-tensor multiplet. We consider the natural
extension of the bosonic correlators to super-correlators of
stress-tensor multiplets and show, in a number of examples, that their light-cone limit exactly reproduces the matching super-amplitudes. Together these results lead to the conjecture that the integrands of the all-loop amplitudes in planar $\cN=4$ SYM  are described by the correlators of stress-tensor multiplets.

In the rest of the introduction we first recall some key points of the proposal  of Refs.~\cite{Eden:2010zz,Eden:2010ce} and then we give a summary of our new proposal for the supersymmetric case.}

\subsection{The bosonic correlators/MHV amplitudes duality}

{The scalar operators used in   \cite{Eden:2010zz, Eden:2010ce} belong to the class of 1/2 BPS (or
 ``short")  operators $\cO_k$ (for reviews see, e.g., \cite{Andrianopoli:1998jh}). These are $k-$linear gauge invariant composites made of the six real scalars of the $\cN=4$ SYM theory and transforming under the representation $[0k0]$ of $SU(4)$.}  In perturbation theory such operators do not undergo renormalization and thus their conformal dimension $d=k$ is protected to all orders. Among them, the simplest example of the bilinear ($k=2$) operator plays a very special role. Its supersymmetric completion is the stress-tensor multiplet \cite{Howe:1981qj}, which includes the stress-tensor and the other conserved currents, as well as the Lagrangian of the $\cN=4$ SYM theory \cite{Eden:1999gh}. In the context of the AdS/CFT correspondence, the 1/2 BPS operators are dual to massive Kaluza-Klein modes in the compactification of type IIB supergravity on an AdS${}_5 \times S^5$ background \cite{Maldacena:1997re}.
{The correlators of 1/2 BPS operators were the subject of numerous studies in the early days of the AdS/CFT correspondence \cite{Penati:1999ba,D'Hoker:1998tz,Lee:1998bxa,Howe:1998zi,Eden:1998hh,GonzalezRey:1998tk,Eden:2000mv,Bianchi:2000hn,Dolan:2000ut}. Of course, no direct comparison between the weakly coupled perturbative results for such correlators and their  strongly coupled duals, the amplitudes of AdS supergravity KK modes, was conceivable. These early studies were limited mainly to qualitative, rather than quantitative results on the correlation functions.

The proposal of  \cite{Eden:2010zz, Eden:2010ce} establishes a duality of a new type, a weak-to-weak coupling relation between the light-cone limit of the correlators of 1/2 BPS operators, on the one hand, and the integrands of the MHV  gluon scattering amplitudes, on the other. {Consider, for instance,  the correlation function of $n$ protected operators of the simplest type $\cO \equiv \cO_2$:}\footnote{The notation $G_{n;0}$ refers to the fact that this correlator is the ``ground state" in a whole supermultiplet of bosonic and fermionic correlators discussed in the present paper.}
\begin{equation}\label{nO}
   G_{n;0}= \vev{\cO(x_1)\cO(x_2)\dots \cO(x_n)}\,.
\end{equation}
As long as we maintain the points $x_i$ (with $i=1,\ldots,n$) in generic positions, this function is well defined and has conformal symmetry. {As a consequence, it is given
by a product of free scalar propagators times some (coupling dependent) function of conformal cross-ratios $x_{ij}^2 x_{kl}^2/(x_{ik}^2x_{jl}^2)$. In perturbation theory this function is expressed in terms of conformally invariant space-time loop integrals.}
Now, consider the limit in which the neighboring points become light-like separated, $x^2_{i,i+1}\to0$, according to \p{moco}. 
The correlator $G_{n;0}$ becomes singular in this limit.
Firstly, at tree level $G^{(0)}_{n;0}$ has poles due to the
propagators $1/x^2_{i,i+1}$. Secondly, the loop integrals develop logarithmic light-cone
divergences $\sim \ln x_{i,i+1}^2$. To deal with
the first problem, it is sufficient to {divide by the
 connected tree-level correlator,} $G_{n;0}/G^{(0)}_{n;0}$. The second problem requires an appropriate regularization.

Two possible regularizations {are:  (i) use  $x^2_{i,i+1}\neq0$ as a cutoff;  (ii) employ  dimensional regularization
and set $x^2_{i,i+1}=0$.} {These two options
were  considered in \cite{AEKMS}, where it was shown that
in both cases the correlation function reduces to a Wilson loop {squared},
\begin{equation}\label{cowll}
    \lim_{x^2_{i,i+1} \to 0} G_{n;0}/G^{(0)}_{n;0}  \propto \left(W[C_n]\right)^2\,.
\end{equation}
{The exact value of the normalization factor in the right-hand side of this
relation depends on the regularization.}
Since $W[C_n]$ is dual to the {planar} MHV gluon amplitude, {we expect that the
ratio of correlators in \p{cowll} is also related to the ratio of amplitudes, $\cA^{\rm MHV}_n/\cA^{\rm MHV\, tree}_n$. }

Another,  more direct way of establishing the relation between correlators and amplitudes, without invoking Wilson loops, was proposed in \cite{Eden:2010zz}. One starts by computing the loop corrections to the correlator by means of Lagrangian insertions. For instance, the one-loop correction
\begin{equation}
   g^2 \frac{\pa}{\pa g^2} G_{n;0} = -i\int d^4 x_0 \,  \vev{ \cL(x_0) \cO(x_1)\ldots {\cal O}(x_n)}^{(0)} \label{g2'}
\end{equation}
is calculated from the {\it Born-level}  $(n+1)$-point correlator with the Lagrangian inserted at the new point $x_0$. This new correlator stays well defined (if divided by the tree $G^{(0)}_{n;0}$) in $D=4$ dimensions, even if we put the outer points $x_i$ on the light cone, while keeping the insertion point $x_0$ in a generic position. The logarithmic singularities originate from the integration over the insertion point in \p{g2'} {over the region where the Lagrangian and the scalar operators $\cO(x_i)$ become light-like separated, $x_{0i}^2=0$}. 

The new proposal of  \cite{Eden:2010zz} was to compare the {\it integrand} in \p{g2'} to that of the momentum loop integrals in the amplitude $\cA^{\rm MHV}_n$, rewritten in the dual space \p{moco}. The claim is that
\begin{equation}\label{1.7}
    \lim_{x^2_{i,i+1} \to 0} G_{n;0}/G^{(0)}_{n;0}  = \left(\cA^{\rm MHV}_n/\cA^{\rm MHV\, tree}_n\right)^2\,,
\end{equation}
in the sense of comparing the integrands on both sides, {evaluated in {\it four dimensions}, that is, after removing the regularization}. The crucial point is that this duality requires no regularization, the integrands of both objects are finite rational functions. 
In \cite{Eden:2010zz, Eden:2010ce} a number of examples at one and two loops were considered. Remarkably, the integrands of the correlators and of the amplitudes coincided exactly, including the parity-odd part found in the Grassmannian approach to amplitudes \cite{ArkaniHamed:2010kv,Boels:2010nw}.

Thus, the correlators of 1/2 BPS operators emerge in a new role. In the light-cone limit they either become light-like Wilson loops, or they are dual to MHV gluon amplitudes. Since the latter are also dual to the former, we may consider the ``triality" relation:
\begin{align}\label{diagram}
\parbox[c]{50mm}{
\psfrag{1}[cc][cc]{$\scriptstyle p_i=x_{i,i+1}$ }
\psfrag{2}[cc][cc]{$\scriptstyle x_{i,i+1}^2\to 0$ }
\psfrag{3}[lc][cc]{$\scriptstyle C_n=p_1\cup\ldots\cup p_n$ }
\psfrag{A}[cc][cc]{$G_n(x_i)$}
\psfrag{B}[cc][cc]{$W[C_n]$}
\psfrag{C}[cc][cc]{$\mathcal{A}^{\rm MHV}_n(p_i)$}
{ \includegraphics[width=48mm]{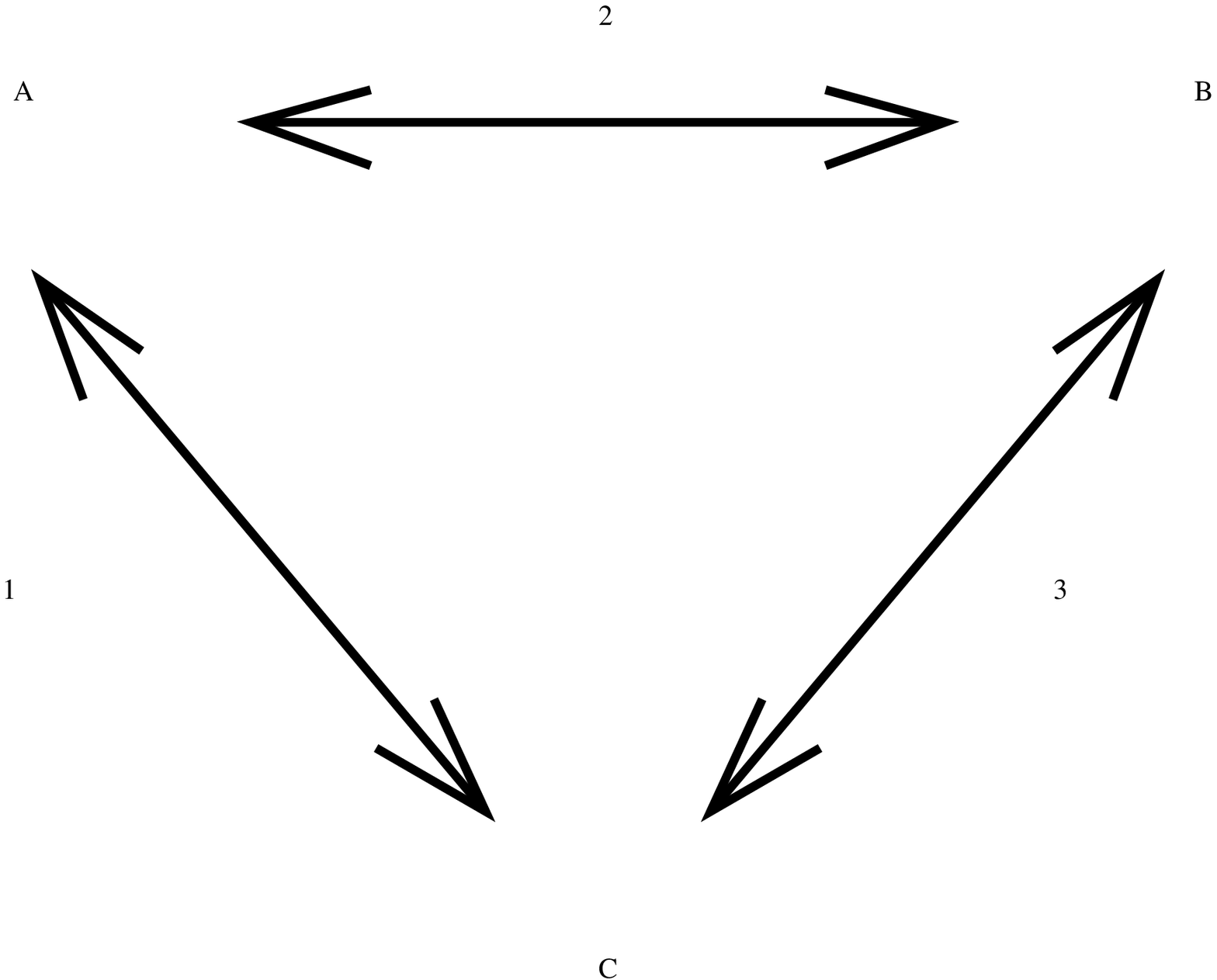} }}
\end{align}
It is important to realize that these relations make use of different
limits of the correlator. The conformally
invariant light-like Wilson loop is defined for
$x^2_{i,i+1}=0$, hence it suffers from cusp
singularities~\cite{P80}. This has the
effect of creating a conformal (or dual conformal, for amplitudes)  anomaly
\cite{Drummond:2007au,Drummond:2007bm}. Thus, the link Correlator
$\leftrightarrow$ Wilson loop relates two divergent objects with broken conformal
symmetry. On the other hand, the link Correlator $\leftrightarrow$ Amplitude involves only rational functions, of the space-time points for the tree-level correlator with insertions, or of the momenta for the integrand of the amplitude. These functions are calculated in four dimensions, without any regulator. In this relation (dual) conformal symmetry remains exact, both in the perturbative computation of the correlators and in the recursive construction of the Grassmannian approach. The third link in \p{diagram}, Wilson loop $\leftrightarrow$ Amplitude, is the least direct one. The two objects are computed in terms of rather different Feynman integrals, so one can only compare their $\ep-$expansions \cite{Drummond:2007aua,Brandhuber:2007yx,Drummond:2007cf,Drummond:2007au,Drummond:2008aq,Bern:2008ap}.\footnote{In the recent papers  \cite{Mason:2010yk,Simon} it is proposed how to establish a direct link between MHV amplitudes and bosonic Wilson loops.}

\subsection{New proposal: Super-correlators/super-amplitudes duality }

The discussion up to this point concerned purely bosonic objects: correlators of scalar operators, Wilson loops and gluon scattering amplitudes, although for all of them the context is that of the maximally supersymmetric $\cN=4$ SYM theory. It is  natural to inquire about the possible supersymmetric extension of the triality relation  \p{diagram}.

The correlators in   \p{diagram} have an obvious
supersymmetrization. The bilinear scalar operator  $\cO=\tr(\phi^2)$ is the
bottom component of a particular 1/2-BPS short multiplet, the stress-tensor
multiplet $\cT$ which contains the
Lagrangian of the theory. This fact will be
very important in view of {the} Lagrangian insertion procedure \p{g2'}. Upgrading $\cO$ and their
correlators \p{nO} to 1/2-BPS short superfields, we can obtain {\it super-correlators}
\begin{align}\label{1.8}
G_n = \vev{\cT(x_1, \q_1,\bq_1) \ldots \cT(x_n,\q_n,\bq_n)} \,,
\end{align}
where $x_i, \q_i, \bq_i$ are points in $\cN=4$
superspace.\footnote{Note that $\cT$ depends on both the chiral $\q$
  and anti-chiral $\bq$ odd superspace coordinates.} The
super-operator $\cT$ is protected, {and} the super-correlators \p{1.8} enjoy the superconformal symmetry $PSU(2,2|4)$. 

The second object in \p{diagram}, the MHV gluon amplitude is part of a {\it super-amplitude}, comprising all types (MHV, NMHV, etc.) of amplitudes of all particles (gluons, gluinos, scalars) of $\cN=4$ SYM. They are described in Nair's on-shell {\it chiral} superspace \cite{Nair:1988bq}. In addition to on-shell supersymmetry, they have a hidden {\it dual superconformal symmetry}  \cite{Drummond:2008vq,BM}. It is made manifest by completing the bosonic T-duality \p{moco} with its fermionic analog $q_{i\, \a} = \q_{i\, \a} - \q_{i+1\, \a}$,
where $q_{i\, \a}$ are the on-shell super-charges and $\q_{i\, \a}$ are the {\it chiral} coordinates of dual superspace. One can show that all tree-level super-amplitudes enjoy the dual superconformal symmetry $PSU(2,2|4)$ acting in the dual superspace \cite{Brandhuber:2008pf,Drummond:2008cr}. At loop level, the super-amplitude divided by the MHV  amplitude {is expected to} conserve its dual conformal symmetry~\footnote{Although this has so far been proven in general at one-loop only~\cite{Brandhuber:2009kh} and verified to two loops by 
explicit calculation of six-particle NMHV amplitude \cite{Kosower:2010yk}.}, while the anti-chiral half of dual Poincar\'e supersymmetry is broken \cite{Korchemsky:2009hm,Bargheer:2009qu}. 

Finally, the third object in \p{diagram}, the light-like Wilson loop can be supersymmetrized by replacing the one-form $dA= dx^\mu A_\mu(x)$ by a super-one-form \cite{Ooguri:2000ps} and by turning the light-like polygonal contour into  a super-contour. 

Now, what happens to the triality relation \p{diagram} if supersymmetry is turned on? Are super-correlators and super-Wilson loops dual to super-amplitudes? In trying to answer this question, we encounter an obvious obstacle. The dual superspace formulation of amplitudes is chiral, reflecting the type of on-shell superspace.  At the same time, the 1/2-BPS conditions on the operator $\cT$ involve half of the chiral {\it and} anti-chiral supercharges, so these objects are {\it not chiral}. Similarly, in chiral superspace there exists no invariant interval needed to replace $dx^\mu$ in the Wilson loop one-form.  However, recalling that at loop level the anti-chiral half of the dual supersymmetry of the amplitudes is broken, one is led to  ``sacrifice" half of the natural full supersymmetry of the super-correlator or of the super-Wilson loop. In other words, these objects may be restricted to their purely chiral sector. If such dualities exist, the full dual supersymmetry of the tree-level super-amplitudes should come as a ``bonus symmetry" of the chiral correlator or Wilson loop.  

In this and in the twin paper \cite{twin} we chose to study
the first of these dualities. We consider the
``half-supersymmetric" (chiral) sector of the super-correlator of
stress-tensor multiplets \p{1.8}. Such correlators with $n\geq 5$
points have a purely bosonic and a nilpotent sectors, the latter
being given by powers $(\q)^{4k}$ (with $k=1,\ldots,n-4$) of the odd
variables. This Grassmann structure is very similar to that of the
super-amplitude rewritten in momentum super-twistor space \cite{hodges,ArkaniHamed:2009dn,Mason:2009qx}. We
claim that in the light-cone limit such super-correlators are  dual to
super-amplitudes, as the direct supersymmetric generalization of the
bosonic duality \p{1.7}:
\begin{equation}\label{1.11}
    \lim_{x^2_{i,i+1} \to 0} G_{n}/G^{(0)}_{n;0}  = \left(\cA_n/\cA^{\rm MHV\, tree}_n\right)^2\,,
\end{equation}
{after the appropriate identification of the variables on both sides.}
In this new relation $G_n$ is the $n-$point super-correlator \p{1.8}, restricted to its chiral sector (all $\bq_i=0$),  and $\cA_n$ is the $n-$particle super-amplitude. 
We consider a number of examples of nilpotent correlators and
demonstrate that they exactly match the corresponding non-MHV
super-amplitudes. In this paper we show that the $(\q)^4$ term of the
$n-$point tree-level correlator is dual to the tree-level NMHV
$n$-point super-amplitude. In the twin paper \cite{twin}  we study the
levels $(\q)^4$ and $(\q)^8$ in the 5- and 6-point correlator, {and $(\q)^4$ in the 7-point correlator,} which are dual to the one- and two-loop MHV${}_4$,  tree-level and one-loop NMHV${}_5$ and NMHV${}_6$, and tree-level NNMHV${}_6$ super-amplitudes. The examples in \cite{twin} show an interesting aspect of the duality: The Lagrangian insertion $\cL(x_0)$ can either generate a new nilpotent $(n+1)$-point correlator at tree level, if the point $x_0$ is part of the light-like polygon; or it can generate loop corrections to the $n-$point correlator (see  \p{g2'}), if  the point $x_0$ is integrated over. 

Finally, what  about the supersymmetrization of the third
object in \p{diagram}, the light-like Wilson loop? This issue has been
addressed in two recent papers. Mason and Skinner \cite{Mason:2010yk}  propose two
scenarios for (half-)supersymmetric Wilson loops. The first uses
twistor space and a twistor version of the $\cN=4$ SYM
Lagrangian. They derive twistor space Feynman rules which yield the MHV rules for super-amplitudes  {(see also \cite{Bullimore:2011ni}). We would like to make the comment that one still needs
to understand the intricate details of the transition from twistor
space back to Minkowski space. In particular, the inverse twistor
transform is supposed to reproduce the inherent cusp singularities of the light-like
Wilson loop, which would require some kind of regulator (dimensional
regularization does not apply to four-dimensional twistors). This important issue requires further investigation}.} Mason and
Skinner also make the first step towards a (half-)supersymmetrization
of the conventional Wilson loop in Minkowski space. The latter
approach is pursued in more detail by Caron-Huot \cite{Simon}, who
gives a general argument why such an object should satisfy the
recursion relations for super-amplitudes recently proposed in
\cite{ArkaniHamed:2010kv,Boels:2010nw}. However, in the absence of
explicit examples how to actually compute  this type of Wilson loop,
one has to remain very cautious  as to the possible unexpected effects
of the cusp singularities.\footnote{ Very recently, in \cite{Belitsky:2011zm} an anomaly mechanism for the chiral half of Poincar\'e supersymmetry  and for the anti-chiral half of the conformal supersymmetry  of the light-like super Wilson loop  was discovered. This indicates that the conventional super Wilson loop
in Minkowski space of Refs.~\cite{Mason:2010yk,Simon}, if carefully treated at the quantum level,  fails to meet the expectations of being dual to super-amplitudes.} In contrast with the Wilson loop,
the correlator/amplitude duality that was proposed {for the bosonic case} in
\cite{Eden:2010zz, Eden:2010ce} and is {extended to the full super-amplitudes in this paper and in the twin paper \cite{twin}},
involves only {\it finite} objects (rational functions) in four
dimensions. 

The paper is organized as follows. Section~\ref{sec:duality} serves as an extended introduction to both this paper and to \cite{twin}. We give a brief overview of  the super-amplitudes and the correlators and formulate the new duality without entering into too much detail. This section will help the reader find his/her way through the formalism gradually built up in the following sections. Section \ref{N4has} describes the formulation of the $\cN=4$ vector and stress-tensor multiplets as 1/2-BPS short multiplets. There we give the vital minimum of information about $\cN=4$ harmonic/analytic superspace, an indispensable tool for the formulation of the new duality. Particular attention is paid to the chiral on-shell Lagrangian as a member of the stress-tensor multiplet. In Section~\ref{cn4stm} we discuss the general structure of the correlators of such multiplets and explain the presence of nilpotent terms in them. We also review the Lagrangian insertion procedure as a way of generating loop corrections to the correlators. Section~\ref{NMtc} contains our first detailed example of the new duality in action. We evaluate the residues at the physical poles of the $n-$particle tree-level NMHV super-amplitudes  and of the tree-level correlator at Grassmann level $(\q)^4$ and demonstrate that they match exactly, thus proving the duality between the two objects. Our further examples are presented in \cite{twin}. Section~\ref{Conclu} formulates some open problems  and lines for future development. Our conventions, definitions and some technical details are collected in three appendices. 
 
\section{The  correlators/scattering amplitudes duality}
\label{sec:duality}

\subsection{Scattering super-amplitudes}

A distinctive feature of $\cN=4$ SYM theory is that all the on-shell asymptotic
states (gluons with helicity $\pm 1$, gluinos with helicity $\pm 1/2$ and six scalars)
can be combined into a single superstate. Then, all $n-$particle scattering amplitudes in the theory are described by a single super-amplitude
$\cA_n=\cA_n(\lambda_i,\bl_i,\eta_i)$ depending on  the particle light-like momenta, $p^{\da\a}_i =  \bl_{i}^{\da}\lambda_{i}^{\a}$ (with $i=1,\ldots,n$), and on the odd chiral variables $\eta^A_i$ with an $SU(4)$ index $A=1,2,3,4$. The super-amplitude can be decomposed 
into a sum of terms corresponding to the possible values of the total helicity of the $n$ scattered particles,
\begin{align}
\cA_n =  \cA_n^{\rm MHV} +  \cA_n^{\rm NMHV}+\ldots+ \cA_n^{\rm N^{n-4}MHV}\,,
\end{align}
where $\cA_n^{\rm N^{k}MHV}$ describes all $n-$particle scattering amplitudes of total helicity $4-n+2k$ (with $k=0,\ldots,n-4$).
The top term $\cA_n^{\rm N^{n-4}MHV}$ corresponds to the $\overline{\rm MHV}$ amplitude (the PCT conjugate to the MHV amplitude), written in the chiral on-shell superspace.

 In planar $\cN=4$ SYM the super-amplitudes have the following general form
\begin{align}\label{supa}
\cA_n^{\rm N^{k}MHV} =  i (2\pi)^4\frac{\delta^{(4)}(p^{\da\a })\
    \delta^{(8)} (q^{\a A})}{\vev{1\, 2}\vev{2\, 3}\ldots\vev{n\, 1}}\, {\R}_{n;k}(\la,\bl,\eta; a)\,,
\end{align} 
where $p^{\da\a}=\sum_{i=1}^n p_i^{\da\a}$ and $q^{\a A}=\sum_{i=1}^n\lambda_{i}^\a\, \eta_i^A$ are the total momentum and supercharge of the $n$ particles.
Here the delta functions  encode momentum and supercharge conservation,
$p\,\cA_n^{\rm N^{k}MHV} =q \,\cA_n^{\rm N^{k}MHV} =0$, while the product of factors $\vev{i,i+1}\equiv \lambda_i^\a\lambda_{i+1,\a}$ in the denominator gives the correct helicity  weight to each partial amplitude in the $\eta$ expansion.  The functions ${\R}_{n;k}(\la,\bl,\eta; a)$ are homogeneous polynomials of degree $4k$ in the odd variables $\eta_i^A$. They are invariant under helicity rescalings,
$\lambda_i \to h_i \lambda_i\,, \ \bl_i \to h_i^{-1} \bl_i\,, \ \eta_i \to h_i^{-1} \eta_i$, 
and admit a perturbative expansion in powers of the `t Hooft coupling,
\begin{align}\label{2.3}
 {\R}_{n;k} = \sum_{\ell\ge 0} a^\ell {\R}_{n;k}^{(\ell)}(\la,\bl,\eta)\,,
 \qquad  a=\frac{g^2 N_c}{4\pi^2}\,.
\end{align}
At tree level,   one has ${\R}_{n;0}^{(0)}=1$ for the MHV amplitude ($k=0$),
while   for N${}^k$MHV amplitudes  ${\R}_{n;k}^{(0)}$ are non-trivial {rational functions of the external momenta}.

The tree-level super-amplitudes are well defined in $D=4$ dimensions and, as a consequence, they inherit the superconformal symmetry of the $\cN=4$ SYM Lagrangian \cite{Witten:2003nn}. At loop level, $\cA_n^{\rm N^{k}MHV}$ suffer from infrared divergences. They break  the superconformal ($k$, $s$ and $\bar s$) symmetry but preserve Poincar\'e supersymmetry. 
In addition to this, the super-amplitudes \p{supa} possess yet another, hidden symmetry, the so-called dual superconformal symmetry \cite{Drummond:2008vq}. It becomes manifest after rewriting the super-amplitude in the dual chiral superspace with  coordinates
\begin{align}
& x^{\da\a}_i - x^{\da\a}_{i+1} = \bl_{i}^{\da}\, \lambda_{i}^{\a}  \,,\qquad  
\q^{\a\, A}_i -  \q^{\a\, A}_{i+1} = \lambda_{i}^{\a}\, \eta^A_i\,, \label{duth}
\end{align}
with the identification $x_{n+1} \equiv x_1$ and $\q_{n+1} \equiv \q_1$.  Alternatively, we may say that the dual variables satisfy the light-cone conditions
\begin{align}\label{2.5}
x^2_{i,i+1}=0\,,\qquad \q_{i, i+1}^{A \a}(x_{i,i+1})_{\a\da} = 0\,,
\end{align}
whose solution is given by \p{duth}.   We remark that the fermionic constraint in \p{2.5} is the superpartner of the bosonic one under the action of the anti-chiral generator of dual Poincar\'e supersymmetry: $\bar Q^A_\da x^2_{i,i+1} = 2i \q_{i, i+1}^{A \a}(x_{i,i+1})_{\a\da}$. We will call the set of points in superspace satisfying the light-cone conditions  \p{2.5} a {\it light-like super-polygon}.  

The functions ${\R}_{n;k}(\la,\bl,\eta; a)$ introduced in \p{supa} have zero helicity weight, therefore they can be rewritten entirely  as functions on the dual superspace $(x,\q)$, without any manifest presence of the on-shell superspace variables $\la,\bl,\eta$:
\begin{align}\label{R-R}
{\R}_{n;k}(\la,\bl,\eta; a) = {\R}_{n;k}(x,\q; a)\,.
\end{align} 

The dual superspace coordinates have simple homogeneous transformation properties under dual superconformal symmetry. In particular, under conformal inversion they transform as follows:\footnote{The conformal weight of $\la$ in \p{2.6} is chosen to fit the momentum super-twistor parametrization of dual superspace. In a real Minkowski space the weight of $\la_i$ depends on both points $i$ and $i+1$ (see \cite{Drummond:2008vq}). }
\begin{align}\label{2.6}
I: \qquad x_i^{\da\a} \ \to\ (x_i^{-1})_{\a\da}\,, \quad \q^{A}_{i\, \a} \ \to \
(x_i^{-1})^{\da\a} \q^{A}_{i\, \a}\,, \quad \ \la_i^\a \to \  \la_{i}^{\a}(x_i)_{\a\da}\,.
\end{align}
It proves convenient to introduce yet another set of dual variables, the so-called momentum supertwistors \cite{hodges,ArkaniHamed:2009dn,Mason:2009qx}:
\begin{align}\label{2.7}
\la^\a_i\,, \qquad \mu_{i \da} = \la_{i}^{\a}(x_i)_{\a\da}\,, \qquad \chi^A_i =  \la_{i}^{\a} \q^{A}_{i\, \a}\,.
\end{align}
Notice that the coordinates of this superspace carry the same helicity weight, i.e., degree of homogeneity under the rescaling of $\la$. The advantage of the parametrization \p{2.7} is that the conformal transformations become linear. In particular, under inversion we have
$\la^\a_i \ \leftrightarrow \ \mu_{i \da}$ and $\chi^A_i  \ \leftrightarrow \ \chi^A_i $.  This allows one to rewrite the super-amplitude \p{R-R} in yet another form, 
\begin{align}\label{lath}
{\R}_{n;k}(x,\q; a)=  {\R}_{n;k}(\la,\mu,\chi;a)\,,
\end{align} 
in which dual conformal symmetry becomes obvious. 
 
{ The full dual {super}conformal symmetry is a property only of the tree-level super-amplitudes. This means that the tree-level functions ${\R}_{n;k}^{(0)}$ are invariant under dual Poincar\'e supersymmetry with generators $Q$ and $\bar Q$ and under
the conformal inversions (\ref{2.6}). This in turn yields invariance under the rest of the dual superconformal  algebra $PSU(2,2|4)$, i.e. the generators of conformal symmetry ($K$) and special conformal supersymmetry ($S$ and $\bar S$).  

At loop level, the dual conformal symmetry of the MHV amplitudes is revealed via the duality with light-like Wilson loops $W_n=\vev{P\exp\lr{i\oint_{C_n} dx\cdot A(x)}}$ \cite{am07,Drummond:2007aua,Brandhuber:2007yx},
\begin{align}
\ln  {\R}_{n;0} = \ln W_n + O(1/N_c) + O(\epsilon)\,,
\end{align}
where the integration contour $C_n$ is a closed light-like polygon in dual space, with sides determined by the particle momenta $p_i= x_{i, i+1}$ and with cusps located at $x_i$. More precisely, the Wilson loop describes the loop corrections to the MHV amplitude, but the tree-level factor in \p{supa} is left out. This is a duality between two  bosonic objects, so neither conventional nor dual supersymmetry manifest themselves. 

It is important to realize that the MHV amplitudes and the light-like Wilson loops are divergent objects, suffering from infrared and ultraviolet singularities, respectively. So, a regularization is needed to make them well defined. The commonly used dimensional regularization takes the theories away from four dimensions, $D=4-2\ep$ (with $\ep<0$ for infrared and $\ep>0$ for ultraviolet divergences). The presence of divergences has another inevitable consequence - the conformal symmetry of the Wilson loop and the matching dual conformal symmetry of the MHV amplitude become anomalous. Due to the universality of the infrared divergences, the dual conformal anomaly cancels in the ratio of the functions ${\R}_{n;k}/{\R}_{n;0}$, turning it into an exact dual conformal invariant to all loops \cite{Drummond:2007au}
\begin{align}
K^{\da\a}  \lr{ {\R}_{n;k}/{\R}_{n;0}} = 0\,.
\end{align}
This ratio function also exhibits exact chiral Poincar\'e ($Q$) and anti-chiral special conformal ($\bar S$) supersymmetry. 
However, the other halves  of these supersymmetries ($\bar Q$ and $S$) are broken beyond tree level. The origin of this symmetry breaking can be traced back to the so-called holomorphic anomaly \cite{Cachazo:2004by,Korchemsky:2009hm,Bargheer:2009qu}. {Unlike the $K-$anomaly,
the corresponding $\bar Q-$ and $S-$anomalies are only known at one-loop level. }

\subsection{The bosonic correlators/MHV amplitudes duality }\label{Dbcma}

{Recently,   an alternative duality, still for MHV amplitudes, was proposed. {This duality has already been mentioned in the introduction, here we provide some details, in order to prepare the ground for the generalization to the supersymmetric case.} It can be formulated in two ways. The first formulation  \cite{AEKMS} compares two divergent objects with anomalous (dual) conformal symmetry,  the light-cone limit of the correlators of protected operators, and the scattering amplitudes. The second formulation \cite{Eden:2010zz,Eden:2010ce} matches  two {\it finite} objects, both of them defined in the (dual) {\it four-dimensional} space-time \p{duth} and having manifest and unbroken (dual) conformal symmetry.  These are the {\it integrands} of the loop corrections to the correlators and to the amplitudes.  }

Consider the $n-$point correlator of gauge invariant scalar operators 
\begin{align}\label{4.6}
\cO^{IJ}= \tr(\phi^I\phi^J) - \frac1{6} \delta^{IJ} \tr(\phi^K\phi^K)\,.
\end{align}
Here $I,J= 1,\ldots,6$ are the $SO(6) \sim SU(4)$ indices of the  six  scalars in the $\cN=4$ SYM theory. The operator \p{4.6} is in the representation $\mathbf{20'}$ of $SU(4)$ and it is the lowest-weight state of the so-called $\cN=4$ stress-tensor multiplet containing, among others, the stress tensor and the Lagrangian of the theory (for a more detailed review see Section~\ref{N4tm}). This multiplet is the simplest representative of the class of the so-called 1/2 BPS multiplets, whose important property is that the scaling dimensions
of the operators in the multiplet are {protected from perturbative corrections.} Consequently, the operator  \p{4.6} has well-defined conformal properties, with fixed conformal weight two. The correlator we are discussing is
\begin{align}\label{2.10}
G_{n;0} = \vev{\cO(x_1) \ldots \cO(x_n)}\equiv \sum_{\ell=0}^\infty a^\ell G_{n;0}^{(\ell)}  \,,
\end{align}
where for the sake of simplicity we do not display the $SO(6)$ indices of the operators.
The first subscript in $G_{n;0}$ refers to the number of points and the second 
subscript `$0$' indicates that this is the lowest term in a whole collections of correlators of bosonic and fermionic composite operators, corresponding to various states in the stress-tensor multiplet (see Section~\ref{cn4stm}).  

In virtue of conformal symmetry, the correlator (\ref{2.10}) is expressed in terms of a product of scalar propagators, $1/(x_{12}^2\ldots x_{n1}^2)$, which gives the correlator the required conformal weight at each point, and 
a weightless function. The latter is decomposed into various $SO(6)$ tensor structures, each accompanied by a non-trivial function of the conformal cross-ratios
$x_{ij}^2 x_{kl}^2/(x_{ik}^2x_{jl}^2)$ and of the coupling constant $a$. 
We wish to compare the MHV amplitude with this correlator in the {\it light-cone limit} $x^2_{i,i+1} \to 0$, that is, in the limit where the operators at adjacent positions become light-like separated. In this limit some of the cross-ratios
vanish and the expression for the correlator simplifies significantly.
Namely, the leading asymptotic behavior of the correlator involves
only one $SO(6)$ structure to all loops. Then, examining the ratio of the correlation
function and its tree-level value, $G_{n;0}/G_{n;0}^{(0)}$, we find that, in the limit
$x_{i,i+1}^2\to 0$, the pole singularities due to the propagator factors cancel  in this ratio and, most importantly, it is an $SO(6)$ singlet. The main claim of  \cite{AEKMS} is that the light-cone limit of this ratio of correlators is related to the function ${\R}_{n;0}$ describing the perturbative corrections to
the MHV amplitude,  as follows:
\begin{align}\label{mhvdu}
\lim_{x^2_{i,i+1} \to 0} \frac{G_{n;0}(x)}{G^{(0)}_{n;0}(x)}  = ({\R}_{n;0}(\la,\bl;a))^2\,.
\end{align}
By definition, ${\R}_{n;0}(\la,\bl;a)$ does not depend on the Grassmann variables $\eta_i$ and carries zero helicity weight for each
particle. As a consequence, it can be expressed as a function of the dual $x-$variables
only, which according to the conjecture (\ref{mhvdu}), is related to the  ratio 
of the correlators in the light-cone limit. 

The quantities on both sides of (\ref{mhvdu}) are divergent and require regularization. For the scattering amplitude, the infrared divergences of ${\R}_{n;0}(\la,\bl;a)$ can be regularized using dimensional regularization. 
Then, the multi-loop corrections to the ratio function $\R_{n;0}$, are given by multiple momentum space, or equivalently, dual space integrals:
\begin{align}\label{R-gen}
{\R}_{n;0}(\la,\bl;a)  = 1+ \sum_{\ell=1}^\infty a^\ell \int \prod_{i=1}^\ell  d^{4-2\ep} x_{0_i} I^{(\ell)}_\epsilon (x_{0_1}, \ldots, x_{0_\ell}; x_1, \ldots, x_n) \,,
\end{align}
with some rational integrand $I^{(\ell)}_\epsilon$ depending on the $n$ external $4-$dimensional points $x_i$ and $\ell$ internal $(4-2\epsilon)-$dimensional
points $x_{0_i}$ (with $\ep<0$). In this expression all external distances $x^2_{i,i+1} = p^2_i=0$ (i.e., all scattered particles are massless), hence the necessity to regularize the integrals. 

For the correlators in the left-hand side of (\ref{mhvdu}), the limit $x_{i,i+1}^2\to 0$ is singular because the multi-loop integrals in the perturbative corrections diverge logarithmically when the external points become light-like separated.  
One possible regularization procedure, considered in \cite{AEKMS}, consists in computing the correlator $G_{n;0}$ in dimensional regularization in $D=4-2\ep$ dimensions (with $\ep>0$), which allows us to set all $x^2_{i,i+1}=0$ from the start. The general expression for the ratio
of correlators defined in this way looks similar to (\ref{R-gen}) (except for the sign of the regulator $\ep$):
\begin{align}\label{G-gen}
\lim_{x^2_{i,i+1} \to 0} \frac{G_{n;0}}{G^{(0)}_{n;0}} =1+ \sum_{\ell=1}^\infty a^\ell \int \prod_{i=1}^\ell  d^{4-2\ep} x_{0_i} G^{(\ell)}_\epsilon (x_{0_1}, \ldots, x_{0_\ell}; x_1, \ldots, x_n) \,,
\end{align}
where the external points $x_i$ refer to the positions of the local operators and the integration points $x_{0_i}$ correspond to the interaction vertices. 
Before taking the light-cone limit, the Feynman integrals contributing to the correlator $G_{n;0}$ are finite and manifestly conformally covariant. However, putting the outer points at light-like separations makes the integrals diverge. Still, the  {\it integrand} in \p{G-gen} remains a finite function, even in the light-cone limit. 
The integrand
$G^{(\ell)}_\epsilon$, computed in dimensional regularization and 
for $x_{i,i+1}^2=0$, is made of interaction vertices and
free propagators that scale as $1/(x^2)^{1-\epsilon}$. Notice that, unlike  the integrand for the amplitude $I^{(\ell)}_\epsilon$ in (\ref{R-gen}), the integrand for the correlator $G^{(\ell)}_\epsilon$ is not a rational function of the $x$'s for $\epsilon\neq 0$. 

The duality relation (\ref{mhvdu}) states that the integrals in the right-hand sides
of (\ref{R-gen}) and (\ref{G-gen}) coincide after the appropriate redefinition of the
regularization parameters.%
\footnote{More precisely, this relation holds between the logarithms of the two quantities,  
up to $O(\epsilon)$ terms:
\begin{align}\label{log}
\lim_{x^2_{i,i+1} \to 0}\lr{ \ln  {G_{n;0}}/{G^{(0)}_{n;0}} } = 2\ln {\R}_{n;0}\,.
\end{align}
Taking the logarithm has the effect of eliminating a number of complicated (parity-odd, $\mu$, etc.) terms from the amplitude \cite{Bern:2006vw}.}
General physical reasons for the duality relation (\ref{mhvdu})
were given in \cite{AEKMS}, explaining why this limit transforms the correlator into a light-like Wilson loop. The latter is known to be dual to the MHV amplitude, hence the duality relation \p{mhvdu}. The square in \p{mhvdu} is a simple consequence of the fact that the fields $\phi$ in \p{4.6}, like all fields in the $\cN=4$ vector multiplet, are in the adjoint representation of the gauge group, while the Wilson loop is in the fundamental.

In \cite{Eden:2010zz,Eden:2010ce} it was proposed to reformulate  
the duality (\ref{mhvdu}) directly in terms of the {\it integrands} in the two perturbative expressions, $I^{(\ell)}_\epsilon$ and $G^{(\ell)}_\epsilon$. They are finite  functions of the external and the integration points, so they need no regularization and have a smooth limit as $\epsilon\to 0$. Then, the duality relation \p{mhvdu} matches the integrand of the correlator with the {\it square} of the integrand of the amplitude. This yields an infinite set of relations between the 
four-dimensional integrands for the correlators and amplitudes
\begin{align}
G^{(1)}_{\epsilon=0} = 2I^{(1)}_{\epsilon=0} \,,\qquad
G^{(2)}_{\epsilon=0} = 2I^{(2)}_{\epsilon=0} + [I^{(1)}_{\epsilon=0} ]^2\,,\quad 
\ldots
\end{align}
In \cite{Eden:2010zz,Eden:2010ce} it was shown  in a number of one- and two-loop examples  that 
these relations hold exactly, both for the parity-even and parity-odd terms. Notice that 
the latter are total derivatives and hence vanish upon integration.%

For the scattering amplitudes, the four-dimensional integrand $I^{(\ell)}_{\epsilon=0}$ can be obtained from the BCFW recursion relations proposed in \cite{ArkaniHamed:2010kv,Boels:2010nw}. The integrand $G^{(\ell)}_{\epsilon=0}$ can be extracted from the
explicit results for the four-dimensional correlator $G_n$, Eq.~(\ref{2.10}),
in the light-cone limit $x_{i,i+1}^2\to 0$. The procedure for 
computing $G^{(\ell)}_{\epsilon=0}$, which was extensively used in \cite{Eden:1999kw,Eden:2000mv,Howe:2000hz}, makes use of the well-known fact that the loop corrections to the correlator \p{2.10} are obtained by making Lagrangian insertions (see Section~\ref{Lip}):
\begin{align}\label{2.13}
G^{(\ell)}_{\epsilon=0} (x_{0_1}, \ldots, x_{0_\ell}; x_1, \ldots, x_n) =
\lim_{x^2_{i,i+1} \to 0}
\frac{\vev{\cL(x_{0_1}) \ldots \cL(x_{0_\ell}) \cO(x_1) \ldots \cO(x_n)}^{(0)}}{\vev{\cO(x_1) \ldots \cO(x_n)}^{(0)}}\,,
\end{align}
where the insertion points $x_{0_i}$ are in general positions, $(x_{0_i}-x_{0_j})^2 \neq 0$, $(x_{0_i}-x_{j})^2\neq 0$. Here  $\cL(x_{0i})$ is the chiral part of the on-shell Lagrangian of $\cN=4$ SYM (see Section \ref{N4tm}) 
and  the superscript `$(0)$' indicates that the correlators in the right-hand side are computed at tree level.%
\footnote{The tree-level $(n+\ell)-$point correlator with $\ell$ Lagrangian insertions is already of order $a^\ell$ in the coupling, thus matching the perturbative level of the left-hand side in \p{2.13}.}  The main advantage of the insertion procedure  \p{2.13} is that the computation of the ratio of the correlators in four dimensions is greatly facilitated by the powerful superconformal symmetry of the  $\cN=4$ theory.  

\subsection{New proposal: the super-correlators/super-amplitudes duality}\label{dscsa}

The scalar operator  $\cO(x)$, Eq.~(\ref{4.6}), and the chiral part of the on-shell  $\cN=4$ Lagrangian  $\cL(x)$ belong to the same 1/2 BPS multiplet of $\cN=4$ supersymmetry, the stress-tensor multiplet.
They appear as coefficients in the expansion of the 1/2 BPS superfield operator $\cT(x,\theta,\bar\theta)$ in powers of the odd variables  (see Section~\ref{N4tm}).
Consequently, the correlator of bosonic operators  $G_{n;0}(x_i)$, Eq.~\p{2.10}, can be boosted to a correlator of superfield operators, that is, to a {\it super-correlator} 
\begin{align}\label{super-cor}
\Gamma_n =\vev{\cT(x_1,\theta_1,\bar\theta_1)\cT(x_2,\theta_2,\bar\theta_2)\ldots \cT(x_n,\theta_n,\bar\theta_n)}
\end{align}
depending on the chiral $\q^A_\a$ and anti-chiral $\bq^\da_A$ odd variables of $\cN=4$ superspace. The bosonic correlator  $G_{n;0}$ is just the bottom component in the  Grassmann expansion of $\Gamma_n$,
\begin{align}\label{}
G_{n;0}(x_i) = \Gamma_n\lr{x_i,\theta_i=0,\bar\theta_i=0}\,.
\end{align}
By construction, the super-correlator $\Gamma_n(x_i,\theta_i,\bar\theta_i)$ is invariant under $\cN=4$ Poincar\'e supersymmetry,
\begin{align}\label{2.20}
Q^\a_A\, \Gamma_n = \bar Q^A_\da\, \Gamma_n = 0\,.
\end{align} 
Moreover, as long as the operators are not light-like separated, $(x_i-x_j)^2\neq 0$, it also enjoys  the full superconformal symmetry $PSU(2,2|4)$.

A natural question arises: What will happen if we apply the limiting procedure from  \p{mhvdu} to the super-correlator $\Gamma_n$ rather than to its lowest component   $G_{n;0}$? Can we expect that it will be dual to the supersymmetric extension of the MHV amplitude, i.e., to the complete super-amplitude \p{supa} rewritten in the dual superspace   \p{duth}? To answer this question we have to look more closely at the structure of the two superspaces, that for the amplitude and that for the correlator.

\subsubsection{Chirality vs Grassmann analyticity}

{ An immediate corollary of the light-cone condition $x^2_{i,i+1} =0$ and the supersymmetry \p{2.20} of the correlator $\Gamma_n$ is the generation of two new conditions on the odd superspace variables:
\begin{align}\label{2.23}
x^2_{i,i+1} =0  \ 
\begin{array}{cc}  
& (x_{i,i+1})_{\a\da}  (\bq_{i,i+1})_A^\da = 0 \,,
\\[-3mm]
{}^{Q}\! {\nearrow} &  
\\[5mm]  
{}_{\bar Q}\! \searrow & 
\\[-3mm] & (x_{i,i+1})^{\da\a}  (\q_{i,i+1})^A_\a = 0\,.
\end{array} 
\end{align}
Compared to \p{2.5}, they define a {\it non-chiral} light-like super-polygon.

On the other hand, }the super-amplitude \p{supa} is by construction a {\it chiral} (holomorphic) object, in the sense that it only involves the odd variables $\eta^A$ in the fundamental irrep of $SU(4)$, but not their conjugates $\bar\eta_A$. This is reflected in the chiral nature of the dual superspace in \p{duth}. Unlike the super-amplitude, the 1/2 BPS supermultiplet $\cT(x,\theta,\bar\theta)$ to which the scalar operator $\cO(x)$ belongs, and hence the super-correlator $\Gamma_n(x_i,\theta_i,\bar\theta_i)$, are {\it not chiral}, but are {\it Grassmann analytic} objects \cite{Howe:1995aq,Howe:1996rk}. This means that they involve half  of the odd variables (hence the name ``1/2 BPS"), but both of the chiral and anti-chiral types. 

Postponing  the detailed discussion of Grassmann analytic superspace to Section~\ref{N4has}, here we can make the following rather suggestive analogy with momentum super-twistor space $(\lambda^\a,\mu_\da,\chi^A)$, Eq.~\p{2.7}. The latter is obtained from the chiral superspace with coordinates $(x_{\a\da}, \q^A_\a)$ by introducing an auxiliary variable, the commuting spinor $\la_\a$, and by projecting with it both coordinates, $\mu_\da =\lambda^\a x_{\a\da}$ and $\chi^A=\lambda^\a \theta^A_\a$. As a result, all three coordinates in momentum super-twistor space $(\la,\mu,\chi)$ carry the same helicity weight and the only coordinate with an undotted Lorentz index is $\la^\a$. We may say that the role of $\la$ is to convert the undotted  $SL(2,\mathbb{C})$ indices into helicities, i.e. into weights of the ``little group"  $GL(1) \subset SL(2)$. Then, the non-trivial fact is that the super-amplitudes \p{supa} (more precisely, the ratios $\hat\cA$) can be rewritten only in terms of  the projected variables, see \p{lath}.

The main idea behind analytic superspace is rather similar, but this time we project the $SU(4)$ index of $\q^A_\a$, rather than its Lorentz index. To this end,
we introduce a set of auxiliary commuting variables $u^{+a}_A$ (called ``harmonic variables"  \cite{hh}). Apart from the $SU(4)$ index $A$,
they carry an index $a=1,2$ and charge `$+$' of the ``little group" $U(2) \subset SU(4)$. Making use of the harmonics, we define projected variables
$
\q^{+a}_\a = \q^A_\a u^{+a}_A\,,
$
 thus converting the $SU(4)$ index into a $U(2)$ index. 
In addition to the chiral  $\q^{+a}_\a$, we define a similar projection of the {\it anti-chiral} odd variable, $\bq^\da_{-a'} = \bq^\da_A \bar u^A_{-a'}$, obtained with a different harmonic variable carrying an $SU(2)'$ index. 
Then, the 1/2 BPS shortening conditions on  the stress-tensor multiplet can be translated into the simple, but very non-trivial statement that the corresponding superfield  depends only on the projected odd variables, $\cT=\cT(x,\q^{+},\bq_{-},u)$, 
but not on their conjugates. In this sense $\cT$ is {\it Grassmann analytic}.\footnote{The notion of Grassmann analyticity in extended supersymmetry, as opposed to chirality, was first introduced in the context of the on-shell $\cN=2$ hypermultiplet in \cite{Galperin:1980fg}. Its off-shell version, employing $SU(2)$ harmonic variables for the first time, was given in \cite{hh}. The same ideas made it possible to formulate the $\cN=3$ SYM theory off shell, this time using $SU(3)$ harmonics \cite{Galperin:1984bu}. The generalization of the notion of analytic superspace to $\cN=4$ SYM was first proposed in \cite{Howe:1995aq}.}
Thus,  the super-correlator \p{super-cor} is defined on the analytic superspace with coordinates $(x,\q^{+},\bq_{-},u)$,
\begin{align}\label{2.19}
\Gamma_n = \vev{\cT(1) \ldots \cT(n)} = \Gamma_n(x_i, \q^+_i, \bq_{-\, i}, u_i)\,.
\end{align}
Unlike momentum super-twistor space \p{2.7}, analytic superspace is {\it not chiral}. 

Then, how can we possibly compare the chiral super-amplitude, Eqs. \p{supa} and \p{2.3}, with the analytic (but  non-chiral) correlator \p{2.19}? To answer this question, let us first compare the 
symmetries of the two quantities. As was already mentioned, the super-amplitudes in planar $\cN=4$ SYM have dual superconformal symmetry whose generators
act  in the dual superspace $(\la,x,\theta)$ and are given by standard expressions.
At tree level this symmetry is exact (modulo contributions localized on singular collinear configurations 
of the external momenta), whereas at loop level some of the symmetries become anomalous due to infrared singularities:
\begin{align}\label{R-anomaly}
 Q^\a_A\, {\R}_{n;k}=\bar S^\da_A\, {\R}_{n;k}=0\,, \qquad \bar Q^A_\da\, {\R}_{n;k}  \,, \  S^A_\a\, {\R}_{n;k}   \,, \  K_{\a\da}\, {\R}_{n;k} \neq 0\,.
\end{align}
For the correlator the situation is different. As long as we do not impose additional
conditions on its arguments, the super-correlator $\Gamma_n(x_i, \q^+_i, \bq_{-\, i}, u_i)$ enjoys
the full superconformal symmetry. To match \p{R-anomaly} we have to break part
of these symmetries. To begin with, by making the neighboring operators in  the 
correlator light-like separated, $x_{i,i+1}^2=0$, we generate additional
light-cone singularities which break the (super)conformal $K-$ and $S-$symmetries. If in addition we set all $\bq=0$, we break the $\bar Q-$supersymmetry of the super-correlator. 
This suggests that the object dual to the super-amplitude should be related to the correlator \p{2.19}  restricted to its chiral sector:
\begin{align}\label{npt}
G_n= \Gamma_n(x_i, \q^+_i, 0, u_i) \equiv \vev{\cT(x_1,\theta_1^+,0,u_1) \ldots \cT(x_n,\theta_n^+,0,u_n)}
\,.
\end{align}
The correlator $G_{n}$ defined in this way and localized on the light-like super-polygon \p{2.5}, is expected to have  the same symmetries  as the amplitude in \p{R-anomaly}:
\begin{align}\label{G-anom}
Q^\a_A\, G_{n}=\bar S^\da_A\, G_{n}=0\,, \qquad \bar Q^A_\da\, G_{n}  \,, \  S^A_\a\, G_{n}  \,, \  K_{\a\da}\, G_{n} \neq 0\,.
\end{align}

{We remark that our decision to set all $\bq=0$ eliminates the first of the conditions in \p{2.23} and we are left with the chiral light-like super-polygon defined in \p{2.5}. Still, what is surprising is that  \p{2.5} was obtained through the ``sacrificed" generator $\bar Q$. This puzzle may be partially answered by realizing that on the light cone the tree-level chiral correlator $G^{(0)}_n$ ``miraculously" recovers the $\bar Q$ half of supersymmetry, like the ``bonus" $\bar Q$ symmetry of the non-MHV tree-level amplitudes.}

\subsubsection{Formulation of the new duality} \label{2.3.2}

In this paper we propose to extend the duality described in Section~\ref{Dbcma} to the chiral sector $G_n$, Eq.~\p{npt}, of the  super-correlator  \p{2.19} and to the complete super-amplitude \p{supa}. This duality is supposed to work 
at all loop orders and for all types of amplitude (MHV, NMHV, etc.). Schematically, we claim
\begin{align}\label{eq:17}
\lim_{x^2_{i,i+1} \to 0} \frac{G_{n}(x_i, \q^+_i, u_i)}{G^{(0)}_{n;0}(x_i,u_i)} =   \left( \sum_{k=0}^{n-4} a^k\,
{\R}_{n;k}(\la_i,\mu_i,\chi_i) \right)^2\,,
\end{align}
where the functions ${\R}_{n;k}$ in the right-hand side describe the N${}^k$MHV all-loop super-amplitudes \p{supa}. {In this relation the limit $x^2_{i,i+1} \to 0$ is understood in the sense of the light-like super-polygon defined in \p{2.5}.} For the MHV amplitude the duality relation \p{eq:17} reduces to \p{mhvdu}.
As before, dividing by the tree-level bosonic correlator $G^{(0)}_{n;0}$, Eq.~\p{2.10}, in the left-hand side of \p{eq:17} removes the poles $1/(x_{12}^2\ldots x_{n1}^2)$ due to the propagator factors in $G_{n}$. Notice the appearance of additional powers of the coupling
constant inside the sum in the right-hand side of \p{eq:17}, as compared with \p{supa}. 
The reason for this will be explained in a moment.

In the duality relation \p{eq:17} we have explicitly indicated the type of variables that each object depends on. In the bosonic sector these are the four-dimensional space-time coordinates $x_i^{\a\da}$ (with $x_{i,i+1}^2=0$) for the correlator and the bosonic momentum twistor variables $\la_i^\a$ and $\mu_{i\da}$ for the amplitude. The relation between them is given by 
\begin{align}
 \mu_{i\da}=\la_i^\a (x_i)_{\a\da}\,, \qquad x_i^{\da\a} = \frac{\mu_{i-1}^\da\la_i^\a-  \mu_{i}^\da\la_{i-1}^\a}{\vev{i-1\,i}}\,.
\end{align}
In the fermionic sector, the super-correlator  depends on $\q^{+a}_\a= \q^A_\a u_A^{+a}$ (with $a,\a=1,2$) and  the super-amplitude depends on the momentum supertwistors $\chi^A = \la^\a \q^A_\a$ (with $A=1,2,3,4$).  Each of them has four components, so the total number of odd variables on both sides of eq.~\p{eq:17} is $4n$. Further, both the correlator and the amplitude are invariant under $Q$ and $\bar S$ supersymmetry, Eqs.~\p{R-anomaly} and \p{G-anom}, with a total of 16 odd generators. These symmetries can be used to gauge away 16 of the odd variables, so the two objects effectively depend on $4(n-4)$ odd variables. This implies that the expansion of the super-correlator in powers of $\q^+$  (see  Section~\ref{cn4stm}) is very similar to that of the super-amplitude \p{supa}:
\begin{align}\label{2.24}
G_n &= \sum_{k=0}^{n-4} a^k G_{n;k}(x_i,\q_i^+,u_i)\,,
\end{align}
where $G_{n;k}$ is a homogeneous polynomial of degree $4k$ in  the odd variables $\q^+$ and the sum terminates at  $k=n-4$ due to $Q$ and $\bar S$ supersymmetry.
Of course, to be able to compare super-correlators with super-amplitudes, we need the change of variables which relates $\q^{+a}_\a$ to $\chi^A$ (see Section~\ref{As}). 

{ Further, we remark that the correlators in the left-hand side of \p{eq:17} depend on the harmonics $u_i$, whereas the right-hand side  is free from them. 
We recall that the harmonic dependence encodes the elaborate $SU(4)$ tensor structure of the correlators. If we restrict the duality relation \p{eq:17} to the zeroth order in the Grassmann expansion, Eq.~\p{G-gen}, then the leading asymptotic behavior of the correlators $G_{n;0}$ and $G_{n;0}^{(0)}$  in the light-cone limit $x_{i,i+1}^2\to 0$ involves the same $SU(4)$ tensor structures, which cancel in their ratio. However, when we go away from this lowest Grassmann level, the emergence of $U(1)$ charged harmonic-projected analytic variables  $\q^+$ in the expansion of $G_n$ changes the balance of the $SU(4)$ structures in the numerator and in the denominator of  \p{eq:17}, even in the light-cone limit. The charges of the odd variables in  $G_n$ have to be compensated by some residual harmonic factors. On the other hand, in the right-hand side of  \p{eq:17} we can replace the chargeless momentum super-twistor variables $\chi$ by charged analytic $\q^+$, and this change produces harmonic factors as well. Remarkably, these factors exactly match, which makes the relation \p{eq:17} possible. We illustrate this important point by the explicit example in Section~\ref{NMtc} and by a series of other examples in \cite{twin}. }

To get a better feel for the duality relation \p{eq:17}, let us first consider the simplest, tree-level version of it. The tree-level super-amplitude has the form
\begin{align}
  \label{eq:18}
 \cA^{\rm (0)}_n = \cA^{\rm MHV\, tree}_n \left(1 + R_n^{\mathrm{NMHV}} + R_n^{\mathrm{NNMHV}} + \ldots +
    R_n^{\overline{\mathrm{\rm MHV}}}\right)\, ,
\end{align}
{or, equivalently,  in terms of the function $\widehat{\cA}_{n;k}$ introduced in \p{supa}
\begin{align}
\widehat{\cA}^{\rm (0)}_{n;k} = R_n^{\mathrm{N^{k}MHV}}  \,.
\end{align}
Each term in the sum} in the right-hand side of \p{eq:18} is a rational function of $x, \la$ and $\q^A$ (or equivalently of $\la,\mu,\chi$), homogeneous in the odd variables  of the corresponding Grassmann degree (for the explicit form of $R_n^{\mathrm{NMHV}}$ see Section~\ref{Rims}). 
Then the conjectured duality \p{eq:17} reads
\begin{align}\notag
\lim_{x^2_{i,i+1} \to 0} \frac{G^{(0)}_{n}}{G^{(0)}_{n;0}} 
&=\left(1 + a  R_n^{\mathrm{NMHV}} + a^2 R_n^{\mathrm{NNMHV}} + \ldots +
   a^{n-4} R_n^{\overline{\mathrm{\rm MHV}}}\right)^2
\\\label{3.11}
&= 1 + 2a R_n^{\mathrm{NMHV}} + a^2\left[2 R_n^{\mathrm{NNMHV}}+ (R_n^{\mathrm{NMHV}})^2\right]
     + \dots \,,
\end{align}
where $G^{(0)}_{n}$ is the lowest perturbative order (tree- or Born-level) expression for the super-correlator \p{npt} and  $R_n^{\mathrm{N^kMHV}}$ denotes the tree-level expression for ${\R}_{n;k}$ in \p{eq:17}. {Comparing terms of equal Grassmann degree, we get  
\begin{align}
    \lim_{x^2_{i,i+1} \to 0} \frac{G^{(0)}_{n;1}}{G^{(0)}_{n;0}}  &=   2  
     R_n^{\mathrm{NMHV}}\,, \label{3.12}\\\notag
     \lim_{x^2_{i,i+1} \to 0} \frac{G^{(0)}_{n;2}}{G^{(0)}_{n;0}}  &=   2
   R_n^{\mathrm{NNMHV}}+ (R_n^{\mathrm{NMHV}})^2 \,,
   \\
   &\ldots  \nt
     \lim_{x^2_{i,i+1} \to 0} \frac{G^{(0)}_{n;n-4}}{G^{(0)}_{n;0}}  &=   \sum_{k=0}^{n-4}  R_n^{\mathrm{N}{}^k \mathrm{MHV}} R_n^{\overline{\mathrm{N}{}^k \mathrm{MHV}}} \ .  \label{2.28}
\end{align}

Somewhat surprisingly, the coupling $a$ appears in \p{3.11} and not in \p{3.12}-\p{2.28}. The reason for this is that the various nilpotent contributions to the {\it tree-level} correlator $G^{(0)}_{n}$ are of different orders in the coupling:
\begin{align}\label{coupdep}
G^{(0)}_{n} = \sum_{k=0}^{n-4}  a^k G^{(0)}_{n;k}  \ .
\end{align}
The explanation of this phenomenon} requires looking a little bit deeper in the structure of the stress-tensor multiplet $\cT(x,\theta^+,\bar\theta_-,u)$. When restricted to its chiral sector, it has the following expansion (for details see Section~\ref{N4tm}):
\begin{align}\label{2.29}
\cT(x,\theta^+,0,u) = \cT_0 +  \q^+ \cT_1  +(\q^+)^2\, \cT_2 +(\q^+)^3 \, \cT_3+ (\q^+)^4 \, \cT_4\,, 
\end{align}
where $\cT_k=\cT_k(x,u)$ are composite gauge invariant operators carrying $U(1)$ charge $4-k$.
The bottom component is the scalar operator \p{4.6} (or rather its harmonic projection), 
$\cT_0=\cO$, while at the top we find the chiral form of the on-shell $\cN=4$ SYM Lagrangian, $\cT_4 =\frac13 \cL$:
\begin{align}\notag
 \cT_0 &= \tr(\phi^{++}\phi^{++}) \,,
 \\\label{T4}
 \cT_4 & = \frac{1}{3} \tr \left\{- \frac12  F_{\alpha\beta}F^{\alpha\beta}  + {\sqrt{2}}  g \psi^{\alpha A} [\phi_{AB},\psi_\alpha^B] - \frac18 g^2 [\phi^{AB},\phi^{CD}][\phi_{AB},\phi_{CD}] \right\}\,.
\end{align}
The explicit expressions for the remaining components are given below in \p{t4c}.

Substituting \p{2.29} into \p{npt} and expanding the correlator in powers
of $\theta^+_i$, we arrive at the expression for $G_n$ of the general form \p{2.24} with $G_{n;k}$ given by a sum of correlators of the form 
\begin{align}\label{G=sum}
 G_{n;k} = \sum_{k_1+\ldots+k_n=4k}  
 (\q^+_1)^{k_1}\ldots (\q^+_n)^{k_n}
 \vev{\cT_{k_1}(1)   \ldots \cT_{k_n}(n)}  =  a^k G_{n;k}^{(0)} + a^{k+1} G_{n;k}^{(1)} + \ldots\,, 
\end{align}
with  $\cT_{k_i}(i)\equiv \cT_{k_i}(x_i,u_i)$. For instance, for $k=1$ the terms in the right-hand side of \p{G=sum} with, e.g., $k_1=\ldots=k_{n-1}=0$, $k_n=1$,  contain the correlator of the {on-shell} Lagrangian $\cL$ with $(n-1)$ scalar operators $\cO$. 
It is easy to see from \p{T4} that such a correlator, {computed with the standard Feynman rules following from the off-shell Lagrangian \p{A.1}}, even at the lowest perturbative level (tree- or Born level) necessarily involves interaction vertices, hence $G^{(0)}_{n;1} = O(a)$, etc.\footnote{The correlators $\vev{\cT_{k_1}(1)   \ldots \cT_{k_n}(n)} $ do not necessarily involve the Lagrangian. Nevertheless, for all such components of the super-correlator the minimal perturbative level is always $a^k$. }   At the same time, all the tree-level amplitudes (MHV, NMHV, etc.) are of the same order in $a$, and their ratios $\R^{(0)}_{n;k}$ are independent of $a$.  This is the reason why in \p{eq:17} we rescaled every level in the Grassmann expansion of the super-amplitude by the appropriate power of the coupling. {The careful comparison of the normalizations shows that the expansion parameter in \p{eq:17} is precisely $a = g^2 N_c/4\pi^2$, as stated in \p{2.3}.}
The same phenomenon, but in the context of the supersymmetric Wilson loop, was observed in \cite{Simon}. Since on both sides of the duality relation \p{eq:17} we have homogeneous polynomials of degree $4k$  in the odd variables, the overall factor $a^{-k}$  can be absorbed into a rescaling of, say, the momentum supertwistors, $\chi \ \to a^{1/4} \chi$. 

\subsubsection{Iterative structure of the Lagrangian insertions}

In order to test the conjectured duality relation \p{eq:17} beyond tree level, we have to find an efficient way of computing the loop
corrections to the correlator of stress-tensor multiplets \p{npt}. As mentioned earlier, the Lagrangian insertion is a very efficient procedure for generating loop corrections. In doing this, we observe an interesting iterative structure.

Since the Lagrangian itself is a member of the stress-tensor multiplet $\cT$ (see \p{2.29}), each insertion is equivalent to adding a new operator $\int d^4 x\, \cL(x) = \frac1{4}\int d^4x d^4\q^+ \cT(x,\q^+)$
 inside the correlator $G_n$. In other terms, this amounts to creating a new point in $G_n$ and passing to $G_{n+1}$. To be more explicit, suppose that we start with, e.g., the component $G^{(0)}_{n;0}$ of Grassmann degree $0$ of the $n-$point tree-level correlator, Eq.~\p{2.10}. The one-loop correction to $G_{n;0}=\vev{\cO(1) \ldots \cO(n)}$ is then given by a single Lagrangian insertion,
\begin{align}\notag
  G^{(1)}_{n;0} &= -i\int d^4x_{n+1} \ \vev{\cO(1) \ldots \cO(n) \cL(n+1)}^{(0)}
\\\label{iter}
  &= -\frac{i}{4} \int d^4x_{n+1} d^4\q^+_{n+1}\ \vev{\cO(1) \ldots \cO(n) \cT(n+1)}^{(0)} \,.
\end{align}
Here the correlator in the right-hand side contains an additional operator and  
is evaluated at tree level. 
{Despite the fact that the correlator in  \p{iter} involves an insertion of the chiral on-shell Lagrangian \p{LSD}, this correlator itself is computed with the standard Feynman rules following from the full off-shell $\cN=4$ Lagrangian \p{A.1}.}
The relation \p{iter} can be easily extended to take account of all components of the super-correlator $G_n$
\begin{align} \notag
 G^{(1)}_{n} & =-\frac{i}{4} \int d^4x_{n+1} d^4\q^+_{n+1}\ \vev{\cT(1) \ldots \cT(n) \cT(n+1)}^{(0)}_{\bq_i=0} 
\\\label{iter-super}
&= -\frac{i}{4} \int d^4x_{n+1} d^4\q^+_{n+1}\ G^{(0)}_{n+1}\,.
\end{align}
Notice that both sides of this relation involve the same type of correlator but for
different numbers of points and loop orders. 

The substitution of \p{2.24} into \p{iter-super} leads to an iterative structure of the loop corrections to the super-correlators $G_{n;k}$. Namely,  the one-loop corrections $G^{(1)}_{n;k}$ are controlled by the tree-level correlator $G^{(0)}_{n+1;k+1}$ with one extra point and at the next Grassmann level $(\q^+)^{4(k+1)}$. The superspace integration over this new point removes the extra power of the odd variables and  yields the one-loop correction $G^{(1)}_{n;k}$. It is straightforward to extend the recursion to higher loops.

The correlator  $G^{(1)}_{n;k}$ obtained as the integral of  $G^{(0)}_{n+1;k+1}$ should be compared with the one-loop correction to the $n-$point N${}^k$MHV amplitude. However, we have the option to leave the new superspace point $(x_{n+1}, \q^+_{n+1})$ in   $G^{(0)}_{n+1;k+1}$ unintegrated, keeping the extra power of $(\q^+_{n+1})^4$ and inserting the extra point as a new cusp of the light-like super-polygon, together with the other $n$ points.  In this case the newly created correlator  $G^{(0)}_{n+1;k+1}$ is to be matched with the $(n+1)-$point N${}^{k+1}$MHV tree-level amplitude. 

Inversely, starting with a given tree-level correlator $G^{(0)}_{n;k}$, we can obtain a number of super-amplitudes, depending on what we do with  the points of the correlator. If we put all of the   points on a light-like super-polygon and we keep all powers of $\q^+$, we obtain an object which matches the $n-$point N${}^k$MHV tree-level amplitude (more precisely, the square of the sum of all amplitudes up to the level $k$). If we pull a superspace point away from the light-like super-polygon and integrate over it (both over $x$ and $\q^+$), we get an object to be matched with the  $(n-1)-$point N${}^{k-1}$MHV one-loop amplitude. This process of integrating over points can be continued until we remove all odd variables, thus obtaining a correlator which matches the  $(n-k)-$point MHV $k-$loop amplitude.

In Section~\ref{NMtc} of this paper we prove this new duality in the case of the $n-$point correlator with one insertion and the $n-$point NMHV tree-level amplitude.  In the twin paper \cite{twin} we work out a number of other examples. Starting with the simplest correlator $G_4$, in which we make one or two Lagrangian insertions, we reproduce  a number of super-amplitudes, namely MHV${}_4$ at one and two loops, NMHV${}_5$ and NMHV${}_6$ at  tree level and one loop, and NNMHV${}_6$ at tree level.

\section{The $\cN=4$  stress-tensor multiplet  in analytic\\ superspace}\label{N4has}

In this section we review some basic facts about $\cN=4$ harmonic superspace, needed for the formulation of the $\cN=4$ stress-tensor multiplet $\cT$  as a Grassmann analytic (or 1/2 BPS short) superfield.  We start with a discussion of the chiral vector multiplet as the elementary building block for the chiral sector of the stress-tensor multiplet. We then introduce the $\cN=4$ harmonic variables and explain how they are used to describe the chiral vector and the stress-tensor multiplets as Grassmann analytic superfields. We point out the special role played by the chiral part of the $\cN=4$ SYM on-shell Lagrangian. 

\subsection{The $\cN=4$ chiral vector multiplet as a 1/2 BPS  multiplet}\label{N4vm}

The $\cN=4$ vector (or SYM) multiplet consists of a gauge field $A_\mu$, six pseudo-real scalars $\phi^{AB} = -\phi^{BA} = \overline{\phi_{AB}} = \frac1{2} \ep^{ABCD} \phi_{CD}$ and four fermions $\psi^A_\alpha$, $\bar\psi_A^{\dot\alpha}$,  all  in the adjoint representation of the gauge group $SU(N_c)$. Their supersymmetry transformations are given in Appendix \ref{a.3}. It is well known that the $\cN=4$ vector multiplet is {\it on shell}, meaning that the supersymmetry algebra
\begin{align}\label{3.1}
&\{Q^\a_{A}, \bar Q^{B\da}\} =2\delta_A^B \tilde\sigma^{\da\a}_\mu P^\mu \nt
&\{Q^\a_{A}, Q^\b_B\} = 0\nt
&\{Q^{A \da}, \bar Q^{B\da}\} = 0
\end{align} 
closes only on the shell of the field equations.\footnote{In addition, the closure of the algebra involves a compensating gauge transformation with a field-dependent parameter.} Unlike the cases $\cN=1$ and $\cN=2$, where finite sets of auxiliary fields can be added to the vector multiplet and the algebra can be closed off shell, it is not known how to do this for the $\cN=4$ theory. Working with on-shell symmetries in the context of a quantum theory is always delicate, one has to make sure that the relevant Ward identities are satisfied.

Fortunately, for the purpose of constructing the {\it chiral} sector
of the $\cN=4$ stress-tensor multiplet we do {not} need the full vector multiplet, but only its {\it chiral} (or self-dual) half. It consists of  the self-dual part $F^{\a\b} =F^{\b\a}= - \half F_{\mu\nu} (\sigma^\mu \tilde\sigma^\nu)^{\a\b}$ of the Yang-Mills curvature $F_{\mu\nu} = \pa_\mu A_\nu - \pa_\nu A_\mu -ig [A_\mu, A_\nu]$, of the chiral gauginos $\psi^A_\alpha$ and of the scalras $\phi^{AB}$.\footnote{Due to their pseudo-reality, the scalars belong to both the chiral and anti-chiral multiplets.} Their {\it chiral} supersymmetry transformations are (see Appendix \ref{a.3}):
\begin{eqnarray}
Q^\a_{A}\, \phi^{BC} &=&2 i \sqrt{2} \delta^{[B}_{A} \psi^{C] \alpha}\,,  \nonumber
\\[1.2mm]
Q^\a_{A} \,\psi_{\beta}^{B}\phantom{b} &=& \delta^{B}_{A} F^\a_{\beta} + ig \delta_{\beta}^\a \lbrack \phi^{BC}, \phi_{CA} \rbrack\,,   \nonumber
\\[1.2mm]
Q^\a_{A} \, F_{\b\gamma} &=&  2\sqrt{2} g \delta^{\a}_{(\b} [\phi_{AB}, \psi^{B}_{\gamma)}] \,, \label{vemu}
\end{eqnarray}
where $[BC]$ and $(\b\gamma)$ denote weighted antisymmetrization and symmetrization, respectively.

The important difference between the full vector multiplet in \p{Qsusy} and its chiral truncated version   \p{vemu} is that the latter is {\it off shell}, in the sense  that the chiral part of the algebra \p{3.1}, $\{Q^\a_{A}, Q^\b_{B}\}=0$, closes without the use of field equations. This is not true for the former, even if restricted to the chiral algebra only.  We call the multiplet  \p{vemu} ``chiral", since only fields with undotted (chiral) Lorentz spinors appear in it. Note that we do not mean setting the anti-self-dual part $\tilde F^{\da\db}$ of the curvature to zero, otherwise the gauge field would be on shell. In fact, none of the fields in \p{vemu} is supposed to satisfy its equation of motion.

The chiral vector multiplet is conveniently described by the $\cN=4$ superfield  (or rather a ``half-superfield", since we are not considering the dependence on the anti-chiral $\bq-$variables) \footnote{This superfield has the geometric meaning of the super-curvature in the anticommutator of two spinor covariant derivatives of the same chirality.} 
\begin{align}\label{coexp}
W^{AB}(x,\q) = -W^{BA} = \phi^{AB}(x) + 2i\sqrt{2} \q^{\a\, [A} \psi^{B]}_\a(x) + i\sqrt{2}\q^{[A}_\a \q^{B]}_\b F^{\a\b}(x) + \ldots\,,
\end{align}
where the ellipsis  stands for non-Abelian terms proportional to the coupling constant.
This   superfield satisfies constraints  which restrict its content to the physical fields shown in \p{coexp}. For instance, the term linear in $\q$ contains only a fermion in the fundamental  representation of $SU(4)$, but not the most general possibility $\q^C\psi^{AB}_C$. The absence of this and other components is encoded in the superspace constraint
\begin{align}\label{chic}
D^\a_C W^{AB} = -\frac{2}{3} \delta^{[A}_C D^\a_D W^{B]D}\,,
\end{align}
where $D^\a_A = \pa/\pa\q^A_\a$ is the chiral spinor derivative.\footnote{More precisely, $D^\a_A = \pa/\pa\q^A_\a + i \bq_{\da\, A} \pa/\pa x_{\a\da}$, but here we set $\bq=0$.}  Thus, we can say that the superfield constraint \p{chic}  defines the $\cN=4$ chiral vector multiplet.\footnote{The complete  $\cN=4$ SYM super-curvature $W^{AB}(x,\q, \bq)$ depends on both the chiral $\q$ and anti-chiral $\bq$ odd variables and contains the full vector multiplet. In addition to \p{chic}, it satisfies the constraint $\bar D^{\da (C} W^{A)B}=0$. The combination of the chiral and anti-chiral constraints imply the field equations for all component fields \cite{Sohnius:1978wk}. This is another way to see that the full $\cN=4$ vector multiplet is on shell.  }

The constraint \p{chic} admits the very important interpretation of a {\it Grassmann analyticity} condition. To see this we need to (temporarily, this will be repaired in Section \ref{N4ha}) break $SU(4)$ down to its subgroup $SU(2)\times SU(2)' \times U(1)$.  In doing so, each index of the (anti)fundamental representation splits into $A=(+a, -a')$, where $\pm$ indicates the $U(1)$ charge and $a,a'=1,2$ are $SU(2)\times SU(2)'$ indices. Then we take the following particular projection of the $SU(4)$ indices in \p{chic}: 
\begin{align}\label{3.4}
D^\a_{-c'} W^{+a+b}= \ep^{ab} D^\a_{-c'} W^{++}= 0\,,
\end{align}
{where $W^{+a+b}=-W^{+b+a}=\ep^{ab}\,W^{++}$.}
The meaning of this constraint is that the projection $W^{++}$, corresponding to the highest weight in the six-plet $W^{AB}$,
is annihilated by half of the chiral spinor derivative,
\begin{align}\label{3.5}
D^\a_{-a'} W^{++}= \frac{\pa}{\pa\q^{-a'}_\a}  W^{++}= 0 \quad \Rightarrow \quad  W^{++}= W^{++}(x,\q^{+a}_\a) \,.
\end{align}
Such a superfield is called {\it Grassmann analytic} \cite{Galperin:1980fg} (or analytic for short), because it depends on half of the Grassmann variables.\footnote{The complete projected super-curvature $W^{++}(x,\q, \bq)$ satisfies two Grassmann analyticity constraints, $D^\a_{-a'} W^{++}=\bar D^{+a}_{\da} W^{++}=0$, implying that it depends on half of the chiral and of the anti-chiral odd variables, $W^{++}= W^{++}(x,\q^{+a},\bq_{-a'})$.} Alternatively, the multiplets of this type are known under the name of ``1/2 BPS short" multiplets, in the sense that they are annihilated by half of the supercharges (see, e.g., \cite{Ferrara:1999ed}).\footnote{Another, rather misleading name for such operators is ``chiral primaries", or CPO. Its origin can be traced back to their realization in terms of $\cN=1$ chiral matter superfields. As we recalled here, the relevant notion in extended supersymmetry is not chirality, but Grassmann analyticity.}

Of course, presented like this, Grassmann analyticity seems to be a property of only one projection (the highest weight) $W^{++}$ of  the six-plet $W^{AB}$, the others do not have it. Also, the price for exhibiting this property of $W^{++}$ was breaking $SU(4)$. These two problems can be solved at once by introducing auxiliary  commuting variables, the so-called $SU(4)$ harmonic variables $u$. This step extends the $\cN=4$ superspace $(x,\q)$ to a harmonic superspace $(x,\q,u)$, in which the property of Grassmann analyticity of the vector multiplet becomes manifest, and $SU(4)$ is maintained intact. We review the basics of harmonic superspace in the next subsection.

\subsection{$\cN=4$ harmonic variables}\label{N4ha}
 
The harmonic variables are introduced in order to covariantly decompose any object in the fundamental representation of $SU(4)$ like $\q^A$, etc., into two halves with quantum numbers of a subgroup. This can be done in two equivalent ways.

In the first case, in the so-called ``harmonic superspace"  approach to extended supersymmetry (see \cite{hh} for the $\cN=2$, \cite{Galperin:1984bu} for the $\cN=3$ and \cite{Hartwell:1994rp}  for  the $\cN=4$ versions), we introduce a harmonic matrix belonging to  $SU(4)$,
\begin{align}\label{su4}
(u^{+a}_A\,, \ u^{-a'}_A) \ \in \ SU(4)\,.
\end{align}
The first index $A=1,2,3,4$ of this matrix transforms under the anti-fundamental representation of $SU(4)$. The second index splits into two halves according to the subgroup $SU(2)\times SU(2)' \times U(1) \subset SU(4)$, with indices $a,a'=1,2$ in the  fundamental representations of $SU(2)$ and $SU(2)'$, and a $U(1)$ charge $\pm1$, respectively.  With the help of this matrix we project the chiral odd coordinate of $\cN=4$ superspace $\q^A_\a$   as follows:
\begin{align}\label{1.4}
 \q^A_\a \quad \Longrightarrow \quad \q^{+a}_\a = \q^A_\a u^{+a}_A\,, \qquad   \q^{-a'}_\a = \q^A_\a u^{-a'}_A\,.
\end{align}
The first projection in \p{1.4} transforms as a doublet of $SU(2)$ and a singlet of $SU(2)'$, and vice versa for the second. 
Notice that each projection carries a $U(1)$ charge which is identified with {the so-called
R charge of  
$\cN=2$ supersymmetry} upon the reduction $\cN=4 \ \to \ \cN=2$. 

The $\cN=4$ harmonic variables defined in this way parametrize the coset
\begin{align}\label{cos2}
{\rm Gr}(4,2)\ = \  \frac{SU(4)}{SU(2)\times SU(2)' \times U(1)}\,,
\end{align}
which coincides with  the Grassmannian manifold ${\rm Gr}(4,2)$, that is the space of all two-dimensional linear subspaces of ${\mathbb C}^4$ \cite{Chern}. It has $15-3-3-1=8$ real or 4 complex dimensions. It is important to realize that the action of the coset denominator on the harmonics is {\it local}, i.e. it depends on the point in harmonic space. At the same time, the coset numerator acts globally. By gauge fixing all the local transformations one obtains a coordinate description of the coset, but this implies breaking the global $SU(4)$ symmetry down to global $SU(2)\times SU(2)' \times U(1)$. By maintaining the harmonic variables in the matrix form \p{su4} we are able to do the projections \p{1.4} {\it without breaking $SU(4)$.}

The unitarity conditions for the harmonic $SU(4)$ matrix $u$ and its conjugate $\bar u$ are
\begin{align}\label{unit}
u^\dagger u = \mathbb{I}: \qquad \bar u^A_{+a} u_A^{+b} = \delta^b_a\,, \qquad \bar u^A_{-a'} u_A^{-b'} = \delta^{b'}_{a'}\,, \qquad \bar u^A_{-a'} u_A^{+b} = \bar u^A_{+a} u_A^{-b'} = 0\,,
\end{align}
leading to the completeness relation
\begin{align}\label{}
 u_A^{+a} \bar u^B_{+a} + u_A^{-a'} \bar u^B_{-a'}= \delta^B_A\,.
 \end{align}
It allows us to reconstruct $\q^A$ from its harmonic projections:
\begin{align}\label{rec}
\q^A = \q^{+a} \bar u^A_{+a}  + \q^{-a'} \bar u^A_{-a'}\,.
\end{align}
A corollary of unitarity \p{unit} is the unit determinant condition
\begin{align}\label{}
\frac14 \ep^{ABCD} u_A^{+a} \ep_{ab} u_B^{+b}\ u_C^{-c'} \ep_{c'd'} u_D^{-d'} = 1\,,
\end{align}
or equivalently,
\begin{align}\label{ucon}
\frac12 \ep^{ABCD} u_A^{+a} \ep_{ab} u_B^{+b} = -\bar u^C_{-c'} \ep^{c'd'} \bar u^D_{-d'}\,.
\end{align}

In the second approach, the so-called ``analytic superspace'' (for a review see \cite{Howe:1995md,Heslop:2003xu}), one complexifies the R symmetry group, $SU(4) \ \to \ GL(4,{\mathbb C})$. Here
the two projections of $\q^A$ look asymmetric:
\begin{align}\label{1.10}
 \q^A_\a \quad \rightarrow \quad  \rho^a_\a = \q^a_\a + \q^{a'}_\a y_{a'}{}^a\,,\quad
 \vartheta^{a'}_\a \equiv \q^{a'}_\a\,,
\end{align}
where $y_{a'}{}^b$ is a complex valued  $2\times2$ matrix. The decomposition \p{1.10}
corresponds to the alternative description of the Grassmannian ${\rm Gr}(4,2)$  \cite{Chern}:
\begin{align}\label{cos}
{\rm Gr}(4,2)\ = \ \frac{GL(4,{\mathbb C})}{{\cal P}}\,, 
\end{align}
where ${\cal P}$ is the parabolic subgroup of upper triangular matrices with $2\times2$ blocks. In Eq.~\p{1.10}, $\rho^a$ and $\vartheta^{a'}$ (with $a,a'=1,2$) transform under the subgroup $GL(2)\times GL(2)' \subset {\cal P}$ of the coset denominator. As usual with coset parametrizations, 4 of the (complex, i.e. not Hermitian) generators of  $GL(4)$ act transitively on the coordinates (shifts of $y_{a'}{}^b$), another 8 act homogeneously ($GL(2)\times GL(2)$ rotations of $y_{a'}{}^b$), the rest are realized non-linearly.  {The latter can be obtained by combining a shift of $y$ with the discrete operation of inversion, $y_{a'}{}^b \to y_{a'}{}^b/y^2$ (with $y^2=-\half\ y_{a'}{}^a y_{b'}{}^b \ep^{a'b'} \ep_{ab}$), in close analogy with the action of the conformal group on the Minkowski space coordinates $x_{\a\da}$. }

{The big advantage of this second, analytic approach comes  when considering more generic operators than the half BPS operators discussed here. One can describe all operators of the theory as unconstrained superfields, and their correlation functions can be described in this formalism in a manifestly superconformal manner by allowing supergroup indices~\cite{Heslop:2001dr,Heslop:2001gp,Heslop:2001zm,Heslop:2003xu}.  }   

The two equivalent descriptions of the manifold ${\rm Gr}(4,2)$ show two of its features \cite{Chern}. The harmonic description \p{cos2} makes its compactness manifest, the analytic description \p{cos} shows that it is holomorphic. In practice, to establish the relation between the two pictures, we replace the  unitary matrix $u$, Eq.~\p{su4},
and its hermitian conjugate $\bar u$ by a lower triangular  $GL(4)$ matrix and its inverse,
respectively,  
\begin{align}\label{3.17}
(u^{+a}_B,\,  u^{-a'}_B) \ \Longrightarrow \ \left(
\begin{array}{rr}
  \delta_{b}{}^a&   0  \\
  y_{b'}{}^a&    \delta_{b'}{}^{a'}
\end{array}
  \right)\,,\qquad
 (\bar u_{+a}^B,\, \ \bar u_{-a'}^B)  \ \Longrightarrow \  
  \left(
\begin{array}{rr}
  \delta_{a}{}^b&   0  \\
  -y_{a'}{}^b&    \delta_{a'}{}^{b'}
\end{array}
  \right)\,,
\end{align}
where $B=(+b,-b')$.


Using the decomposition \p{rec} we can express the correlators $G_n(x_i, \q^A_i)$ as functions of the projected odd variables, $\theta_i^{+a}$ and $\theta_i^{-a'}$, and of the harmonics $(u_i)^{+a}_A$ and $(u_i)^{-a'}_A$ (with the subscript $i=1,\ldots,n$ indicating the point in harmonic superspace). We recall that $G_n$ is covariant
under the action of $SU(4)$.  The effect of the harmonic projection is that it becomes invariant under $SU(4)$, but instead transforms covariantly under the subgroup $SU(2)\times SU(2)' \times U(1)$. Thus, the correlator  should depend, in addition to $x$ and $\q$,  on $SU(4)$
invariants built from the harmonics. 

The simplest harmonic invariant can be constructed by
taking four different harmonics $u^{+a}_A$ belonging to the anti-fundamental representation of $SU(4)$ and contracting their $SU(4)$ indices with the totally
antisymmetric tensor $\epsilon^{ABCD}$. For example, with two sets of harmonics at two different points,
$(u_1)^{+a}_A$ and $(u_2)^{+a}_A$,
we can  construct the $SU(4)$ invariant
\begin{align}\label{2w}
(\mathbf{12}) = \frac14 \ep^{ABCD} 1_A^{+a} \ep_{ab} 1_B^{+b}\ 2_C^{+c} \ep_{cd} 2_D^{+d}\,,
\end{align}
where we used the short-hand notation $k^{+a}_A\equiv (u_k)^{+a}_A$. This quantity is invariant not only under $SU(4)$, but also under  $SU(2)\times SU(2)'$ from the coset denominator. However, it carries $U(1)$ charge $+2$ at each point (we recall that the coset denominator acts locally).

With the help of \p{ucon} we can convert, e.g., the harmonics at point 1 into conjugate harmonics:
\begin{align}\label{2w''}
(\mathbf{12}) &= \frac14 \ep^{ABCD} 1_A^{+a} \ep_{ab} 1_B^{+b}\  2_C^{+c} \ep_{cd} 2_D^{+d}
 \notag\\ &
= -\frac12 (\bar 1^C_{-a'} \ep^{a'b'} \bar 1^D_{-b'}\  2_C^{+c} \ep_{cd} 2_D^{+d})
=  -\frac12\ \ep^{a'b'} (\bar 12)_{a'}{}^c \ (\bar 12)_{b'}{}^d \ep_{cd}
= \det\| (\bar 12)\|\,,
\end{align}
where $(\bar 12)$ is the $2\times2$ matrix
\begin{align}\label{y}
(\bar 12)_{a'}{}^b = \bar 1^A_{-a'} 2_A^{+b} \equiv (\bar u_1)^A_{-a'} (u_2)_A^{+b}
\end{align}
with one $SU(2)'$ and one $SU(2)$ index. 
In the holomorphic parametrization \p{cos}, following the substitution rules \p{3.17},  the matrix \p{y} is replaced by
\begin{align}\label{matr}
& (\bar 12)_{a'}{}^b  =   (y_1 - y_2)_{a'}{}^b \equiv (y_{12})_{a'}{}^b \,,
\notag
\\[2mm]
&(\mathbf{12}) =  \det \|y_{12} \|=  -\frac12\ \ep^{a'b'} (y_{12})_{a'}{}^c \ (y_{12})_{b'}{}^d \ep_{cd} =  y_{12}^2\,.
\end{align}
{In the last relation, the local $U(1)$ charge in the left-hand side is translated into  the $GL(4)$ weight under inversion, $y_{12}^2 \to y^{-2}_1 y^{-2}_2 y_{12}^2$,   in the right-hand side.}

\subsection{The $\cN=4$ chiral vector multiplet as an analytic superfield}\label{N4vmas}

The main purpose of introducing the $SU(4)$ harmonic variables is to be able to turn the defining constraint \p{chic} of the chiral vector multiplet into a {\it Grassmann analyticity} condition, {\it without breaking $SU(4)$}. Instead of simply splitting each index in \p{chic} up into halves, as we did in \p{3.4}, we will project the constraint \p{chic} with harmonics:
\begin{align}\label{}
\bar u^C_{-a'} D^\a_C\, W^{AB} u^{+a}_A u^{+b}_B =0\,.
\end{align}
For the moment, we restrict ourselves to the Abelian case ($g=0$), otherwise the spinor derivative would need a gauge connection.
The meaning of this constraint is that the projected superfield 
\begin{align}\label{}
W^{+a+b} = \ep^{ab} W^{++}(x,\q^{A},u) = W^{AB} u^{+a}_A u^{+b}_B 
\end{align}
is annihilated by the projected  chiral spinor derivative  $D^\a_{-a'} = \bar u^A_{-a'} D^\a_A= \bar u^A_{-a'} \partial/\partial \theta^A_\a = {\pa}/{\pa\q^{-a'}_\a}$, \footnote{We recall that the spinor derivative contains also a term $\bq \pa/\pa x$, but here we set all $\bq=0$.}
\begin{align}\label{anco}
D^\a_{-a'} W^{++}= \frac{\pa}{\pa\q^{-a'}_\a}  W^{++}= 0 \quad \Rightarrow \quad  W^{++}= W^{++}(x,\q^{+a},u) \,.
\end{align}
The important difference of this result, compared to the analogous statement in \p{3.5}, is the presence of the harmonics in $W^{++}(x,\q^{+a},u)$. This is what makes the Grassmann analytic superfield manifestly $SU(4)$ covariant: It  is inert under $SU(2)\times SU(2)'$, but it carries $U(1)$ charge $(+2)$. Moreover, this harmonic superfield now encodes the entire solution of the constraint \p{chic}, not just a particular projection of it.  

The analytic superfield depends on half of the odd variables, hence its component expansion is shorter than that of a standard superfield:\footnote{This explains the term ``short multiplet" for denoting the special representations of $PSU(2,2|4)$.} 
\begin{align}\label{exp}
W^{++}(x,\q^{+a},u) &= \phi^{++} + i\sqrt{2} \q^{+a}_\a \ep_{ab} \ep^{\a\b} \psi^{+b}_\b -  i\frac{\sqrt{2}}2\q^{+a}_\a \ep_{ab} \q^{+b}_\b F^{\a\b} + \ldots \,,
\end{align}
where the ellipsis denotes terms proportional to the coupling constant and the notation
was introduced for the $SU(4)$ non-singlet component fields projected with harmonics:
\begin{align}\label{}
\phi^{++}(x,u) = -\frac12 u^{+a}_A \ep_{ab}\, u^{+b}_B \ \phi^{AB}(x)\,, \qquad 
\psi^{+a}_\a(x,u) = u^{+a}_A \psi^A_\a (x)\,.
\end{align}
Using the linearized (or free, with $g=0$) version of the transformations \p{vemu}, it is easy to check that the superfield \p{exp} is closed under the chiral supersymmetry transformations 
\begin{align}
Q^\a_A\, W^{++}(x,\theta^+,u)=0 \,.
\end{align}
Notice that the expansion of a generic  Grassmann analytic superfield satisfying this
relation goes up to the maximal power $(\q^+)^4$, while the expansion of $W^{++}$ in  \p{exp} stops at $(\q^+)^2$, or as one says, $W^{++}$ is an ``ultra-short" superfield. This is due to the low value $(+2)$ of the harmonic charge of the superfield \p{exp}. For $W^{++}$ to have components with $(\q^+)^3$ or $(\q^+)^4$, one would need to balance their charge with negative-charged harmonics $u^-$. This however is forbidden by another constraint, the so-called harmonic analyticity. It states that the superfield \p{exp} should be made of highest-weight states of  $SU(4)$ irreps (see \cite{Heslop:2000af} for more detail). 
 
 The field content of the expansion \p{exp} is precisely that of the chiral vector multiplet from \p{vemu}. If we wish to describe the full vector multiplet, we need to restore the $\bq$ dependence of $W^{++}$. It can be shown that the full superfield $W^{++}(x,\theta^+,\bq_-, u)$, depending on the projection $\bq^\da_{-a'} = \bq^\da_A \bar u^A_{-a'}$, is Grassmann analytic in the anti-chiral sector as well. For this superfield,  analyticity (Grassmann and harmonic) implies that the fields satisfy their equations of motion. {This is another way to say} that the supersymmetry transformations of the full vector multiplet close only on shell.  

\subsection{The $\cN=4$ stress-tensor multiplet as an analytic superfield}\label{N4tm}

The superfield $W^{++}$ is really ultra-short only in the free theory (for vanishing coupling constant). The reason for this is that 
in the interacting (non-Abelian) theory the spinor derivative in the analyticity constraint \p{anco} becomes gauge covariant, involving a spinor super-connection, ${\cal D}^\a_{-a'} = D^\a_{-a'} - ig{\Gamma}^\a_{-a'}(x,\q,u)$. Consequently,  the superfield $W^{++} $ is not manifestly Grassmann analytic anymore and its expansion \p{exp} involves $\q^{-}$ terms. However, all such terms take the form of a  non-Abelian gauge transformation of $W^{++}$ with a field-dependent parameter. In this sense $W^{++}$ is covariantly analytic.  This shows that the gauge invariant composite operators  made of $W^{++}$
are  manifestly analytic objects, even in the interacting theory. The simplest example is the bilinear operator
\begin{align}\label{t4}
\cT(x,\theta^+,\bq_-, u) = \tr(W^{++}W^{++}) \,,\qquad  W^{++} = -\frac12 W^{AB}(x,\q,\bq) u^{+a}_A  \ep_{ab}u^{+b}_B\,.
\end{align}
It is inert under $SU(2)\times SU(2)'$ from the harmonic coset denominator \p{cos2}, but it carries $U(1)$ charge $(+4)$. This allows us to immediately determine the $SU(4)$ representation of its lowest component, once the harmonics are stripped off. 
For instance,  recall that $W^{++}= -\frac12 u^{+a}_A \ep_{ab} u^{+b}_B \ \phi^{AB} + \ldots$ is the highest weight of the $SU(4)$ six-plet, i.e. the representation $[0 1 0]$. Similarly, $\cT$ is the highest weight of the representation $[020]$, or the $\mathbf{20'}$. 

The composite operator \p{t4} is the first in a series of the so-called 1/2 BPS operators $\tr[(W^{++})^k]$ with lowest component in the $[0k0]$ of $SU(4)$ \cite{Howe:1992bv}. They are Grassmann analytic (hence the name ``1/2 BPS") and, most importantly, they do not undergo renormalization in the quantum field theory. Such operators are called ``protected". They have well defined properties under the  $\cN=4$ superconformal group $PSU(2,2|4)$, in particular, they have fixed conformal weight $k$. 

{
For $k=2$, the operator \p{t4} occupies a very special place because it  contains the stress-tensor $T_{\mu\nu}$ (as well as all other conserved currents) of the $\cN=4$ SYM theory \cite{Bergshoeff:1980is}:
\begin{align}\label{3.28}
\cT(x,\theta^+,\bq_-, u)=  \tr(\phi^{++}\phi^{++}) + \ldots + (\q^{+a}\sigma^\mu\bq_{-a'}) (\q^{+}_a\sigma^\nu\bq_-^{a'})\ T_{\mu\nu}(x) + \ldots\,.
\end{align}
For this reason it is called the ``stress-tensor" or the ``supercurrent"  multiplet \cite{Howe:1981qj}. 
In this paper we are interested only in the purely chiral sector of its component expansion, $\cT(x,\q^+,0,u)$. It  can be worked out by successively applying the non-linear chiral supersymmetry transformations \p{vemu} to the lowest component in \p{3.28}:
\begin{align}\label{expp}
\cT(x,\q^+,0,u) &={\rm e}^{(\q^A \,  Q_A)} \cT(x,0,0,u)= {\rm e}^{(\q^+ \,  Q_+) +( \q^- \,  Q_-)} \tr(\phi^{++}\phi^{++}) \nt 
& = {\rm e}^{(\q^+ \,  Q_+)} {\rm e}^{( \q^- \,  Q_-)} \tr(\phi^{++}\phi^{++}) = {\rm e}^{(\q^+ \,  Q_+)}\tr(\phi^{++}\phi^{++}) \,.
\end{align}
Here $(\q^+ \,  Q_+) \equiv \q^{+a}_\a Q^\a_{+a}$, $(\q^- \,  Q_-) \equiv \q^{-a'}_\a Q^\a_{-a'}$ and $Q^\a_{+a} = \bar u^A_{+a} Q^\a_A$, $Q^\a_{-a'} = \bar u^A_{-a'} Q^\a_A$ are the two harmonic
projections of the chiral supersymmetry generator. In the third relation in \p{expp} we have used the fact that $[(\q^+ \,  Q_+), (\q^- \,  Q_-)]=0$, as follows from  the rule $Q^\a_A \,\q^{B}_\b = \delta^\a_\b \delta^B_A$, from the harmonic defining conditions \p{unit} and from the supersymmetry algebra $\{Q^\a_A,Q^\b_B\}=0$. It is important that the transformations  \p{vemu} close {\it off shell}, which allows us to work out the expansion of $\cT$ without using the field equations. Finally, in the last  relation in  \p{expp} we took into account that
\begin{align}\label{}
Q^\a_{-a'}\phi^{++}=-\frac12 \bar u^A_{-a'} \lr{Q^\a_A\phi^{BC}} u^{+b}_B \ep_{bc} u^{+c}_C =0\,,
\end{align}
as follows from \p{vemu}. This last step clearly explains why the superfield $\cT$ is indeed Grassmann analytic, i.e. it depends on $\q^+$ only. 

The resulting $\q^+$ expansion of $\cT$ has the form}   
\begin{align}\notag
\mathcal{T} & (x,\q^+,0,u)= \tr \big(\phi^{++}\phi^{++}\big)
  +2\sqrt{2} i \theta^{+a}_\alpha\tr\big(
\psi^{+\alpha}_a \phi^{++}\big)
\\[2mm]\notag
&+  \theta_\alpha^{+a}\epsilon_{ab}\theta_\beta^{+b} \tr \left(
 \psi^{+c(\alpha}\psi_c^{+\beta)}-
i {\sqrt{2}}  F^{\alpha\beta}\phi^{++}\right)  
\\[2mm]\notag
&- \theta_\alpha^{+a}\epsilon^{\alpha\beta}\theta_\beta^{+b} \tr   \left(\psi^{+\gamma}_{(a}\psi^{+}_{b)\gamma}
- {g}{\sqrt{2}} [\phi_{(a}^{+C},\bar\phi_{C +b)}]\phi^{++}\right) 
\\ \label{t4c}
&-\frac{4}3 (\theta^+)^{3\, a}_{\ \alpha}\, \tr\left(F_\beta^\alpha \psi_a^{+\beta}
+  ig [\phi_a^{+B},\bar\phi_{BC}]\psi^{C\alpha}\right)
  +\frac{1}3(\q^{+})^4\ \cL(x) \,,
\end{align}
where $\cL(x)$ is the chiral form of the $\cN=4$ SYM Lagrangian given in \p{LSD} below and the notation was introduced for 
\begin{align}\label{3.33}
(\q^{+})^3{}_{\a}^{\, a} = \q^{+b}_\a \q^{+\b}_b \q^{+a}_\b\,, \qquad (\q^{+})^4 = \q^{+b}_\a \q^{+\b}_b \q^{+c}_\b \q^{+\a}_c \,,
\end{align}
and, e.g., $ \psi^{+\,\a}_a = \ep_{ab} u^{+b}_A \psi^A$, $\phi^{+\, B}_a =\ep_{ab}  u^{+b}_A\phi^{A B} $. It is easy to see that $(\theta^+)^4$ involves the product of four $\theta$'s and, therefore, it is proportional to the Grassmann delta function
\begin{align}
(\theta^+)^4 = 12\prod_{a,\a=1,2} \theta_\a^{+a} = 12\, \delta^{(4)}(\theta^+)\,,\qquad \int d^4\theta^+\, \delta^{(4)}(\theta^+) =1\,,
\end{align}
leading to
\begin{align}\label{S-T}
\mathcal{L}(x) =\frac1{4} \int d^4\theta^+\,  \mathcal{T} (x,\theta^+)\,.
\end{align}

By construction, each component in the expansion \p{t4c} is annihilated by $Q^\a_{-a'}$, while $Q^\a_{+a}$ transforms a given component into the one at the next expansion level. Consequently,  the top component $\cL$ is invariant under both projections,  so
\begin{align}\label{Q-SD}
Q^\a_A\, \cL(x) =0\,.
\end{align}
It carries no $U(1)$ charge, i.e. it is an $SU(4)$ singlet and therefore it is independent of the harmonics. 
In fact, this is the chiral form of the $\cN=4$ SYM {\it on-shell} Lagrangian
\begin{align}\label{LSD}
\cL= \tr \left\{ -\frac12  F_{\alpha\beta}F^{\alpha\beta}  + {\sqrt{2}}  g \psi^{\alpha A} [\phi_{AB},\psi_\alpha^B] - \frac18 g^2 [\phi^{AB},\phi^{CD}][\phi_{AB},\phi_{CD}] \right\}\,.
\end{align} 
Notice the absence of kinetic terms for the fermions and the scalars, they have been replaced by interaction terms via the field equations. However, we insist once more on the fact that the fields in the {\it off-shell}  multiplet \p{t4c} and hence in the chiral Lagrangian \p{LSD} are not supposed to satisfy their equations of motion.  The Lagrangian \p{LSD} is not a proper scalar, for example, the chiral gluon term $F^{\a\b} F_{\a\b}$ contains a pseudo-scalar total derivative (topological) term.  The parity conjugate anti-chiral Lagrangian $\bar\cL$ is the top component in the anti-chiral sector of the stress-tensor multiplet $\cT(x,0,\bq_-,u) $.

{ 
According to \p{Q-SD}, the Lagrangian $\cL(x)$ is invariant under the chiral half of Poincar\'e supersymmetry.\footnote{The fact that the $\cN=4$ SYM on-shell Lagrangian is a member of the stress-tensor multiplet has been known for a long time \cite{Eden:1999gh}. In \cite{Simon} the Lagrangian \p{LSD} was called ``chiral" because of its property  to be annihilated by the chiral half of the supersymmetry generators. In reality, as we have explained, the Lagrangian \p{LSD} belongs to the {\it chiral sector} of a Grassmann analytic on-shell multiplet. This is an exceptional property of the $\cN=4$ SYM Lagrangian; the $\cN=1$ and the $\cN=2$ SYM Lagrangians belong to genuine chiral {\it off-shell} multiplets. } It can also be shown that it transforms through a total space-time derivative under the {action of anti-chiral $\bar Q$ supersymmetry.}\footnote{Consider, for example, the free theory case ($g=0$). There $\bar Q^{\dot\gamma A} F^{\a\b} = -2 i \pa^{\dot\gamma(\a} \psi^{\beta) A}$, so the variation of $\mathcal{L}= -\frac12 \tr ( F_{\alpha\beta}F^{\alpha\beta}) $ is,  $\bar Q^{\dot\gamma A} \mathcal{L} =2 i\pa^{\dot\gamma\a} \tr (F_{\a\b} \psi^{\b A})$, up to the field equation for $F_{\a\b}$. }  Consequently, one can write down the {\it on-shell} superspace $\cN=4$ SYM action as the analytic superspace integral of the stress-tensor superfield \cite{Eden:1999gh}:\footnote{We are grateful to Paul Howe for a discussion on this point.}
\begin{align}\label{n4act}
S_{\rm on-shell} &=  \int d^4x\, \cL(x)  = \frac1{4} \int d^4x\, d^4\q^+\, \cT(x, \q^+,\bq_-,u)\nt  
 &=  \int d^4x\, \bar\cL(x)  =  \frac1{4} \int d^4x\, d^4\bq_-\, \cT(x, \q^+,\bq_-,u)  \,.
\end{align}
Here we do not have to set $\bq_-=0$ (in the first line) or $\q^+=0$ (in the second line) by hand, the Grassmann integration does it automatically. Notice that the $U(1)$ charge of the superspace measure balances that of the integrand, as expected from the $SU(4)$ singlet action.  The odd part of the superspace measure $d^4\q^+$ (or $d^4\bq_-$) involves only one quarter of the full $\cN=4$ superspace measure $d^8\q\, d^8\bq$. Usually, integration over less than half of a superspace does not produce supersymmetric invariants, but in this case it does, due to the very specific constraints satisfied by the integrand. This is an exceptional supersymmetric invariant, an example of a so-called ``superaction" \cite{Howe:1981xy}. 

The equivalence of the two forms, chiral and anti-chiral,  of the on-shell action \p{n4act} can be shown by repeating the procedure from Appendix \ref{a.2} (see \p{A.11}), but this time subtracting the total derivative term \p{A.10},  $\bar\cL = [\cL_{\cN=4} - \Delta\cL]_{\rm on-shell}$. This has the effect of eliminating the chiral terms from the Lagrangian.  Clearly, the total  derivatives drop out under the space-time integral in \p{n4act} (in a topologically trivial background).  

The existence of two equivalent forms of the $\cN=4$ on-shell action, related to each other by PCT conjugation, has an important consequence for the duality correlators/amplitudes. The imaginary part of  $\cL$ is a pseudo-scalar, which is a total space-time derivative (see \p{A.13}). In what follows we will use $\cL$ to make Lagrangian insertions into the correlators, and we will match this to the integrand of the loop corrections to the scattering amplitudes. These integrands do indeed have a parity-odd part which is a total derivative \cite{ArkaniHamed:2010kv}. 

}

\section{Correlators of the $\cN=4$ stress-tensor multiplet }\label{cn4stm}

In the previous section we defined the stress-tensor multiplet  $\cT(x, \q^+,\bq_-,u)$. We now turn to   the correlation functions of such operators, restricted to their chiral sector, Eq.~\p{npt}. We recall that, according to the conjectured duality relation \p{eq:17},  the light-cone limit $x_{i,i+1}^2\to 0$ of these super-correlators should match the scattering super-amplitudes in $\cN=4$ SYM.

\subsection{Bosonic correlators at tree level and beyond}

To begin with, let us consider the correlator \p{npt} at ``tree level" (or Born level), i.e. at the lowest order in the coupling constant. {The expansion of the correlator
$G_n$ in powers of $\q^+$ has the form \p{2.24} and, as explained in Section~\ref{2.3.2}, the
perturbative expansions of  the various components $G_{n;k}$ start at different orders in the coupling. }
For the purely bosonic component $G_{n;0}$ (with all $\q^+_i=0$) this order is $a^0$, i.e. the tree-level bosonic correlator is computed in the free theory (with $a=0$).  In this case,  the correlator \p{npt} becomes
\begin{align}\label{G-tree}
G_{n;0}^{(0)} = \vev{\tr \left(W^{++}(1)W^{++}(1)\right)\ldots \tr \left(W^{++}(n)W^{++}(n)\right)}^{(0)}\,,
\end{align} 
and it reduces to a product
of $n$ free scalar propagators with harmonic projections. Indeed, examining the expression for the chiral vector multiplet  superfield $W^{++}$, Eq.~\p{exp}, we observe that the only field in the
multiplet which has a non-zero propagator is the scalar $\phi^{++}$. 
Therefore, the free propagator of the superfield $W^{++}$ is given by%
\footnote{{To simplify the formulae, we do not display the $SU(N_c)$ color indices of the fields and, in addition, we normalize the scalar propagator to be $\vev{\phi^{AB}(x)\phi^{CD}(0)}=\epsilon^{ABCD}/(4\pi^2 x^2)$.} }
\begin{align}\label{2w'} 
\vev{W^{++}(1) W^{++}(2)}^{(0)} &= \vev{\phi^{++}(1)\phi^{++}(2)}^{(0)}
 \nt[2.3mm]
&= \frac14 \lr{1^{+a}_A \ep_{ab} 1^{+b}_B}\lr{ 2^{+c}_C \ep_{cd} 2^{+d}_D} \vev{\phi^{AB}(x_1) \phi^{CD}(x_2)}^{(0)} 
 \nt 
&= \frac14 \lr{1^{+a}_A \ep_{ab} 1^{+b}_B 2^{+c}_C \ep_{cd} 2^{+d}_D} \frac{ \ep^{ABCD}}{4\pi^2 x_{12}^2} 
  =  \frac{(\mathbf{12})}{4\pi^2 x^2_{12}} = \frac{y^2_{12}}{4\pi^2 x^2_{12}}\,,
\end{align}
where in the last relation we used \p{2w''} and \p{matr}.  The absence of odd variables in  \p{2w'}  is in agreement with the fact that the complete propagator $\vev{W^{++}(1) W^{++}(2)}^{(0)}$ only depends on   the product $\q\bq$, which vanishes if we set $\bq=0$.  

With the help of \p{2w'} we compute the tree-level correlator \p{G-tree} for  $n=2,3$ 
as
\begin{align}\label{4.7}
& G_2=  \vev{\cT(1) \cT(2)}_{\bq_i=0} = \frac{N^2_c-1}{2(4\pi^2)^2}  \left( \frac{(\mathbf{12})}{x^2_{12}}\right)^2\,, 
\nt
&  G_3=\vev{\cT(1)\cT(2) \cT(3)}_{\bq_i=0} =  \frac{N^2_c-1}{(4\pi^2)^3} \frac{(\mathbf{12})(\mathbf{23})(\mathbf{31})}{x^2_{12}x^2_{23}x^2_{31}} \,,
\end{align}
where the prefactor is a product of combinatorial and $SU(N_c)$ color factors. 
It is well known that the two- and three-point correlators for 1/2 BPS operators like \p{4.7} are protected from quantum corrections \cite{Penati:1999ba,D'Hoker:1998tz,Lee:1998bxa,Howe:1998zi}. This implies that the 
relations \p{4.7} are exact to all orders in perturbation theory, and at strong coupling, { that is $G_{n}=G_{n;0}^{(0)}$ for $n=2,3$.}

The first non-trivial correlator which receives loop corrections is $G_4$. In this
case, we have at tree level and at level $(\q^+)^0$
\begin{align}\label{G4-tree}
 G_{4;0}^{(0)}=& \vev{\cT(1)\cT(2) \cT(3)\cT(4)}^{(0)}_{\q^+=\bq_-=0} =  \frac{N^2_c-1}{(4\pi^2)^4}
 \nt[2mm]
& \times 
\bigg[ 
  \frac{(\mathbf{12})(\mathbf{23})(\mathbf{34})(\mathbf{41})}{x^2_{12}x^2_{23}x^2_{34}x^2_{41}}
+\frac{(\mathbf{13})(\mathbf{32})(\mathbf{24})(\mathbf{41})}{x^2_{13}x^2_{32}x^2_{24}x^2_{41}}
+\frac{(\mathbf{12})(\mathbf{24})(\mathbf{43})(\mathbf{31})}{x^2_{12}x^2_{24}x^2_{43}x^2_{31}}\bigg]\,.
\end{align}
The three terms have different harmonic dependence and correspond to different Wick contractions of the superfields.%
\footnote{The propagator structures involving squares of propagators, e.g., $((\mathbf{12})(\mathbf{34})/(x^2_{12}x_{34}^2))^2$, describe disconnected Green's functions and are not taken into account. }  
 The relation \p{G4-tree} holds for arbitrary $x_{ik}^2$. We notice 
that in the light-cone limit $x_{i,i+1}^2\to 0$ the leading asymptotic behavior
of the correlator is dominated by the contribution of the first term inside the square 
brackets, while the contribution of the remaining terms is suppressed by a power of 
$x_{i,i+1}^2$. 

It is easy to see that the same simplification takes place in the light-cone
limit for the $n-$particle tree-level correlator, restricted to the lowest level in its $\q$ expansion:
\begin{align}\label{lowl0}
G^{(0)}_{n;0} \ 
 \stackrel{x_{i,i+1}^2\to 0}{\longrightarrow} \ 
\frac{N^2_c-1}{(4\pi^2)^n}  \frac{(\mathbf{12})(\mathbf{23})\ldots(\mathbf{n1})}{x^2_{12}x^2_{23}\ldots x^2_{n1}} \,.
\end{align}
Here the expression in the right-hand side is formed by a cyclic chain of 
propagators \p{2w'}. It carries the $U(1)$ charges and conformal weights required 
at each point. The other possible propagator structures are obtained by non-cyclic 
permutations of the points. Among all permutations, the cyclic chain in \p{lowl} has 
the leading singularity in the light-cone limit $x^2_{i,i+1} \to 0$. Hence it is the only 
propagator structure which appears in the duality with amplitudes \p{mhvdu} and \p
{eq:17}. 

{Beyond tree level we find that,  in the light-cone limit, the bosonic component of the correlator,
i.e. the $(\q^+)^0$ term in the Grassmann expansion \p{2.24}, has the same $SU(4)$ tensor structure as in \p{lowl0}. The only change is that a non-trivial coupling-dependent coefficient function appears: }
\begin{align}\label{lowl}
G_{n;0} \ 
 \stackrel{x_{i,i+1}^2\to 0}{\longrightarrow} \ 
 \frac{N^2_c-1}{(4\pi^2)^n}  \frac{(\mathbf{12})(\mathbf{23})\ldots(\mathbf{n1})}{x^2_{12}x^2_{23}\ldots x^2_{n1}} F_{n;0}(x;a) \,.
\end{align}
The propagator factor in \p{lowl} carries the necessary conformal weights at each point. Therefore the coefficient function $F_{n;0}$ is conformally invariant, i.e. it depends on $x_i$ through cross-ratios ${x^2_{ij} x^2_{kl}}/{x^2_{ik} x^2_{jl}}$. Similarly, the (local) $U(1)$ harmonic charges  in \p{lowl} are carried by the propagator factor, hence  the coefficient function $F_{n;0}$ is chargeless, i.e. an $SU(4)$ invariant. This makes it independent of the harmonics $u$ because of the so-called harmonic analyticity.%
\footnote{
 The operator $\tr(\phi^{++}\phi^{++})$ is the highest weight of the $SU(4)$ irrep $[020]$. Consequently, it  is a polynomial in the complex harmonic variables $y$ of degree four, as determined by its $U(1)$ charge, i.e., by the Dynkin label. The same applies to the correlator \p{lowl}, it is a polynomial of degree four in the set of $n$ harmonic variables $y_i$. Then any function like $F_{n;0}(x;a)$ with vanishing $U(1)$ charges must be a polynomial of degree zero, i.e., harmonic independent. }

The drastic simplification of the harmonic structure of \p{lowl} in the light-cone limit has the important consequence that the dependence on the harmonics cancels  in the ratio
\begin{align}\label{pubo}
\lim_{x^2_{i,i+1}\to 0} \frac{G_{n;0}}{G^{(0)}_{n;0}} = \lim_{x^2_{i,i+1}\to 0} F_{n;0}(x;a)\,.
\end{align}
The light-cone limit of the $SU(4)$ singlet coefficient function $F_{n;0}$ can then be successfully compared to the light-like Wilson loops and MHV amplitudes, also $SU(4)$ singlets \cite{AEKMS,Eden:2010zz,Eden:2010ce}.

\subsection{General Grassmann structure of the correlators} 
 
In Section~\ref{2.3.2} we stated that the $\q^+-$expansion of $G_n$  has the general form \p{2.24} and it goes in powers of $(\q^+)^4$.  The reason for this has to do with  the $\mathbb{Z}_4$ center of $SU(4)$ \cite{Eden:1999gh}. It acts on
the odd variables as  $\q \to e\q$, where $e$ (with $e^4=1$) is the generator of $\mathbb{Z}_4$. Thus, the $\mathbb{Z}_4$ invariants are of the form $(\q^+)^{4k}$ with $k=0,1,2,\ldots$.  The maximal possible value of $k$ is determined by the number $n$ of points, but in fact the  $\cN=4$ superconformal symmetry $PSU(2,2|4)$ reduces it down to $k=n-4$, as stated in Eq.~\p{2.24}.

To show this, we recall that two of the $PSU(2,2|4)$ generators, $Q^\a_A$ and $\bar S^\da_A $, act like shifts of the $\q$'s:
\begin{align}\label{QbS}
Q^\a_A \,\q^{+a}_\b = \delta^\a_\b u^{+a}_A\,, \qquad \bar S^\da_A\, \q^{+a}_\b = x^\da_\b u^{+a}_A\,.
\end{align} 
Altogether, the two generators have $2\times2\times4=16$ components, so with their help we can gauge away four analytic $\q_i^{+}$ (assuming that the matrix $x^\da_\b$ in the $\bar S$ transformation is invertible, i.e., $x^2\neq 0$). For example, we may fix the following 
\begin{align}\label{qsga}
\mbox{$(Q+\bar S)-$gauge}: \qquad (\q_1)^{+a}_\a=(\q_2)^{+a}_\a=(\q_3)^{+a}_\a=(\q_4)^{+a}_\a=0\,.
\end{align} 
In this gauge the correlator $G_n$ effectively depends on the $(n-4)$ remaining $\q$'s, which explains its maximal Grassmann degree $4(n-4)$. {It has a non-vanishing lowest  
component $G_{n;0}$ even in the free theory  (for $a=0$), while}
in order to see the non-trivial Grassmann structure, i.e. the components $G_{n;k}$ with $k>0$ in Eq.~\p{npt}, we need to switch on the interactions, even at tree level.  

Let us first consider the simplest case $n=4$.  It is very special since, according to \p{qsga} 
and \p{2.24}, the super-correlator $G_{4}$ does not depend on the Grassmann 
variables $\theta^+$ and, therefore, it is reduced to its purely bosonic part (again, 
we recall that this is only true in the chiral sector $\bq_i=0$),
\begin{align}\label{4ptco}
G_4 \equiv G_{4;0} &=\vev{\cT(1)\ldots  \cT(4)}_{\q_i=\bq_i=0} \nt[2mm]
& =\vev{\cO(1) \ldots \cO(4)}
= \frac{N^2_c-1}{(4\pi^2)^4}\frac{(\mathbf{12})(\mathbf{23})(\mathbf{34}) (\mathbf{41})}{x^2_{12}x^2_{23}x^2_{34} x^2_{41}}F_{4;0}(x;a) + \ldots \,,
\end{align}
where the ellipsis denote the remaining terms obtained through noncyclic permutation of the indices
and needed to restore the Bose symmetry of the correlator. 
Here the loop corrections are encoded in the conformally and $SU(4)$ invariant function $F_{4;0}(x;a) = \sum_{\ell=0}^\infty  a^\ell F_{4;0}^{(\ell)}(x) $ depending on the coupling $a$.  If this bosonic correlator is known, the complete dependence on the analytic odd coordinates $\q^+$ (together with $\bq_-$) can be restored by a finite  superconformal transformation.

The first case where the correlator can depend on $\q^+$ 
is $n=5$. Here we can use the $Q$ and $\bar S$  gauge \p{qsga}
and expand the correlator in $(\q^{+}_5)^4$. In this way, we obtain in the gauge  \p{qsga}
\begin{align}
G_5=\vev{\cO(1)\cO(2)\cO(3)\cO(4) \cT(5)} = G_{5;0}+ G_{5;1}\,,
\end{align}
where $G_{5;0}$ and $G_{5;1}$ are of Grassmann degree  $(\q_5^+)^0$ and $(\q_5^+)^4$, respectively.
The function $G_{5;0}$ is given by the correlator of five scalar operators $\cO$. 
In the function $G_{5;1}$ the same operators appear at the first four points, while at point 5 we find  the top component in the chiral sector of the stress-tensor supermultiplet \p{t4c}, the chiral on-shell Lagrangian:
\begin{align}\label{5p'}
 G_{5;0}  &=  \vev{\cO(1) \ldots \cO(5)} \nt & \qquad\qquad\qquad
 =  \frac{N^2_c-1}{(4\pi^2)^5}\frac{(\mathbf{12})(\mathbf{23})(\mathbf{34})(\mathbf{45})(\mathbf{51})}{x^2_{12}x^2_{23}x^2_{34} x^2_{45}  x^2_{51}}F_{5;0}(x;a) + \ldots \\[3mm]
\label{5p}
 G_{5;1} &=  \frac13 (\q^{+}_5)^4 \vev{\cO(1) \ldots \cO(4) \cL(5)} \nt & \qquad \qquad\qquad
 =   \frac{N^2_c-1}{(4\pi^2)^5}(\q^{+}_5)^4 \left(\frac{(\mathbf{12})(\mathbf{23})(\mathbf{34})(\mathbf{41})}{x^2_{12}x^2_{23}x^2_{34} x^2_{41}}F_{5;1}(x;a) + \ldots   \right)\,. 
\end{align}
  We may say that \p{5p} corresponds to the insertion, at  point 5, of the Lagrangian \p{LSD} into the four-point correlator \p{4ptco}. As before, $F_{5;0}$ and $F_{5;1}$ are harmonic
independent functions of the conformal cross-ratios made out of the five  $x_i$. We recall
that $F_{5;0}=1+O(a)$.

{ In the light-cone limit  the ratio
\begin{align}\label{4.15}
\lim_{x^2_{i,i+1}\to 0} \frac{G_{5;1}}{G^{(0)}_{5;0}} =   \lim_{x^2_{i,i+1}\to 0} \ \left[ (\q^{+}_5)^4  \frac{(\mathbf{41})}{(\mathbf{45})(\mathbf{51})}\  \frac{x^2_{45}  x^2_{51}}{x^2_{41}}F_{5;1}(x;a)\right]
\end{align}
simplifies considerably, all other harmonic structures from \p{5p} being subleading. Unlike the purely bosonic case \p{pubo}, here the remaining harmonic structure is not completely trivial. The harmonic factor in \p{4.15} neutralizes the $U(1)$ charge of the Grassmann factor $(\q^{+}_5)^4$ (we recall that at point $5$, the harmonic factors $(\mathbf{45})$ and $(\mathbf{51})$ have twice the charge of $\q^+_5$). It is highly non-trivial that the matching 5-point super-amplitude, once its native momentum super-twistor variables $\chi^A$ are replaced by the analytic superspace variables $\q^+$, contains exactly the same Grassmann and harmonic structure. Likewise, the $x-$space factor in \p{4.15} balances the conformal weight of  $(\q^{+}_5)^4$. The limit in the right-hand side of \p{4.15} is non-vanishing because the function $F_{5;1}(x;a)$ has poles at $x^2_{45} =  x^2_{51}=0$. Both of these aspects are illustrated by explicit calculation in Section~\ref{esc} at tree level and in the twin paper \cite{twin} at loop level.}

For $n\geq 6$ the dependence of the correlator $G_n$ on $\q^+$ becomes much more involved. For example, for $n=6$, in the gauge \p{qsga} we are left with $\q^{+}_5$ and $\q^{+}_6$, so that 
\begin{align}
G_6 =\vev{\cO(1)\cO(2)\cO(3)\cO(4) \cT(5) \cT(6)}  = G_{6;0} + G_{6;1} +G_{6;2} \,, 
\end{align}
where $G_{6;k}$ are homogenous polynomials in $\q^{+}_5$ and $\q^{+}_6$ of degree
$(4k)$. Replacing the superfields $\cT(5)$  and $\cT(6)$ by their explicit expressions, Eqs.~\p{2.29} and \p{t4c}, we obtain 
\begin{align}
 G_{6;0}   &=  \vev{\cO(1) \ldots \cO(6)}   =  \frac{N^2_c-1}{(4\pi^2)^6}\frac{(\mathbf{12})(\mathbf{23})(\mathbf{34})(\mathbf{45})(\mathbf{56})(\mathbf{61})}{x^2_{12}x^2_{23}x^2_{34} x^2_{45} x^2_{56} x^2_{61}}F_{6;0}(x;a) + \ldots \notag \\[2mm]
G_{6;1}   &= (\q^{+}_5)^4 G^{(4,0)}+(\q^{+}_6)^4 G^{(0,4)}    \nt 
& \qquad + (\q^{+}_5)^3 \q^{+}_6 G^{(3,1)}
+ \q^{+}_5(\q^{+}_6)^3 G^{(1,3)}  + (\q^{+}_5)^2 (\q^{+}_6)^2 G^{(2,2)} \label{6p} 
  \\[2mm] 
G_{6;2} &=  \frac{1}{9} (\q^{+}_5)^4  (\q^{+}_6)^4 \vev{\cO(1) \ldots \cO(4) \cL(5)  \cL(6)} \nt
&  =   \frac{N^2_c-1}{(4\pi^2)^6}  (\q^{+}_5)^4 (\q^{+}_6)^4  \left(\frac{(\mathbf{12})(\mathbf{23})(\mathbf{34})(\mathbf{41})}{x^2_{12}x^2_{23}x^2_{34} x^2_{41}}F_{6;2}(x;a) + \ldots   \right)\,.  \label{6p'}
\end{align}
In \p{6p} the notation was introduced for the correlators 
\begin{align}
G^{(k,4-k)} = \vev{\cO(1) \cO(2) \cO(3) \cO(4) \cT_k(5) \cT_{4-k}(6)}
\end{align}
involving the bosonic operators
$\cT_0=\cO$ at the first four points and the mixed operators $\cT_k$ (see \p{2.29})  at the remaining two
points (we recall that $\cT_4=\frac13\cL$ is the chiral Lagrangian \p{LSD}).
We notice that  $G^{(4,0)}$ and $G^{(0,4)}$ in the first line of \p{6p} are given by the
correlators of five bosonic operators and  the chiral Lagrangian inserted at points 5 and 6, respectively. They have a  form similar to \p{5p}. 
The expressions for the correlators in the second line of \p{6p} are more involved, e.g.
\begin{align}
 G^{(1,3)} \sim  \vev{\cO(1) \ldots \cO(4)\ \tr[\psi^+\phi^{++}](5)\ \lr{\tr[F\psi^+](6)+O(g)}} \,.
\end{align}

{ Let us consider the light-cone limit of the ratio
\begin{align}\label{}
\lim_{x^2_{i,i+1}\to 0} \frac{G_{6;2}}{G^{(0)}_{6;0}} =  \lim_{x^2_{i,i+1}\to 0} \,  \left[ (\q^{+}_5)^4 (\q^{+}_6)^4 \frac{(\mathbf{41})}{(\mathbf{45})(\mathbf{56})(\mathbf{61})}\  \frac{x^2_{45} x^2_{56} x^2_{61}}{x^2_{41}}F_{6;2}(x;a)  \right]\,.
\end{align}
As before in \p{4.15}, we see a harmonic and an $x-$space factor which balance the $U(1)$ charge and conformal weight of the Grassmann factor. We give an explicit example of this limit in the twin paper \cite{twin}. }

For arbitrary  $n\geq 6$ we find the same pattern for $G_n$. Namely, the lowest $(G_{n;0})$ and the highest  $(G_{n;n-4})$ terms in the expansion of the correlator \p{2.24} have a
much simpler form as compared to other terms. As before,  $G_{n;0}$ is given by
the correlator of $n$ bosonic operators \p{2.10}. 
The two examples \p{5p} and \p{6p'} are typical for the maximally nilpotent correlators. In all cases $G_{n;n-4}$, in the gauge  \p{qsga} we have just as many $\q$'s left as needed to form the term of maximal Grassmann degree $4(n-4)$. All of them then have to come from Lagrangian insertions:
\begin{align}\label{}
 G_{n;n-4}   =  (1/{3})^{n-4} (\q^{+}_5)^4 \ldots (\q^{+}_n)^4\ \vev{\cO(1) \ldots \cO(4) \cL(5)\ldots  \cL(n)}\,.
\end{align}
 According to the duality conjecture \p{eq:17}, \p{2.28},  such correlators are dual to the maximally nilpotent super-amplitudes $\R^{4(n-4)}_n = \cA_n^{\overline{\rm MHV}}/\cA_n^{\rm MHV\, tree}$, plus quadratic combinations of all super-amplitudes of lower degree. 
 
 \subsection{Lagrangian insertion procedure}\label{Lip}

The Lagrangian insertions that we have seen appearing in the examples above are a characteristic feature of the nilpotent correlators of stress-tensor multiplets. They are intimately related to the generation of loop corrections, as explained below. 
 
The  correlator $G_n$ \p{npt}  is defined by the functional integral 
\begin{align}\notag
    G_n  &= \int {\cal D}\Phi \ e^{i\int d^4x \cL_{\cN=4}}\  \cT(x_1,\q^+_1, 0,u_1)\ldots {\cal T}(x_n,\q^+_n,0, u_n) \\
    &= G^{(0)}_n + a G^{(1)}_n + a^2 G^{(2)}_n + \ldots\ ,\label{patco}
\end{align}
with the full Lagrangian \p{A.1}. The loop corrections to this correlator can be obtained by differentiating  the functional integral \p{patco} with respect to the coupling $a \propto g^2$.\footnote{This standard quantum field theory procedure has been successfulely used in calculations of  $\cN=2$ correlators \cite{Eden:1999kw,Eden:2000mv} (for a recent review of the procedure see Appendix A in \cite{Eden:2010zz}).}   Prior to differentiating, one rescales all the fields in the Lagrangian by a factor of $g^{-1}$. As a result, the coupling disappears from inside $\cL_{\cN=4}$, but instead reappears as a prefactor, 
\begin{align}\label{}
\cL_{\cN=4} \ \to \ g^{-2} \cL_{\cN=4}\,.
\end{align}
Then the derivative gives
\begin{align}\label{4.220}
a \frac{d}{d a}  G_n &= aG^{(1)}_n + 2a^2 G^{(2)}_n +O(a^3)  \nt
& = -i\int d^4 x_{n+1} \  \vev{\cL_{\cN=4}(x_{n+1}) \cT(x_1,\q^+_1,  u_1)\ldots {\cal T}(x_n,\q^+_n, u_n)}\,,
\end{align}
which amounts to inserting the Lagrangian as an extra point in the $n-$point correlator $G_n$.

In writing this relation we have neglected the effect of the rescaling,  $\cT \ \to g^{-2}\cT$ and  $G_n \ \to \ g^{-2n} G_n$. The derivative $d/da$ sees this overall scaling factor and reproduces the correlator itself. It can be shown that this ``spurious" effect of the differentiation is exactly canceled by contact term contributions to the $(n+1)-$point correlator under the integral in \p{4.220} (see \cite{Eden:2000mv} and \cite{Simon}). These contact terms originate from the kinetic terms of the fields in the inserted Lagrangian. If we agree to put such spurious terms aside, as we have done in \p{4.220},  we are allowed to use the {\it on-shell} Lagrangian \p{LSD} (see Appendix~\ref{a.2}). In other words, 
 up to contact terms, under the space-time integral in \p{4.220} we may make the replacement
\begin{align}\label{4.25}
\int d^4 x_{n+1} \  \cL_{\cN=4}(x_{n+1}) \ \to \     \int d^4 x_{n+1} \  \cL(x_{n+1}) \,.
\end{align}
Then  Eq.~\p{4.220} becomes (recall \p{n4act}) 
\begin{align}\label{2.17}
a \frac{d}{d a}  G_n
& = -i\int d^4 x_{n+1} \  \vev{\cL(x_{n+1}) \cT(x_1,\q^+_1,  u_1)\ldots {\cal T}(x_n,\q^+_n,  u_n)}\nt
& = -\frac{i}{4}\int d^4 x_{n+1} d^4\q^+_{n+1}\  \vev{\cT(x_{n+1}, \q^+_{n+1},  u_{n+1}) \cT(x_1,\q^+_1,  u_1)\ldots {\cal T}(x_n,\q^+_n,  u_n)}\nt
&= -\frac{i}{4}\int d^4 x_{n+1} d^4\q^+_{n+1}\  G_{n+1}(1,2,\ldots,n,n+1)\,.
\end{align}
This relation expresses the $\ell-$loop correction to the  $n-$point correlator of stress-tensor multiplets through the $(n+1)-$point correlator, calculated at $\ell-1$ loops. In particular, the one-loop correction to $G_n$ is given by the {\it tree-level} $G_{n+1}$:
\begin{align}\label{2.18}
 G^{(1)}_n = -\frac{i}{4}\int d^4 x_{n+1} d^4\q^+_{n+1}\  G^{(0)}_{n+1}\,.
\end{align}
We repeat once again that in the computation of $G_{n+1}$ all   contact terms should be neglected, in accord with our convention not to display the spurious effects of the differentiation. 

The relation \p{2.18} exhibits one of the key features of the duality correlators/amplitudes. The same tree-level correlator $G^{(0)}_{n+1}$ can play multiple roles in the duality. If we put all of its $(n+1)$ space-time points at the vertices of a light-like polygon, then it is dual to the tree-level $(n+1)-$particle super-amplitude. If instead we  integrate over one point, as in \p{2.18}, we generate the one-loop correction to the $n-$point correlator. Putting these $n$ points on a light-like polygon, we find the dual to the $n-$particle one-loop amplitude.

Here we recall the coupling dependence of  the tree-level correlator $G^{(0)}_{n+1}$, relative to its $\q$ expansion, see \p{coupdep}. 
While the purely bosonic tree-level correlator $G^{(0)}_{n+1;0}$ is of order $a^0$ (it is made of free propagators), the nilpotent terms require non-trivial interactions between different component operators from the expansion \p{t4c}. Thus, the Grassmann integral in \p{2.18} picks the Lagrangian component at the new point $x_{n+1}$, which interacts with the operators at the other points, producing a factor of $a$. To be more specific, let us restrict relation \p{2.18} to the lowest $\q-$component in the left-hand side, i.e., let us set $\q^+_1=\ldots=\q^+_n=0$. Then, after doing the Grassmann integral, we find:
\begin{align} 
&  G^{(1)}_{n;0}= -\frac{i}{4}  \int d^4 x_{n+1} d^4\q^+_{n+1}\  G^{(0)}_{n+1;1}\,, 
\end{align}
or equivalently
\begin{align}\label{oo1}
  \vev{\cO(x_1) \ldots \cO(x_n)}^{(1)} =  -i\int d^4 x_{n+1} \  \vev{\cL(x_{n+1}) \cO(x_1) \ldots \cO(x_n)}^{(0)} \,.
\end{align}
The right-hand side contains the $n-$point correlator of scalar bilinears with a Lagrangian insertion at point $x_{n+1}$. This $(n+1)-$point correlator is of order $a\sim g^2$, in accord with the left-hand side (see Section~\ref{esc} for an explicit example of a tree-level calculation). 

Now, let us repeat the differentiation \p{2.17} twice:
\begin{align}\label{4.21}
  a^2 \frac{d^2}{d a^2}  G_n &= 2 a^2 G^{(2)}_n + O(a^3)   \nt
& =   \lr{-\frac{i}{4}}^2 \int d^4 x_{n+1} d^4 x_{n+2}  d^4\q^+_{n+1}  d^4\q^+_{n+2}\  \vev{\cT({n+1}) \cT({n+2})\cT(1)\ldots {\cal T}(n)}\nt
&= -\frac1{16} \int d^4 x_{n+1} d^4 x_{n+2}  d^4\q^+_{n+1}  d^4\q^+_{n+2}\   G_{n+2}\,.
\end{align}
Here we have neglected the single-insertion term originating from the overall scaling factor $g^{-2(n+1)}$ of $G_{n+1}$. As explained earlier, such ``spurious" terms are compensated by contact terms in $G_{n+2}$. They are due to the kinetic terms in the full Lagrangian, which we have eliminated by the substitutions \p{4.25}.  Again, restricting to the lowest $\q-$component, we obtain
\begin{align}\label{}
\vev{\cO(x_1) \ldots \cO(x_n)}^{(2)} =  -\frac12\int d^4 x_{n+1}  d^4 x_{n+2} \  \vev{\cO(x_1) \ldots \cO(x_n)\cL(x_{n+1}) \cL(x_{n+2}) }^{(0)} \,.\end{align}

In general, the $\ell-$loop correction to $G_n$ is obtained from the tree-level correlator $G^{(0)}_{n+\ell}$, integrated over $\ell$ points:
\begin{align}\label{2.22}
 G^{(\ell)}_n =   \frac{(-i/4)^\ell}{\ell !}\int \prod_{i=1}^\ell d^4 x_{n+i}  d^4\q^+_{n+i}\   G^{(0)
}_{n+\ell}\,.
\end{align} 
In the right-hand side, the odd integral picks the term of Grassmann degree $4\ell$ in the tree-level correlator, which is of order $O(a^\ell)$ in the coupling. Inversely, given the tree-level correlator $G^{(0)}_n$, we can use it to generate loop corrections to all correlators $G_k$ with $4\le k\le n-1$:
\begin{align}\label{}
G^{(0)}_n \quad \Rightarrow \quad  &    G^{(1)}_{n-1} = -\ft{i}{4} \int d^4 x_{n}  d^4\q^+_{n}\   G^{(0)
}_{n} \nt
&    G^{(2)}_{n-2} =  \half \lr{-\ft{i}{4}}^2\int d^4 x_{n}  d^4 x_{n-1}  d^4\q^+_{n}  d^4\q^+_{n-1}\   G^{(0)
}_{n} \nt
& \ldots \nt
&  G^{(n-4)}_{4} = \ft{1}{(n-4) !}\lr{-\ft{i}{4}}^{n-4}\int \prod_{i=0}^{n-3} d^4 x_{n-i}  d^4\q^+_{n-i}\    G^{(0)
}_{n} \,.
\end{align}
In the right-hand side of the last line we have reached the maximal Grassmann degree $4(n-4)$ in the correlator   $G^{(0)}_n$. At the same time, in the left-hand side we have reached the minimal number of points, for which a correlator of 1/2 BPS operators can have non-trivial perturbative corrections, $G_4$. This simple fact explains why the two- and three-point functions \p{4.7} are protected from quantum corrections  \cite{Penati:1999ba,D'Hoker:1998tz,Lee:1998bxa,Howe:1998zi}.

\section{The NMHV tree as a super-correlator}\label{NMtc}

In \cite{Eden:2010zz,Eden:2010ce} it has been demonstrated how the duality relation \p{eq:17} works
for the simplest, MHV amplitudes. In that case, the loop corrections to the correlation 
function of $n$ bosonic operators do not involve Grassmann variables and 
match the loop corrections to the $n-$particle MHV amplitude. In this section, we shall 
verify the duality relation for the NMHV tree super-amplitude, Eq.~\p{3.12}. This is the 
first time when the dependence on the Grassmann variables enters into consideration.
 
The duality relation \p{3.12} involves the $n-$point tree-level correlator $G^{(0)}_{n;
1}$. By definition, this correlator describes the lowest-order perturbative contribution to the
super-correlator $G_n$, Eqs.~\p{npt} and \p{coupdep},  at the first non-trivial 
nilpotent level $(\q^+)^4$.  According to \p{6p}, even in the simplest case $n=6$ it is given by  a sum of
five correlators  involving various components of the stress-tensor multiplet \p{t4c}.
Each of these correlators vanishes in the free theory (for $a=0$) and receives
the first non-trivial contribution at order $O(a)$ due to the interaction. As a consequence,
the explicit evaluation of the complete correlator $G^{(0)}_{n;1}$ is a rather complicated task.  

Fortunately,  there exists a shortcut. Viewed as a function of the space-time coordinates $x_i$ 
of the operators, the correlator $G^{(0)}_{n;1}$ has singularities at $x_{ik}^2=0$
corresponding to the null-distance separation of the operators at points $x_i$ and $x_k$. The asymptotic behavior
of the correlator for $x_{ik}^2\to 0$ can be analyzed with the help of the operator product expansion (OPE). To lowest order in the coupling,  the OPE produces
poles in $1/x_{ik}^2$, implying that $G^{(0)}_{n;1}$ is a meromorphic function of the distances
$x_{ik}^2$ (logarithms of $x_{ik}^2$,
contributing to the anomalous dimensions of the Wilson operators in the right-hand side of the OPE,  only appear at  higher loops). For $k=i+1$, when two neighboring operators in the correlator
become null separated, the poles of  $G^{(0)}_{n;1}$ in $1/x_{i,i+1}^2$ are cancelled by dividing by the bosonic tree-level correlator $G^{(0)}_{n;0}$. 
So, the ratio $G^{(0)}_{n;1}/G^{(0)}_{n;0}$ is completely determined by the residues at all singularities $x^2_{ik}=0$ with $|i-k|\geq 2$. As we will show
in this section, the latter can be easily computed using the standard Feynman graph technique. 

For the NMHV super-amplitude, written in dual space, the asymptotic behavior of the 
ratio function $\R^{\rm NMHV}$ for $x_{ik}^2\to 0$ corresponds to singularities of 
the amplitude in the limit where the invariant mass of several particles vanishes, $x_{ik}^2=
(p_i+\ldots+p_{k-1})^2\to 0$. To lowest order in the coupling, the ratio function
is known to be a meromorphic function of $x_{ik}^2$ and the corresponding residues were computed in \cite{Korchemsky:2009hm} using the unitarity properties of the scattering amplitudes.
 
In this section we compare the residues of both sides of the duality relation \p{3.12} at $x_{ik}^2=0$
and find perfect agreement, thus proving the duality relation for the tree-level NMHV super-amplitude. 

\subsection{The NMHV tree-level super-amplitude and its residues}

In this subsection, we start by recalling some basic facts about the NMHV tree-level super-amplitude and then we work out its residues in a particular supersymmetry gauge.  

\subsubsection{$R$ invariants}\label{Rin}
 
The function $R^{\rm NMHV}_n$ defined as the ratio
of the NMHV and MHV $n-$particle tree-level super-amplitudes,
${\cA_n^{\rm NMHV\, (0)}}   \equiv  {\cA_n^{\rm MHV\, (0)}} R^{\rm NMHV}_n$,
is given \cite{Drummond:2008vq} by the sum of all so-called ``$R$ invariants" of dual superconformal symmetry with {\it equal coefficients},
\begin{align}\label{ra}
 R^{\rm NMHV}_n = \sum_{r+2 \leq s < t+1 \leq r} R_{rst}\,,
\end{align}
where the choice of the first label $r$ is arbitrary and all indices are defined modulo $n$. The $R$ invariants have the expression
\begin{align}\label{R}
R_{rst} = \frac{\vev{s-1\; s} \vev{t-1\; t} \delta^{(4)}(\Xi_{rst})}{x^2_{st} \vev{r|x_{rs} x_{st}|t-1} \vev{r|x_{rs} x_{st}|t} \vev{r|x_{rt} x_{ts}|s-1} \vev{r|x_{rt} x_{ts}|s}}\,,
\end{align}
where the Grassmann delta function is defined by  
\begin{align}\label{}
\delta^{(4)}(\Xi_{rst}) =  \frac1{4!} \ep_{ABCD} \Xi_{rst}^A\Xi_{rst}^B\Xi_{rst}^C\Xi_{rst}^D\,,
\end{align} 
with
\begin{align}\label{}
\Xi_{rst}^A  &= x^2_{st} \vev{r \q_r^A} + \vev{r|x_{rs} x_{st}|\q_t^A}  + \vev{r|x_{rt} x_{ts}|\q_s^A}\nt
& = \vev{r|x_{rs} x_{st}|\q_{rt}^A}  + \vev{r|x_{rt} x_{ts}|\q_{rs}^A} \,.
\end{align}
The $R$ invariants are homogeneous functions of degree four of the dual superspace odd variables $\q^A_i$, and are rational functions of the dual space points $x_i$. 

Eq.~\p{R} is manifestly invariant under the full dual $PSU(2,2|4)$. In addition, it is also invariant under the ordinary superconformal symmetry of the amplitude, acting non-locally in dual superspace \cite{Korchemsky:2009jv}. The combination of dual and ordinary superconformal symmetry has a Yangian algebraic structure \cite{Drummond:2009fd}. It can be shown \cite{Korchemsky:2010ut} that  Eq.~\p{R} are the unique (up to a constant factor) representations of the Yangian of Grassmann degree 4.

An important feature of the ratio \p{ra} is its singularity structure. Each individual $R$ invariant \p{R} has two types of singularities, physical and spurious. The former correspond to the multi-particle discontinuities of the amplitude and occur at $x^2_{st}=0$, with $|s-t| \geq 2$. The latter correspond to any of the four brackets in the denominator of \p{R} vanishing. The amplitude should be free from spurious singularities and indeed, it can be shown \cite{Korchemsky:2009hm} that the sum of $R$ invariants with {\it equal coefficients} in \p{ra} is the unique combination with this property. So, in conclusion,  the ratio \p{ra} is a meromorphic function of $x_{st}^2$ fully characterized by its poles at $x^2_{st}=0$, with $|s-t| \geq 2$. Knowing the residues at these poles allows us to unambiguously reconstruct the ratio function. 

\subsubsection{$R$ invariants and momentum supertwistors}\label{Rims}

The $R$ invariants can be rewritten in terms of the so-called momentum super-twistors \cite{hodges,ArkaniHamed:2009dn,Mason:2009qx}. The bosonic momentum twistor is defined by
\begin{align}\label{}
Z^{M}_i = 
\left(
\begin{array}{r}
  \ket{i}   \\
  x_i\ket{i}  
\end{array}
\right) = 
\left(
\begin{array}{r}
  \ket{i}   \\
  x_{i+1}\ket{i}  
\end{array}
\right) \,.
\end{align}
Here $M =1,2,3,4$ is an index in the fundamental representation of the conformal group $SU(2,2)$ or rather, of its complexification $SL(4,\mathbb{C})$. Four such twistors, $Z^{M}_i$ with $i=1,2,3,4$, can form an $SL(4)$ (i.e., conformal) invariant given by the determinant of the $4\times 4$ matrix:
\begin{align}\label{}
(1234) = \ep_{MNPQ} Z^{M}_1 Z^{N}_2 Z^{P}_3 Z^{Q}_4\,.
\end{align}  
The fermionic part of the supertwistor is defined by 
\begin{align}\label{}
\chi^A_i = \vev{i \q^A_i}=\vev{i \q^A_{i+1}} \,.
\end{align}
The advantage of the momentum supertwistors notation is the linear action of the dual superconformal group $SL(4|4)$, which makes the symmetry properties of the $R$ invariants more transparent. In \cite{Mason:2009qx} the $R$ invariants \p{R} were rewritten in terms of momentum supertwistors with 5 labels $r,s-1,s,t-1,t$ instead of the 3 labels $r,s,t$ in \p{R}:
\begin{align}\label{NMH}
R_{rst} 
&= \frac{\delta^{(4)}(\Sigma_{r, s-1, s, t-1, t})}{(s-1,s,t-1,t)(s,t-1,t,r)(t-1,t,r,s-1)(t,r,s-1,s)(r,s-1,s,t-1)}\,,
\end{align} 
where
\begin{align}\label{5.8}
\Sigma_{r, s-1, s, t-1, t} &= (s-1,s,t-1,t)\chi_r  + (s,t-1,t,r)\chi_{s-1}  + (t-1,t,r,s-1)\chi_{s} \nt &  + (t,r,s-1,s)\chi_{t-1}   + (r,s-1,s,t-1)\chi_{t} \,.
\end{align}

\subsubsection{$R$ invariants and analytic superspace} \label{As}

For the purpose of comparing amplitudes with correlators, we need to express both of them in terms of the same odd variables. We recall that
the scattering super-amplitudes depend on $4n$  momentum supertwistor odd variables  $\chi_i^A$ (with $i=1,\ldots,n$ and $A=1,2,3,4$), while the correlators depend on the same number of analytic superspace
variables $(\q^{+}_{i})^a_\alpha$ (with $a,\alpha=1,2$). To establish the relation
between the two quantities, we have to find a change of variable relating the $\q^+$'s to the $\chi$'s. 

 This is done by decomposing the $SU(4)$ index of $\chi^A_i$ in the basis given by the harmonics at points $i$ and $i+1$:
\begin{align}\label{5.10}
\chi_i^A \quad \to \quad \chi_{i/i}^a =\chi_i^A (u_i)^{+a}_A \,,\qquad
\chi_{i/i+1}^a =\chi_i^A (u_{i+1})^{+a}_A\,.
\end{align}
The variables $\chi_{i/i}^a$ and  $\chi_{i/i+1}^a$ defined in this way carry the
helicity of the $i$th particle and the $U(1)$ harmonic charges at points $i$ and $i+1$,
respectively. Then, we identify them with the analytic superspace variables for the correlators,
projected onto the helicity state of the particle $i$,
\begin{align}\label{chi-p}
\chi_{i/i}^a  := \vev{i\theta_i^{+a}}\,,\qquad 
\chi_{i/i+1}^a := \vev{i\theta_{i+1}^{+a}}
\end{align}
and inversely
\begin{align}\label{5.11}
\q^{+a}_{i\, \a} := \frac{\chi^a_{i-1/i}}{\vev{i-1\ i}}\  \la_{i\, \a} +  \frac{\chi^a_{i/i}}{\vev{i\ i-1}}\  \la_{i-1\, \a}\,.
\end{align}
Eqs.~\p{chi-p} and \p{5.11} establish the relation between the $\q^+-$ and $\chi-$variables. In addition, 
we obtain from \p{chi-p} (the $SU(2)\times SU(2)'$ indices are suppressed)
\begin{align}\label{chit}
\chi^A_i &= \chi_{i/i+1}   (\bar i\ i+1)^{-1} \bar i^A +  \chi_{i/i} (\overline{ i+1} \ i)^{-1}\overline{ i+1}^A \notag\\
&= \vev{i \q^+_{i+1}} (\bar i\ i+1)^{-1} \bar i^A +  \vev{i \q^+_{i}}  (\overline{ i+1} \ i)^{-1}\overline{ i+1}^A \,,
\end{align}
where we used the notation from \p{y}. To prove this relation, we multiply it by $i_A^{+a}$ and by $(i+1)_A^{+a}$  and use the fact that, e.g., $\bar i_{-a'}^A i_A^{+a}=0$.

We see from \p{chit} that each variable $\chi_i^A$ is equivalent to a pair of analytic superspace variables at two adjacent points, $ \vev{i \q^{+a}_{i}}$ and $\vev{i \q^{+a}_{i+1}}$. Substituting \p{chit} into \p{NMH} we find that the $R$ invariant $R_{r, s-1, s, t-1, t}$ depends on a set of eight points in analytic superspace and the resulting expression is rather cumbersome. 
We can however use $Q$ and $\bar S$ supersymmetry to simplify the $R$ invariants. We recall
that these generators transform $\chi_i$ linearly and we can make use
of them to gauge away 16 components of the $\chi$'s.  Previously, we used the gauge  \p{qsga} which eliminates the odd variables at four points in  analytic superspace. As we will see in a moment, for our purposes in this section it is convenient to choose the gauge 
\begin{align}\label{Cg}
\mbox{$Q+\bar S$ gauge:} \qquad     (\q_s)^A_\a = (\q_{t-1})^{+a}_\a = (\q_{t+1})^{+a}_\a = 0\,, 
\end{align}
or equivalently,
\begin{align}\label{gauge1}
 \chi_{s-1}^A=\chi_s^A=\chi_{t-2/t-1}^a=
\chi_{t-1/t-1}^a=\chi^a_{t/t+1}=\chi^a_{t+1/t+1}=0\,,
\end{align}
where the projected variables $\chi_{i/j}$ were defined in \p{5.10}, \p{chi-p}.

\subsubsection{Residues of the $R$ invariants}\label{rri}

Making use of the identity $(s-1,s,t-1,t) = x_{st}^2 \vev{s-1\; s}\vev{t-1\; t}$,
we find that the $R$ invariant \p{NMH} has a pole at $x_{st}^2=0$. To evaluate the 
residue at this pole, it is convenient to introduce the new spinors 
\begin{align}\label{J}
  x_{st}= \sum_{i=s}^{t-1} |i] \bra{i} \equiv |J] \bra{J}  \,.
\end{align}
This leads to the following expressions for the four non-vanishing factors in the denominator in \p{NMH}:
\begin{align}\label{5.3}
&(s,t-1,t,r)  = \vev{J\; s} \vev{t-1\; t} [J|x_{rs}\ket{r}   \nt 
&(t-1,t,r,s-1)  = \vev{J\; s-1} \vev{t-1\; t} [J|x_{rs}\ket{r}    \nt 
&(t,r,s-1,s)  = \vev{J\; t} \vev{s-1\; s} [J|x_{rs}\ket{r}    \nt
&(r,s-1,s,t-1) = \vev{J\; t-1} \vev{s-1\; s} [J|x_{rs}\ket{r}\,.
\end{align}
Then,  the residue of the $R$ invariant  \p{NMH} (in the gauge \p{gauge1}) takes the following very simple form (see Appendix~\ref{dofeq})
\begin{align}\label{NMfin}
 {\rm Res}_{x^2_{st}=0}\ R_{rst}  =  \frac{(\q^+_t)^4}{12}\ \frac{\vev{s-1\; s}\vev{J\; t-1} \vev{J\; t}}{\vev{t-1\; t} \vev{J\; s-1} \vev{J\; s} } \frac{(\mathbf{t-1\; t+1})}{(\mathbf{t-1\; t}) (\mathbf{t+1\; t})} 
 \,.
\end{align}   
Notice that the residue does not depend on point $r$.
We can rewrite it in terms of space-time variables by multiplying the numerator and the denominator by $[t-1\; t] [s-1\; J] [s\; J]$:
\begin{align}\label{NMx}
 &{\rm Res}_{x^2_{st}=0}\ R_{rst} =
  \frac{(\q^+_t)^4}{12}\ \frac{\tr(x_{s-1,s} \tx_{s,s+1}x_{s+1,t-1} \tx_{t-1,t} x_{t,t+1} \tx_{t+1,s-1})}{x^2_{t-1,t+1} x^2_{s-1,t} x^2_{s+1,t}} \frac{(\mathbf{t-1\; t+1})}{(\mathbf{t-1\; t}) (\mathbf{t+1\; t})} 
 \,,
\end{align}  
where the explicit expression for the trace can be found in \cite{Eden:2010ce}.
This residue has manifest (dual) conformal symmetry. It has zero conformal weight at points $s-1$ and $t+1$, and has weight $(-1)$ at points $s$ and $t$, matching the weight of the singular factor $1/x^2_{st}$ that has been pulled out. 

\subsection{Residues of the correlator}

In this subsection we study the behavior of the correlator $G^{(0)}_{n;1}$  in the singular regime  $x^2_{st}\to0$. This drastically simplifies the calculation, reducing it to an elementary Feynman graph in terms of component fields.

\subsubsection{The origin of the residues of the correlator}

As explained in the beginning of this section, to prove the duality \p{3.12} it is sufficient to show that the residues of both sides are the same,
\begin{align}\label{cjt}
   {\rm Res}_{x^2_{st}=0}\ \lim_{x^2_{i,i+1} \to 0} \frac{G^{(0)}_{n;1}}{G^{(0)}_{n;0}}  =   2\, {\rm Res}_{x^2_{st}=0} R^{\rm NMHV}_n =   2\, {\rm Res}_{x^2_{st}=0} R_{rst} \,.
\end{align} 
Here in the last relation we applied \p{ra} and took into account that the residue of $R^{\rm NMHV}_n$ only receives a nonzero contribution from $R_{rst}$.

The practical question is how to calculate the residue of the ratio of correlators in the left-hand side of \p{cjt}. We recall that the limit $x_{st}^2\to 0$ corresponds to the situation
where two of the operators in the correlator $G^{(0)}_{n;1}$, located at points $x_s$ and $x_t$, become null separated. Were  these points completely independent, one would say that the asymptotic behavior of the correlator for $x_{st}^2\to 0$ follows from the operator product expansion of these operators, regardless of their neighbors. However, points $x_s$ and $x_t$ are already null separated from their neighbors, which may cause interference with other operators. Here we argue that the $(Q+\bar S)-$supersymmetry  gauge \p{Cg} allows us to avoid this effect. 

In  the gauge \p{Cg} we have $\theta_s^+=\q^+_{t-1} = \q^+_{t+1}=0$, so
the stress-tensor multiplets  $\cT$ at points $s$, $t-1$ and $t+1$ are reduced to their lowest bosonic component  $\cO=\tr(\phi^{++}\phi^{++})$. Further, the gauge condition $\q^A_s=0$ in \p{Cg} and the fermionic light-like condition in \p{duth} imply $\q^{+a\,\a}_{s-1}= \la^\a_{s-1} \eta^a_{s-1/s-1}$ and $\q^{+a\,\a}_{s+1}= \la^\a_{s} \eta^a_{s/s+1}$. This cuts the expansion \p{t4c} by half, e.g.,
\begin{align}\label{5.21}
\cT(s+1)&=   \tr(\phi^{++}\phi^{++})
  \nt[1.2mm] &
-2i \sqrt{2} \eta^a_{s/s+1}\ \tr( \vev{s\, \psi^{+}_a}\phi^{++})\nt[1.2mm]
&+  (\eta_{s/s+1})^2\ \tr \big\{  i \sqrt{2}\bra{s}F\ket{s} \phi^{++}  +  \vev{s\, \psi^{+c}}  \vev{s\, \psi^{+}_c} \big\} \,.
\end{align} 
Finally, at point $t$ the stress-tensor multiplet has its full expansion \p{t4c}. 

We need to find out how the various operators at points $s-1,s,s+1$ and $t-1,t,t+1$ can interact with each other, in order to contribute to  $G^{(0)}_{n;1}$  at the lowest level $g^2$ in the coupling.  These interactions should produce poles at $x^2_{i,i+1} = 0$ (to be canceled by those of $G^{(0)}_{n;0}$ in \p{cjt}) and  at $x^2_{st} = 0$, whose residue we want to compute. We are only interested in terms of Grassmann degree $(\q)^4$. In the gauge \p{Cg} and in the singular regime $x^2_{st} \to 0$ the odd variables are available at points $s-1$, $s+1$ and $t$. 

Suppose that we try to collect odd variables from points $s-1$ or $s+1$. The operators there need to talk to their scalar neighbors  at the more distant points $s-2$ or $s+2$, in order to close the frame of free propagators $1/x^2_{i,i+1}$. This leaves the fields $F$ and $\psi$ at points $s-1$ or $s+1$ the possibility to  talk to the scalars at point $s$ via cubic couplings of order $O(g)$. The third field from such a cubic coupling will have to talk to a matching field from the operators at points  $t-1,t,t+1$, in order to produce the required pole at $x^2_{st}=0$. But this is not all, at points  $t-1,t,t+1$ we still need fields not engaged  in the interaction, which could talk to their neighbors and restore the frame of propagators. It is easy to see that this scenario is not possible at order $O(g^2)$. Thus, the odd variables cannot come from points $s-1$ or $s+1$, all four of them have to appear at point $t$, accompanied by the chiral on-shell Lagrangian $\cL(x_t)$.

We conclude that in the gauge \p{gauge1} and at  order $O(g^2)$, the residue of the correlator $G_{n;1}^{(0)}$ at $x_{st}^2=0$ 
originates from the interaction of $\cL$ at point $x_t$ with a frame of  scalar operators $\cO$ at all points $x_i\neq x_t$:
\begin{align}\label{specor}
 {\rm Res}_{x^2_{st}=0} G^{(0)}_{n;1} = \frac13 (\q^+_t)^4\ {\rm Res}_{x^2_{st}=0}\ \vev{\cO(1) \ldots  \cO(t-1) \cL(t) \cO(t+1) \ldots  \cO(n)}^{\rm (0)}\,.
 \end{align}
More precisely, we will look for singular exchanges between the Lagrangian at point $x_t$ and the scalars at points $s-1,s,s+1$ under the additional light-cone condition $x_{i,i+1}^2\to 0$.

\subsubsection{Evaluating the singularities of the correlator}\label{esc}


In the light-cone limit $x_{i,i+1}^2\to 0$, we can view the $n-$point correlator 
in the right-hand side of \p{specor}
as obtained from the $(n-1)-$point scalar correlator
\begin{align}\label{de}
\cG^{(0)}_{n-1} = \vev{\cO(1) \ldots  \cO(t-1)   \cO(t+1) \ldots  \cO(n)}^{(0)}
\end{align}
by making a Lagrangian insertion at point $t$.
Notice the gap between $t-1$ and $t+1$ in \p{de}, which implies that these two points are not light-like separated, $x_{t-1,t+1}^2\neq 0$. 

We start with the tree-level expression for the correlator \p{de} and for the one in the denominator of \p{cjt}. They are given by products of scalar propagators \p{2w'}:
\begin{align}
\cG^{(0)}_{n-1} &=  (N^2_c-1) \prod_{i=1}^{t-2} \frac{(\mathbf{i\; i+1})}{4\pi^2 x^2_{i,i+1}}\ \frac{(\mathbf{t-1\; t+1})}{4\pi^2 x^2_{t-1,t+1}}  \   \prod_{j=t+1}^{n} \frac{(\mathbf{j\; j+1})}{4\pi^2 x^2_{j,j+1}}+ \ldots \label{5.29} \\
G^{(0)}_{n;0} &=  (N^2_c-1) \prod_{i=1}^{n} \frac{(\mathbf{i\; i+1})}{4\pi^2 x^2_{i,i+1}}+ \ldots\,,   \label{5.30}
 \end{align}
where the ellipses denote terms subleading as $x_{i,i+1}^2\to 0$.
In the light-cone limit one of the propagators in \p{5.29} remains finite, $x_{t-1,t+1}^2\neq 0$, while  the expression \p{5.30} contains an additional singular factors $1/(x_{t-1,t}^2 x_{t,t+1}^2)$,
\begin{align}\label{missing}
G^{(0)}_{n;0}/ \cG^{(0)}_{n-1}  =\frac1{4\pi^2} \frac{x^2_{t-1,t+1}}{x_{t-1,t}^2 x_{t,t+1}^2} \frac   {(\mathbf{t-1\; t})(\mathbf{t\; t+1})}{(\mathbf{t-1\; t+1})} +\ldots\ .
\end{align}
Thus, for the ratio of the correlators $G_{n;1}^{(0)}/G^{(0)}_{n;0}$ to be finite in the 
light-cone limit, the insertion of the Lagrangian $\cL(t)$ into the correlator \p{de} should produce this missing factor. In addition, since we are looking
for poles at $x_{st}^2=0$, we have to make sure that the insertion $\cL(t)$ produces a
contribution $\sim 1/x_{st}^2$.

Looking at the chiral Lagrangian \p{LSD}, it is easy to see that the tree-level, i.e. the lowest-order contribution to the correlator in \p{specor} is of order  $O(a)$ and it  can only come from the interaction of the gluon term $F^{\a\b} F_{\a\b}$ with the frame of scalar propagators. The other, non-linear terms in   \p{LSD} start contributing at order $O(a^2)$. 
To lowest order in the coupling, each factor $F^{\a\b}(x_t)$ from $\cL=-\frac12 \tr (F_{\a\b}F^{\a\b})+ \ldots$ can 
simultaneously interact with two scalars, say $\phi^{++}(x_i)$ and $\phi^{++}(x_j)$. The
corresponding correlator is known as the (bosonic) T-block \cite{Eden:2000mv,Eden:2010zz}:
\begin{align}\label{T-block}
T^{\alpha\beta}(x_i,x_t,x_j) &= \vev{\phi^{++}(x_i) F^{\alpha\beta}(x_t) \phi^{++}(x_j)} \nt[2mm]
&=  \vev{\phi^{AB}(x_i)F^{\alpha\beta}(x_t) \phi^{CD}(x_{j})}
\big[
(u_i)_A^{+a} \ep_{ab} (u_i)^{+b}_B (u_j)_C^{+c} \ep_{cd} (u_j)^{+d}_D \big]
\nt
& = \frac{g}{(4\pi^2)^2} \frac{(x_{it} x_{jt})^{(\alpha\beta)}}{x_{ij}^2 x_{it}^2 x_{jt}^2}\, (\mathbf{i\; j})\,.
\end{align}
Examining this expression for different values of $i$ and $j$ we notice that for $i=t-1$ and $j=t+1$ the T-block involves the same  two `missing' factors $1/(x_{t-1,t}^2 x_{t,t+1}^2)$ 
as in \p{missing}. Thus, in order to restore the correct light-cone asymptotic of the correlator, the field strength tensor $F^{\a\b}(x_t)$ should interact with the scalars $\phi^{++}(x_{t-1})$ and $\phi^{++}(x_{t+1})$. In addition to this, we have to ensure that the second
factor $F_{\a\b}(x_t)$ inside the chiral Lagrangian produces a singularity at $x_{st}^2=0$. It follows from \p{T-block} that this happens for two different choices of the indices: for $i=s-1$ and $j=s$ and for $i=s$ and $j=s+1$. The corresponding Feynman
diagrams are shown in Fig.~\ref{Fig1}.  

\bigskip

\begin{figure}[h!]
\psfrag{dots}[cc][cc]{$\mathbf{\dots}$}
\psfrag{1}[cc][cc]{$x_{t+1}$}
\psfrag{2}[cc][cc]{$x_{s-1}$}
\psfrag{3}[cc][cc]{$x_s$}
\psfrag{4}[cc][cc]{$x_{s+1}$}
\psfrag{5}[cc][cc]{$x_{t-1}$}
\psfrag{6}[cc][cc]{$x_t$}
\centerline{ \includegraphics[height=50mm]{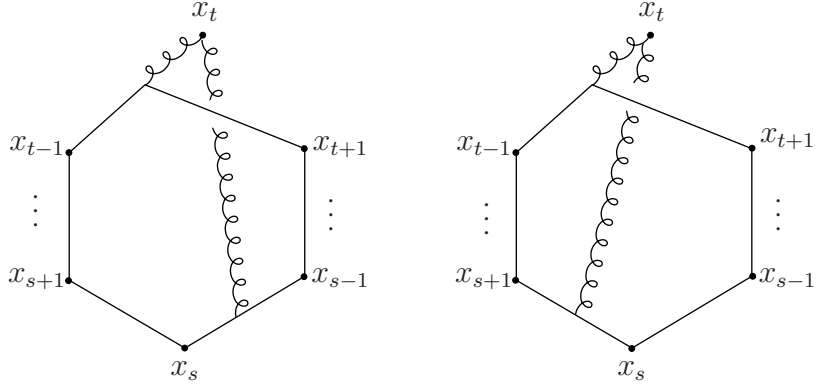} }
\caption{Feynman diagrams providing the dominant contribution to the correlator $\frac13 (\theta_t^+)^4\vev{  \cL(x_t) \prod_{i\neq t} \cO(x_i) }$
 in the double scaling limit, $x_{st}^2\to 0$ and $x_{i,i+1}^2\to 0$, to lowest
order in the coupling. {Solid and cirly lines denote scalar and gauge field propagators, respectively.}
}\label{Fig1}
\end{figure}

Let us start with the diagram shown in the left panel in Fig.~\ref{Fig1}. 
The vertex at the  point $x_t$ describes the insertion of the operator $-\frac16(\theta_t^+)^4  \tr (F_{\a\b}F^{\a\b})$. Then, its contribution is given by
the product of two T-blocks and $(n-3)$ scalar propagators:
\begin{align}
 -\frac16(\theta_t^+)^4 N_c\, T^{\alpha\beta}(x_{t-1},x_t,x_{t+1}) T_{\alpha\beta}(x_{s-1},x_t,x_{s}) \lr{ (N^2_c-1)\prod_{i\neq s-1,t-1,t} \frac{(\mathbf{i\; i+1})}{4\pi^2 x_{i,i+1}^2}}\ .
\end{align}
Adding the contribution of the second diagram in Fig.~\ref{Fig1} and taking into account
the explicit expressions for the T-block \p{T-block}, we find  
\begin{align}\label{expect}
 -\frac{1}{6}  a\,(\theta_t^+)^4
 \frac{1 }{ x_{st}^2 }  \frac{(\mathbf{t-1\; t+1})}{(\mathbf{t-1\; t}) (\mathbf{t\; t+1})} \   \frac{(x_{t-1,t} x_{t,t+1})^{(\alpha\beta)}}{x^2_{t-1,t+1}}\left[ 
 \frac{(x_{s-1,t} x_{t,s})_{(\alpha\beta)}}{ x_{s-1,t}^2  }
 +
  \frac{(x_{st} x_{t,s+1})_{(\alpha\beta)}}{  x_{s+1,t}^2}
 \right]G_{n;0}^{(0)} \,, 
\end{align}
where $G_{n;0}^{(0)}$ is given by \p{5.30} and $a=g^2N_c/(4\pi^2)$. We recall that this relation defines the leading
asymptotic behavior of the correlator in the right-hand side of \p{specor} in the double
scaling limit $x_{st}^2\to 0$ and $x_{i,i+1}^2\to 0$.

The first factor in \p{expect} is the expected pole $1/x_{st}^2$. The factor $G_{n;0}^{(0)}$ is the leading light-cone singular term in the tree-level $n-$point correlator \p{5.30}. Hence,
\begin{align}\notag
 {\rm Res}_{x^2_{st}=0} \ G^{(0)}_{n;1}/G^{(0)}_{n;0} 
 =&  -\frac{1}{6}  \,(\theta_t^+)^4  \, \frac{(\mathbf{t-1\; t+1})}{(\mathbf{t-1\; t}) (\mathbf{t\; t+1})} 
\\\label{racoo} 
&\times
  \frac{(x_{t-1,t} x_{t,t+1})^{(\alpha\beta)}}{x^2_{t-1,t+1}}\left[ 
 \frac{(x_{s-1,t} x_{t,s})_{(\alpha\beta)}}{ x_{s-1,t}^2  }
 +
  \frac{(x_{st} x_{t,s+1})_{(\alpha\beta)}}{  x_{s+1,t}^2}
 \right]  \,.
\end{align}
In the right-hand side of this expression we are now allowed to go to the limit $x^2_{i,i+1} = x^2_{st}=0$.%
\footnote{ The only exception is the boundary case $|s-t|=2$, which is treated in Appendix~\ref{bot}. }
Substituting $x_{st} = \ket{J} [J|$ and using $x_{s-1,t}^2 =(x_{s-1,s}+x_{st})^2 =
\vev{s-1\; J}[J\; s-1]$ and $x_{s+1,t}^2 =(x_{s,s+1}-x_{st})^2 =
-\vev{s\;J}[J\; s]$, we find after some algebra
\begin{align}\label{}
 {\rm Res}_{x^2_{st}=0}\ \lim_{x^2_{i,i+1} \to 0}  \frac{G^{(0)}_{n;1}}{G^{(0)}_{n;0}} = 2 \,\frac{(\theta_t^+)^4}{12}  \frac{(\mathbf{t-1\; t+1})}{(\mathbf{t-1\; t}) (\mathbf{t\; t+1})} \ \frac{\vev{s-1\; s}}{\vev{t-1\; t}} \frac{\vev{t-1\; J}\vev{J\; t}}{\vev{s-1\; J}\vev{J\; s}} \,.
\end{align}
This result is twice the expression in  \p{NMfin},  thus confirming the duality relation \p{cjt}.

\section{Discussion and conclusions}\label{Conclu}

In this and in the twin paper \cite{twin} we have proposed a new duality between supersymmetric correlation functions on the light cone and scattering super-amplitudes. We have shown, in a number of examples, that the super-correlators, computed at tree level, exactly reproduce the structure of the super-amplitudes at tree level and of their integrands at loop level. 

A very important feature of this duality is that we are dealing with finite objects defined in $D=4$ dimensions. At no point we need regularization, avoiding the related delicate issues about anomalies, etc. Thus, we are comparing objects with exact (super)conformal symmetry. 

It should be pointed out, however, that at the starting point of the discussion of super-correlators we made the choice to set all anti-chiral Grassmann variables $\bq_i=0$, in order to match the nature of the super-amplitude formulated in chiral dual superspace. A priori, this would mean that we have deliberately sacrificed the anti-chiral ($\bar Q$) half of Poincar\'e supersymmetry. Yet, quite surprisingly, this is not always the case! The explicit example in Section~\ref{NMtc} shows how the nilpotent tree-level correlator $G^{(0)}_{n;0}$ reproduces the NMHV tree \p{ra}. The latter is made of $R$ invariants \p{R}, which are known to have full $\cN=4$ supersymmetry, despite their manifest chiral appearance. This property is due to the very special Grassmann delta function in the numerator of \p{R}, which suppresses the $\bar Q$ variation of the denominator.  So, contrary to all expectations, the correlator of stress-tensor multiplets $\cT$, restricted to its purely chiral sector and put on the light cone, still has the full supersymmetry ($Q$ and $\bar Q$). At present we have no explanation for this ``miracle", but it shows that our knowledge of the nilpotent invariants in analytic superspace is probably incomplete. 

The next question is what happens to $\bar Q$ supersymmetry at loop level. As conjectured in this paper and confirmed by the examples worked out in \cite{twin}, the {\it integrands} of the super-amplitudes at loop level are given by the same type of tree-level super-correlators of $\cT$. However, this time some of the points of the correlator are not on the light cone, they serve as integration points in the loop integrals. If we attribute the miraculous restoration of $\bar Q$ supersymmetry to the light-like kinematics, we should expect that pulling some points away from the light cone would break $\bar Q$ supersymmetry. Indeed, we learn from \cite{ArkaniHamed:2010kv} that in the Grassmannian approach to the same integrands for loop amplitudes one observes that the $\bar Q$ variation of the integration points does not vanish, but it produces total space-time derivatives. This fits very well with our interpretation of the integration points as insertions of the chiral on-shell Lagrangian $\cL$. The latter is invariant under $Q$, but it transforms into a total derivative under $\bar Q$. 

Still, one might wonder why the complete correlator \p{2.19}, with its full dependence on $\q$ and $\bq$ and unbroken $Q$ and $\bar Q$ supersymmetry, does not play a role in the duality with amplitudes? In it, the purely chiral sector \p{npt} is not invariant under $\bar Q$, but it transforms into the other sectors of the super-function. Could it be that the super-amplitudes admit an alternative, non-chiral description, where what is known now as the purely chiral super-amplitude will be just a small corner? Apart from unbroken $\bar Q$ supersymmetry, such a description will have  another advantage, manifest PCT symmetry. Of course, one would need to find out the meaning of the rest of such a much bigger super-amplitude. 

A related issue is the existence of a second, hidden superconformal symmetry of the correlators. We know that the amplitudes (at least at tree level), in addition to their natural on-shell symmetry, have another, dual superconformal symmetry. The former acts on the amplitude non-locally, while the latter is local. Inverting the roles, the correlator dual to the amplitude has its native superconformal  symmetry, which is the dual symmetry of the amplitude. The obvious question now is: What is the analog of the native non-local symmetry of the amplitude, applied to the correlator, and why is it there? Is this some symmetry enhancement due to the light-cone limit or is there some other hidden reason for it? 

Along the same lines, the Grassmannian approach to amplitudes reveals a recursive structure \cite{ArkaniHamed:2010kv,Boels:2010nw} which allows in principle to obtain any type of amplitude by a finite number of algebraic steps. The amplitude/correlator duality implies the existence of an analogous recursion for the latter. What is the dynamical principle behind it? And, more generally, why are amplitudes dual to light-cone correlators? Is there yet another well hidden symmetry which completely fixes both objects?

{Other related recent developments suggest that it would be interesting to study gauge invariant operators inserted into non-trivial external states. On the one hand, there is evidence of a  duality between form factors (the case with a single operator insertion) and Wilson loops~\cite{Alday:2007he}. On the other hand, applications of BCFW at strong coupling show that vacuum expectation values of operators naturally mix with  operators inserted inside external states~\cite{Raju:2011mp}. Amplitudes and correlation functions are both special cases of such objects.}

We would like to make a comment about the third link in the triality relation  \p{diagram}, the link correlator $\to$ Wilson loop. In \cite{AEKMS} it was shown how  one obtains a light-like Wilson loop by taking the light-cone limit of the correlator in $D=4-2\ep$ dimensions and setting $x^2_{i,i+1}=0$. A general argument was given for the validity of this relation, at least for scalar correlators and bosonic Wilson loops. The next obvious question is what happens if we apply the same limiting procedure to the super-correlator? Should we expect to arrive at the super-Wilson loop of \cite{Mason:2010yk,Simon} or not? We postpone the answer to this question to the future, but we have to stay alert that in this singular limit we ought to handle divergent objects with due care \cite{Belitsky:2011zm}. 

Finally, one should not forget that initially, in the context of the AdS/CFT correspondence, the correlators of 1/2 BPS operators have been studied in view of their duality with AdS supergravity amplitudes (see, e.g., \cite{D'Hoker:1999pj,Arutyunov:2000py}). Now we propose a new duality between the same correlators and amplitudes on the CFT boundary of AdS. It is tempting to try to establish a more profound link between the old and the new dualities, although this is far form obvious. The only  free parameter in AdS supergravity is $N_c$, there is nothing which could match the perturbative  expansion parameter $a$.  Still, we know from \cite{am07} that the string dual to the CFT gluon amplitudes are  the minimal surfaces spanned on light-like polygons. It is perhaps necessary to investigate how these surfaces are related to AdS amplitudes plus possible string corrections.

\section*{Acknowledgements}

ES is grateful to  Sergio Ferrara, Paul Howe  and Raymond Stora {
  and PH is grateful to Valya Khoze and Claude Duhr} for a
number of enlightening discussions.   GK and ES acknowledge discussions
with David Skinner and Simon Caron-Huot. BE and PH acknowledge support by STFC under the rolling grant ST/G000433/1. 

\appendix 
 
\section{Appendix A: $\cN=4$ super-Yang-Mills Lagrangian}

\subsection{Conventions and definitions}\label{a.1}

The dotted and undotted spinor indices, as well as the $SU(2)$ indices, are raised and lowered as
follows:
\begin{eqnarray}\label{2.57}
  &&\psi^\alpha = \epsilon^{\alpha\beta}\psi_\beta\,, \qquad \bar\chi^{\dot\alpha} =
  \epsilon^{\dot\alpha\dot\beta}\bar\chi_{\dot\beta}\,, \qquad \psi_\alpha = \epsilon_{\alpha\beta}\psi^\beta\,, \qquad \bar\chi_{\dot\alpha} =
  \epsilon_{\dot\alpha\dot\beta}\bar\chi^{\dot\beta}\,, \label{2.58}
\end{eqnarray}
where the antisymmetric $\epsilon$ symbols have the properties:
\begin{equation}\label{2.59}
  \epsilon_{12} = \epsilon_{\dot 1\dot 2} = -\epsilon^{12} = -\epsilon^{\dot 1\dot 2} =
  1\,, \qquad \epsilon_{\alpha\beta}\epsilon^{\beta\gamma} =
  \delta_\alpha^\gamma\,,  \qquad \epsilon_{\dot\alpha\dot\beta}\epsilon^{\dot\beta\dot\gamma} =
  \delta_{\dot\alpha}^{\dot\gamma}\,.
\end{equation}
The convention for the contraction of a pair of spinor indices is
\begin{equation}\label{2.60}
  \psi^\alpha\lambda_\alpha\equiv \psi\lambda\,, \qquad \bar\chi_{\dot\alpha}\bar\rho^{\dot\alpha}
  \equiv \bar\chi\bar\rho\,, \qquad \psi^2 \equiv \psi^\alpha\psi_\alpha\,,
  \qquad \bar\psi^2 \equiv \bar\psi_{\dot\alpha}\bar\psi^{\dot\alpha}\,.
\end{equation}

We switch between spinor and vector notation using
\begin{align}\label{spinor-vector1}
&x_{\alpha \dot{\alpha}} =
\sigma^{\mu}_{\alpha \dot{\alpha}} x_{\mu}  \,, \qquad x^{\dot{\alpha}\a} =
\tilde\sigma_{\mu}^{\dot{\alpha}\a} x^{\mu}\nt
&  x_{\mu}=\frac{1}{2}(\sigma_\mu)_{\a\da} x^{\da\a}=\frac{1}{2} \tilde\sigma_{\mu}^{ \dot{\alpha}\alpha} x_{\alpha
\dot{\alpha}} \,, \qquad x^2 = x_{\mu}
x^{\mu} = \frac{1}{2} x_{\alpha \dot{\alpha}}  x^{
\dot{\alpha}\alpha} 
\end{align}
with $\tilde\sigma_{\mu}^{ \dot{\alpha}\alpha} = \ep^{\a\b} \ep^{\da\db} (\sigma_\mu)_{\b\db}$. 
Consequently, we have
\begin{equation}\label{spinor-vector2}
\partial^{\dot{\alpha}\a}x_{\beta\dot{\beta}} = 2
\delta^{\alpha}_{\beta} \delta^{\dot{\alpha}}_{\dot{\beta}}
\,,\qquad x^{ \dot{\beta}\alpha}  x_{\alpha \dot{\alpha}}= x^2
\delta_{\dot{\alpha}}^{\dot{\beta}} \,.
\end{equation}

The Yang-Mills field strength is defined by the commutator of two covariant
derivatives, $[\cD_\mu, \cD_\nu] = -ig F_{\mu\nu}$, with  $\cD_\mu = \pa_\mu -ig A_\mu$. In two-component spinor notation this becomes
\begin{equation}\label{comcor}
 F_{\mu\nu} (\sigma^\mu)_{\alpha\dot\alpha} (\sigma^\nu)_{\beta\dot\beta}
=\epsilon_{\alpha\beta}\bar F_{\dot\alpha\dot\beta} + \epsilon_{\dot\alpha\dot\beta} F _{\alpha\beta}\,,
\end{equation}
where $F_{\a\b} = F_{\b\a}$ and $F_{\da\db} = F_{\db\da}$ are given by  
\begin{align}
& F^\beta{}_\alpha=-\frac12 F_{\mu\nu} (\sigma^\mu\bar\sigma^\nu)_\alpha{}^\beta
\,,\qquad \bar F_{\dot\beta}{}^{\dot\alpha} = -\frac12F_{\mu\nu} (\bar\sigma^\mu\sigma^\nu)^{\dot\alpha}{}_{\dot\beta}
\end{align}
leading to
\begin{align}
F_{\mu\nu} F^{\mu\nu} =-\frac12\lr{F^\beta{}_\alpha F^\alpha{}_\beta +  \bar F_{\dot\beta}{}^{\dot\alpha}  \bar F_{\dot\alpha}{}^{\dot\beta} } =\frac12 \lr{F_{\alpha\beta}F^{\alpha\beta} + 
\bar F_{\dot\alpha\dot\beta}F^{\dot\alpha\dot\beta}} 
\end{align}

\subsection{Different forms of the $\cN=4$ SYM Lagrangian}\label{a.2}

The $\cN=4$ SYM Lagrangian
in Minkowski space,  written in two-component spinor notation,  has the form
\begin{align}\notag
\mathcal{L}_{\cN=4} =& \tr\bigg\{ -\frac1{4} \lr{F_{\alpha\beta}F^{\alpha\beta}+ \bar F_{\dot\alpha\dot\beta}\bar F^{\dot\alpha\dot\beta}} + \frac14 D_{\alpha\dot\alpha}\phi^{AB} D^{\dot\alpha\alpha}\phi_{AB} + \frac18 g^2 [\phi^{AB},\phi^{CD}][\phi_{AB},\phi_{CD}]
\\  
& 
+i \bar\psi_{\dot\alpha A}D^{\dot\alpha \alpha}  \psi_\alpha^A 
-i (D^{\dot\alpha \alpha}\bar\psi_{\dot\alpha A})  \psi_\alpha^A-\sqrt{2} g \psi^{\alpha A} [\phi_{AB},\psi_\alpha^B] +\sqrt{2} g
\bar\psi_{\dot\alpha A} [\phi^{AB},\bar\psi^{\dot\alpha}_B]\bigg\}\,.\label{A.1}
\end{align} 
All fields are in the adjoint representation of the gauge group
$SU(N_c)$, and the generators and the structure constants satisfy the
relations
\begin{equation}
\tr (t^{a} t^{b}) = \frac{1}{2}\delta^{ab}, \qquad f^{abc}f^{abd} = N_c\,\delta^{cd}\,.
\end{equation}
The scalar fields $\phi^{AB}$ satisfy the reality condition
\begin{equation}
\phi_{AB} =  \lr{\phi^{AB}}^\dagger  = \ft12 \ep_{ABCD}
\phi^{CD},
\end{equation}
where $\ep_{1234} = \ep^{1234} = 1$.

In this paper we use another, {\it on-shell and chiral} form of the Lagrangian. It is obtained by adding to \p{A.1} the following total derivative terms:
\begin{align}\label{A.10}
\Delta\cL=\tr\bigg\{- \frac1{4}  \lr{ F_{\alpha\beta}F^{\alpha\beta}- \bar F_{\dot\alpha\dot\beta}\bar F^{\dot\alpha\dot\beta}} -\frac1{4} \partial^{\dot\alpha \alpha}  (\phi^{AB} \cD_{\a\da}\phi_{AB})
- i\partial^{\dot\alpha \alpha} (\bar\psi_{\dot\alpha A} \psi_\alpha^A)\bigg\}\,,
\end{align}
and then using the field equations for the scalars and the fermions to eliminate their kinetic terms. The result is 
\begin{align}\label{A.11}
\mathcal{L}&=\left[ \cL_{\cN=4} + \Delta\cL\right]_{\rm on-shell}\nt
& = \tr \left\{- \frac12  F_{\alpha\beta}F^{\alpha\beta}  + {\sqrt{2}}  g \psi^{\alpha A} [\phi_{AB},\psi_\alpha^B] - \frac18 g^2 [\phi^{AB},\phi^{CD}][\phi_{AB},\phi_{CD}] \right\}\,.
\end{align} 
We note that the imaginary part 
\begin{align}\label{A.13}
{\rm Im}\,\mathcal{L} = {\rm Im}\,\Delta\cL = \tr \left\{ -\frac{i}{4} \lr{ F_{\alpha\beta}F^{\alpha\beta}- \bar F_{\dot\alpha\dot\beta}\bar F^{\dot\alpha\dot\beta}} -  \partial^{\dot\alpha \alpha} (\bar\psi_{\dot\alpha A} \psi_\alpha^A) \right\}
\end{align}
is a total derivative and a {\it pseudo-scalar}. 

\subsection{Supersymmetry transformations}\label{a.3}

The $\cN=4$ supersymmetry transformations that leave the action $S_{\cN=4}= \int d^4x\ \cL_{\cN=4}$  invariant are
\begin{eqnarray}
Q^\a_{A} \phi^{BC} &=& i\sqrt{2} (\delta^{B}_{A} \psi^{C \alpha} - \delta^{C}_{A} \psi^{B \alpha})  \nonumber\\
Q^\a_{A}  A_{\b\db} &=& -2i \delta^\a_\b \bar\psi_{A\db} \nonumber\\
Q^\a_{A} \psi_{\beta}^{B} &=& \delta^{B}_{A} F^\a_{\beta} + ig \lbrack \phi^{BC}, \phi_{CA} \rbrack \delta_{\beta}^\a  \nonumber\\
Q^\a_{A} \bar{\psi}^{\dot{\beta}}_{ B} &=& \sqrt{2} {\cal D}^{\dot{\beta}\a} \phi_{AB} \label{Qsusy}
\end{eqnarray}
together with the conjugate expressions for the action of $\bar Q^A_\da$. From the variation of $A_{\a\da}$ one derives that of the chiral part of the field strength:
\begin{eqnarray}\label{QF}
Q_A^\alpha \, F_{\beta\gamma} &= 
  i \delta^\alpha_{ \beta} D_{\gamma \dot\alpha} \bar\psi_A^{\dot\alpha} + (\b \leftrightarrow \gamma) \,,
\end{eqnarray}
which contains the Dirac operator from the field equation for the fermion. 

The algebra $\{Q^\a_{A} , Q^\b_{B}   \}=0$ of the transformations \p{Qsusy} closes modulo compensating gauge transformations and only on shell. The first effect can be seen from, e.g., the commutator (with $(\ep_i Q)=\ep_{i \a}^A Q^\a_{A}$)
\begin{align}\label{}
[(\ep_1 Q) , (\ep_2 Q) ]A_{\b\db} = i 2\sqrt{2}  \cD_{\b\db} (\ep^{A\a}_1 \ep^B_{2\a} \phi_{AB})\,,
\end{align}  
which has the form of a gauge transformation of $A_{\gamma\dot\gamma}$ with a  field-dependent parameter proportional to $\ep^{A\a}_1 \ep^B_{2\a} \phi_{AB}$. The second effect is present in the anticommutator (for simplicity, we set $g=0$)
\begin{align}\label{}
\{ Q^\a_{A} , Q^\b_{B}\} \psi^{C\gamma} = - i\delta^C_A( \ep^{\a\b} \pa^{\gamma\da} \bar\psi_{B\da}  +  \ep^{\gamma\b} \pa^{\a\da} \bar\psi_{B\da})  + (A \leftrightarrow B,\a \leftrightarrow\b)\,,
\end{align}
which vanishes only if the field equation for $\bar\psi$ is used. 

It is quite remarkable that this algebra closes {\it off shell} on a subset of the fields in \p{Qsusy}:
\begin{eqnarray}
Q^\a_{A} \phi^{BC} &=& i\sqrt{2} (\delta^{B}_{A} \psi^{C \alpha} - \delta^{C}_{A} \psi^{B \alpha})   \nonumber\\
Q^\a_{A} \psi_{\beta}^{B} ~&=& \delta^{B}_{A} F^\a_{\beta} + ig \lbrack \phi^{BC}, \phi_{CA} \rbrack \delta_{\beta}^\a  \nonumber\\
Q^\a_{A} \, F_{\b\gamma} &=& \sqrt{2}g \delta^{\a}_{\b} [\phi_{AB}, \psi^{B}_{\gamma}]  + (\b \leftrightarrow \gamma)   \,, \label{vemu'}
\end{eqnarray}
where the variation of $F$ from \p{QF} has been recast in the new form with the help of the field equation for $\bar\psi$. This off-shell algebra is used in Section~\ref{N4has} (see \p{vemu}). 

\section{Appendix: Derivation of Eq.~\p{NMfin}}\label{dofeq}

We are interested in the residue of the $R$ invariant \p{NMH} in the limit $x^2_{st} \propto (s-1,s,t-1,t) \to 0$. In this limit the momentum super-twistor $\chi_r$ in \p{5.8} drops out. Further,   in the gauge  \p{gauge1} two other odd variables  in \p{5.8} vanish,  $\chi_{s-1}=\chi_{s}=0$. According to \p{chit}, the remaining nonzero odd variables,  $\chi_{t-1}$ and $\chi_{t}$,
are expressed in terms of the two components of the same analytic $(\q^+_t)^{a}_\a$:
\begin{align}\label{} \notag
 \chi_{t-1}^A = \vev{t-1 | \q^+_t} (\overline{t-1}\; t)^{-1} \overline{t-1}^A\,, \qquad 
 \chi_{t}^A = \vev{t | \q^+_t} (\overline{t+1}\; t)^{-1} \overline{t+1}^A\,.
\end{align} 
Consequently, the Grassmann delta function in \p{NMH} becomes
\begin{align}\label{resu'}
&\delta^{(4)}\Big((t,r,s-1,s)\chi_{t-1} + (r,s-1,s,t-1)\chi_{t}\Big)   \nt
&= \delta^{(4)}\Big((t,r,s-1,s)\vev{t-1 | \q^+_t} (\overline{t-1}\; t)^{-1} \overline{t-1} + (r,s-1,s,t-1)\vev{t | \q^+_t} (\overline{t+1}\; t)^{-1} \overline{t+1}\Big)   \nt
& =   [(t,r,s-1,s)(r,s-1,s,t-1)]^2\nt
&\quad \times  \frac{6}{4!}  \ep_{ABCD} \Big[\vev{t-1 | \q^+_t}^a (\overline{t-1}\; t)^{-1}{}_a{}^{a'}\  (\overline{t-1})^A_{a'}\   \vev{t-1 | \q^+_t}^b (\overline{t-1}\; t)^{-1}{}_b{}^{b'}\  (\overline{t-1})^B_{b'}   \nt
 &\quad  \times  \vev{t| \q^+_t}^a (\overline{t+1}\; t)^{-1}{}_c{}^{c'}\  (\overline{t+1})^C_{c'}\   \vev{t | \q^+_t}^d (\overline{t+1}\; t)^{-1}{}_d{}^{d'}\  (\overline{t+1})^D_{d'}\Big] \nt
 &=   [(t,r,s-1,s)(r,s-1,s,t-1)]^2 (\vev{t-1 | \q^+_t})^2 (\vev{t| \q^+_t})^2 \nt
  &\times \frac1{4}\Big[ \ep^{ab} (\overline{t-1}\; t)^{-1}{}_a{}^{a'} (\overline{t-1}\; t)^{-1}{}_b{}^{b'} \ep_{a'b'}\Big] \Big[ \ep^{cd}  (\overline{t+1}\; t)^{-1}{}_c{}^{c'} (\overline{t+1}\; t)^{-1}{}_d{}^{d'} \ep_{c'd'}  \Big] \nt
  &\times \frac{6}{4!} \ep_{ABCD} (\overline{t-1})^A_{e'}\ep^{e'f'} (\overline{t-1})^B_{f'}\  (\overline{t+1})^C_{g'}\ep^{g'h'} (\overline{t+1})^D_{h'} \nt
  &= \frac{(\q^+_t)^4}{12}  [(t,r,s-1,s)(r,s-1,s,t-1)]^2 \   \vev{t-1\; t}^2\  \frac{(\mathbf{t-1\; t+1})}{(\mathbf{t-1\; t}) (\mathbf{t+1\; t})}  \,,
\end{align}
where we have used the notations \p{2w}, \p{3.33} and relation \p{2w''}.   If we use the explicit harmonic coordinates \p{cos}, the harmonic factor in \p{resu'} becomes
\begin{align}\label{}
\frac{(\mathbf{t-1\; t+1})}{(\mathbf{t-1\; t}) (\mathbf{t+1\; t})} =  \frac{y^2_{t-1\; t+1}}{y^2_{t-1\; t}\ y^2_{t+1\; t}}\,.
\end{align}
Inserting this result  in \p{NMH} and using the factorized expressions \p{5.3} for the twistor conformal invariants,  we obtain Eq.~\p{NMfin}.
 
\section{Appendix: Boundary case} \label{bot}

In this appendix we treat the special case $|s-t|=2$ and show that the residues of the NMHV amplitude and of the correlator with one Lagrangian insertion vanish.

The analysis in Section~\ref{rri} of the $R$ invariant and its residue was done for generic values of the labels $s$ and $t$.  More care is needed in the special case $|s-t|=2$. For definiteness, let us take $s=t-2$. We refer the reader to \cite{Korchemsky:2009hm} for the detailed analysis and here just state the result:
\begin{align}\label{vares}
{\rm Res}_{x^2_{t-2,t}=0}\ R_{r,t-2,t}(\q^A_{t-2} = \q^+_{t-1} =\q^+_{t+1} = 0)   =0\,.
\end{align}

This behavior is expected, since $x^2_{t-2,t} \to 0$ is one of the collinear limits of the amplitude. These limits are the same for the NMHV and the MHV amplitudes, hence their ratio should be free from collinear singularities.

Similarly, the residue of the correlator was computed in Section~\ref{esc}   for generic values of the labels $s$ and $t$.  According to \p{vares},  for $s=t-2$ the residue of $R_{r, t-2,t}$ at the pole $1/x^2_{t-2,t}$ vanishes, so we expect this to be the case for the correlator as well. 

However, we have to be careful, because  the denominator $x^2_{s+1,t} =x^2_{t-1,t}$ in the second term in \p{expect} vanishes in the light-cone limit. Let us look in more detail into the potentially singular contribution
\begin{align}\label{}
 &\frac{(x_{t-1,t} x_{t,t+1})^{(\alpha\beta)}}{x^2_{t-1,t+1}}\left[ 
 \frac{(x_{t-3,t} x_{t,t-2})_{(\alpha\beta)}}{ x_{t-3,t}^2  }
 +
  \frac{(x_{t-2,t} x_{t,t-1})_{(\alpha\beta)}}{  x_{t-1,t}^2}
 \right]  \nt
 &\Rightarrow \ \frac{\tr(x_{t-1,t} x_{t,t+1} [x_{t-2,t-1}\; , \;  x_{t,t-1}])}{x^2_{t-1,t+1}   x_{t-1,t}^2}   \nt
 &\Rightarrow \ \frac{(x_{t-1,t} \cdot x_{t,t+1})}{x^2_{t-1,t+1} }\  \frac {(x_{t-1,t} \cdot x_{t-2,t-1})}{  x_{t-1,t}^2}\,.  \label{311}
\end{align} 
The first factor here is non-singular in the light-cone limit, while the second is potentially singular. However, in the present context we are only interested in the combination of the light-cone limit with another singular limit, where the pole $1/x^2_{st}$ appears. In this special limit the two light-like vectors $x_{t-2,t-1}$ and $x_{t-1,t}$ become collinear, therefore the second factor in \p{311} remains finite. 

At the same time, we should remember that in the ratio of correlators \p{specor} there is the nilpotent prefactor $(\q^+_t)^4$. Using the gauge condition $\q^A_s=0$, we find
\begin{align}\label{241}
-\q^A_t = \q^A_{st}=\q^A_{t-2,t} = \ket{t-2} \eta^A_{t-2} + \ket{t-1} \eta^A_{t-1} \,,
\end{align}
hence
\begin{align}\label{}
(\q^+_t)^4 = (\ket{t-2} \eta^{+a}_{t-2/t} + \ket{t-1} \eta^{+a}_{t-1/t})^4 = \vev{t-2\ t-1}^2 ( \eta^+_{t-2/t})^2 ( \eta^+_{t-1/t})^2\,.
\end{align}
The factor $\vev{t-2\ t-1}^2$ makes the whole residue vanish, in accord with the result for the amplitude.

\end{document}